\documentclass[longauth]{aa} 

\usepackage{graphicx}
\usepackage{textcomp,gensymb}
\usepackage{subfigure}
   
\usepackage{upgreek}
\usepackage{enumitem}
\usepackage{epstopdf}
\usepackage{stfloats}
\usepackage{longtable,multirow}
\usepackage[para,online,flushleft]{threeparttable}
\usepackage{pdflscape}
\usepackage{multirow,bigdelim}
\counterwithin*{paragraph}{section}
\usepackage[dvipsnames]{xcolor}
\usepackage{ulem}
\usepackage{cancel}

\let\oldpageref\pageref
\renewcommand{\pageref}{\oldpageref*}
\usepackage[varg]{txfonts}
\usepackage{microtype}

\usepackage{booktabs}
\usepackage{booktabs}
\usepackage{multirow}

\usepackage{footmisc}

\usepackage{environ}         
\usepackage{etoolbox}        
\usepackage{graphicx}        

\newlength{\myl}
\let\origequation=\equation
\let\origendequation=\endequation

\RenewEnviron{equation}{
  \settowidth{\myl}{$\BODY$}                       
  \origequation
  \ifdimcomp{\the\linewidth}{>}{\the\myl}
  {\ensuremath{\BODY}}                             
  {\resizebox{\linewidth}{!}{\ensuremath{\BODY}}}  
  \origendequation
}
\usepackage{hyperref}

\hypersetup{%
  colorlinks=true,
  linkcolor=blue,
  citecolor=blue,
}
\let\oldpageref\pageref
\renewcommand{\pageref}{\oldpageref*}

\let\oldcitet=\citet

\renewcommand{\citet}[1]{\textcolor[rgb]{0,0,1}{\oldcitet{#1}}}
\usepackage{cleveref} 
\crefformat{footnote}{#2\footnotemark[#1]#3}

\newcommand{\CI}{[C\,{\sc i}]}

\newcommand{\CII}{[C\,{\sc ii}]}
\newcommand{\HII}{H\,{\sc ii}}

\def\kms{{\hbox {km\thinspace s$^{-1}$}}} 
\def\Kkms{{\hbox {K\thinspace km\thinspace s$^{-1}$}}} 
 
\def\Ls{{\hbox {$L_{\odot}$}}} 

\def\Ms{{\hbox {$M_{\odot}$}}} 
 
\def\AV{{\hbox{$A_\mathrm{V}$}}}

\begin{document} 


\title{HCN emission from translucent gas and UV-illuminated cloud edges revealed by wide-field IRAM\,30m maps of the Orion~B~GMC}
\subtitle{Revisiting its role as a tracer of the dense gas reservoir for star formation}
\titlerunning{Extended HCN \textit{J}\,=\,1-0 emission in Orion B} 
\authorrunning{Santa-Maria, Goicoechea et al.}


 \author{
 M. G. Santa-Maria\inst{\ref{CSIC}} %
 \and J. R. Goicoechea\inst{\ref{CSIC}} %
 \and J. Pety\inst{\ref{IRAM},\ref{LERMA/PARIS}} %
 \and M. Gerin\inst{\ref{LERMA/PARIS}} %
 \and J. H. Orkisz\inst{\ref{IRAM},\ref{Chalmers}} %
 \and F. Le Petit\inst{\ref{LERMA/MEUDON}} %
 \and L. Einig\inst{\ref{IRAM},\ref{GIPSA-Lab}}
 \and P. Palud\inst{\ref{CRISTAL},\ref{LERMA/MEUDON}}
 \and \mbox{V. de Souza Magalhaes\inst{\ref{IRAM}}} %
 \and I. Be\v{s}li\'c\inst{\ref{LERMA/PARIS}}%
 \and L. Segal\inst{\ref{LERMA/PARIS},\ref{Toulon}}%
 \and S. Bardeau\inst{\ref{IRAM}} %
 \and E. Bron\inst{\ref{LERMA/MEUDON}} %
 \and P. Chainais\inst{\ref{CRISTAL}} %
 \and \mbox{J. Chanussot\inst{\ref{GIPSA-Lab}}} %
 \and P. Gratier \inst{\ref{LAB}} %
 \and \mbox{V. V. Guzm\'an\inst{\ref{Catholica}}} %
 \and \mbox{A. Hughes\inst{\ref{IRAP}}} %
 \and D. Languignon\inst{\ref{LERMA/MEUDON}} %
 \and F. Levrier\inst{\ref{LPENS}} %
 \and \mbox{D. C. Lis\inst{\ref{JPL}}} %
 \and H. S. Liszt\inst{\ref{NRAO}} %
 \and J. Le Bourlot\inst{\ref{LERMA/MEUDON}} %
 \and \mbox{Y. Oya\inst{\ref{DPTok},\ref{RCEU}}} %
 \and \mbox{K. \"Oberg\inst{\ref{CfA}}} %
 \and \mbox{N. Peretto\inst{\ref{UC}}} %
 \and E. Roueff\inst{\ref{LERMA/MEUDON}} %
 \and \mbox{A. Roueff\inst{\ref{Toulon}}} %
 \and A. Sievers\inst{\ref{IRAM}} %
 \and P.-A. Thouvenin\inst{\ref{CRISTAL}} %
 \and \mbox{S. Yamamoto\inst{\ref{DPTok},\ref{RCEU}}} %
}

 \institute{%
 Instituto de Física Fundamental (CSIC). Calle Serrano 121-123, 28006, Madrid, Spain, \email{miriam.g.sm@csic.es} \label{CSIC}  %
 \and IRAM, 300 rue de la Piscine, 38406 Saint Martin d'H\`eres,  France. \label{IRAM} %
 \and LERMA, Observatoire de Paris, PSL Research University, CNRS, Sorbonne Universit\'es, 75014 Paris, France. \label{LERMA/PARIS} %
 \and Chalmers University of Technology, Department of Space, Earth and Environment, 412 93 Gothenburg, Sweden. \label{Chalmers} %
 \and LERMA, Observatoire de Paris, PSL Research University, CNRS, Sorbonne Universit\'es, 92190 Meudon, France. \label{LERMA/MEUDON} %
 \and Univ. Grenoble Alpes, Inria, CNRS, Grenoble INP, GIPSA-Lab, Grenoble, 38000, France. \label{GIPSA-Lab} %
 \and Univ. Lille, CNRS, Centrale Lille, UMR 9189 - CRIStAL, 59651 Villeneuve d’Ascq, France. \label{CRISTAL} %
 \and Université de Toulon, Aix Marseille Univ, CNRS, IM2NP, Toulon, France. \label{Toulon} %
 \and Laboratoire d'Astrophysique de Bordeaux, Univ. Bordeaux, CNRS,  B18N, Allee Geoffroy Saint-Hilaire,33615 Pessac, France. \label{LAB} %
 \and Instituto de Astrofísica, Pontificia Universidad Católica de Chile, Av. Vicuña Mackenna 4860, 7820436 Macul, Santiago, Chile. \label{Catholica} %
 \and Institut de Recherche en Astrophysique et Planétologie (IRAP), Université Paul Sabatier, Toulouse cedex 4, France. \label{IRAP} %
 \and Laboratoire de Physique de l’Ecole normale supérieure, ENS, Université PSL, CNRS, Sorbonne Université, Université de Paris, Sorbonne Paris Cité, Paris, France. \label{LPENS} %
 \and Jet Propulsion Laboratory, California Institute of Technology, 4800 Oak Grove Drive, Pasadena, CA 91109, USA. \label{JPL} %
 \and National Radio Astronomy Observatory, 520 Edgemont Road, Charlottesville, VA, 22903, USA. \label{NRAO} %
 \and Department of Physics, The University of Tokyo, 7-3-1, Hongo, Bunkyo-ku, Tokyo, 113-0033, Japan. \label{DPTok} %
 \and Research Center for the Early Universe, The University of Tokyo, 7-3-1, Hongo, Bunkyo-ku, Tokyo, 113-0033, Japan. \label{RCEU} %
 \and Harvard-Smithsonian Center for Astrophysics, 60 Garden Street, Cambridge, MA 02138, USA. \label{CfA} %
 \and School of Physics and Astronomy, Cardiff University, Queen's buildings, Cardiff CF24 3AA, UK. \label{UC} %
} %

\date{Received 2023; accepted 2023}

  \abstract
{Massive stars form within dense clumps inside giant molecular clouds (GMCs). Finding appropriate chemical tracers of the dense gas ($n$(H$_2$)\,$>$\, several 10$^4$\,cm$^{-3}$ or A$_\mathrm{V}$\,$>$\,8\,mag) and linking their line luminosity with the star formation rate is of critical importance.}
{Our aim is to determine the origin and physical conditions of the HCN-emitting gas and study their relation to those of other molecules.}
{In the context of the IRAM\,30m\, ORION-B large program, we present 5\,deg$^2$ ($\sim$250\,pc$^2$) HCN, HNC, HCO$^+$, and \mbox{CO~$J$=1$-$0} maps of {the} Orion~B  GMC,   complemented with existing wide-field \CI\,492\,GHz maps, as well as new pointed  observations of  rotationally excited HCN, HNC, H$^{13}$CN, and {HN$^{13}$C}  lines. 
We compare the observed HCN line intensities with radiative transfer models including line overlap effects and electron excitation. {Furthermore, we} study the HCN/HNC isomeric 
{abundance} ratio with updated photochemical models.}
{We spectroscopically resolve the HCN $J$\,=\,1–0 hyperfine structure (HFS) components ({and partially resolved} 
$J$=2$-$1 and 3$-$2 components). We detect anomalous HFS line intensity (and line width) ratios almost everywhere in the cloud. 
About 70\% of the total HCN $J$=1--0 luminosity, \mbox{$L'$(HCN $J$\,$=$\,1$-$0)\,$=$\,110\,K\,km\,s$^{-1}$\,pc$^{-2}$}, arises from  $A_\mathrm{V}$\,$<$\,8\,mag. The \mbox{HCN/CO $J$=1$-$0} line intensity ratio, widely used as a tracer of the dense gas fraction, 
shows a bimodal behavior  
with an inflection point at $A_\mathrm{V}$\,$\lesssim$\,3\,mag typical of translucent gas and illuminated cloud edges. We find that most of the  \mbox{HCN $J$\,$=$\,1--0} emission arises from extended gas with  \mbox{$n$(H$_2$)\,$\lesssim$\,10$^4$~cm$^{-3}$},  and even lower {density gas} if the ionization fraction is \mbox{$\chi_e$\,$\geq$\,10$^{-5}$} and electron excitation dominates. This result contrasts with the prevailing view of \mbox{HCN $J$\,$=$\,1--0} emission as a tracer of dense gas and  explains the 
\mbox{low-$A_\mathrm{V}$} branch of the  \mbox{HCN/CO\,$J$\,$=$1--0} intensity ratio distribution. Indeed, the highest HCN/CO ratios ($\sim$\,0.1) at  $A_\mathrm{V}$\,$<$\,3\,mag correspond to regions of high \CI\,492\,GHz/CO\,$J$\,$=$1$-$0 intensity ratios ($>1$) characteristic of low-density {photodissociation regions}. {The low surface brightness ($\lesssim$\,1~\Kkms) and extended  \mbox{HCN} and \mbox{HCO$^+$\,$J$\,$=$\,1--0} emission scale with~$I_{\rm FIR}$ -- a proxy of the stellar far-ultraviolet (FUV) radiation field -- in a similar way. Together with \mbox{CO\,$J$\,$=$\,1--0}, these lines respond to  increasing $I_{\rm FIR}$  up to $G_0$\,$\simeq$\,20. 
On the other hand, the  bright \mbox{HCN\,$J$\,$=$\,1--0} emission ($>$\,6~\Kkms) from dense gas in star-forming clumps  weakly responds to $I_{\rm FIR}$  once the FUV  field becomes too intense ($G_0$\,$>$\,1500).
In contrast, \mbox{HNC\,$J$\,$=$\,1--0} and \CI\,492\,GHz lines  weakly respond to $I_{\rm FIR}$ for all $G_0$.}
The different power law scalings (produced by different {chemistries}, densities, and  line excitation regimes) in a single but spatially resolved GMC  resemble the variety of \mbox{Kennicutt-Schmidt law indexes found in galaxy averages.}}
{{Given the widespread and extended nature of the \CI\,492\,GHz emission,  as well as its spatial correlation with that {of} HCO$^+$, HCN, and $^{13}$CO~$J$\,=\,1--0 lines (in this  order), we argue that the edges of GMCs are porous to FUV radiation from nearby massive stars.}
{Enhanced FUV radiation} favors the formation and excitation of HCN on large scales, not only in dense star-forming clumps, and it leads to a
relatively low value of the dense gas mass to total luminosity ratio,
\mbox{$\alpha$\,(HCN)\,$=$\,29\,M$_\odot$\,/\,(\Kkms pc$^{2}$)} in Orion~B.
As a corollary for extragalactic studies, we conclude that
high \mbox{HCN/CO~$J$=1--0} line intensity ratios do not always imply the presence of dense gas, which
may be better traced by HNC than by HCN.}
 

   \maketitle
%
\section{Introduction}\label{sec:intro}
Massive stars dominate the injection of radiative energy into their interstellar environment through ultraviolet (UV) photons. 
{They} form within cold clumps of dense  gas inside giant molecular clouds \citep[GMCs, e.g.,][]{LadaE92,Lada2003}. Observations reveal that the star formation rate (SFR) is close to linearly proportional to the cloud mass above a visual extinction threshold of $A_\mathrm{V}\!\simeq$\,8~mag \citep{Schmidt1959,Schmidt1963, Kennicutt1998a,Kennicutt1998b,Lada2010, Evans2020}{, which corresponds} 
to an approximate  gas density threshold of $n$(H$_2$)\,$>$\,10$^4$\,cm$^{-3}$ {\citep[e.g.,][]{Bisbas2019}}. 
{In galaxies}, the far-infrared (FIR) {dust luminosity \citep[$L_{\rm FIR}$, defined between 40~$\upmu$m and 500~$\upmu$m, see Sect.~\ref{subsec-Herschel} and][]{Sanders1996}} provides a measure of the {SFR}, especially in starbursts \citep[e.g.,][]{Kennicutt1998a}. 
The {$L_{\rm FIR}$} and the \mbox{HCN\,$J$\,$=$\,1--0} line luminosity ($L_{\rm HCN\,1-0}$) are linearly correlated over a broad range of spatial scales and galaxy types, from spatially resolved star-forming clumps  ($L_\mathrm{FIR}\!\simeq$10$^4$~$L_\odot$) to ultraluminous infrared galaxies (ULIRGs, with $L_\mathrm{FIR}\!\geq$10$^{11}$~$L_\odot$) \citep{Solomon1992,Gao2004b,Wu2010}. These studies
suggest that $L_{\rm HCN\,1-0}$ is a good tracer of the dense star-forming gas mass. 
However,  the $L_{\rm CO\,1-0}$  to $L_\mathrm{FIR}$ luminosity ratio in ULIRGs
 is lower than in normal  galaxies. This leads to a superlinear 
relationship \mbox{$L_{\rm FIR}$\,$\propto$\,$L_{\rm CO}^{N\,>\,1}$}
 \citep[e.g.,][]{Kennicutt1998b,Gao2004a}.
The above relations are observational proxies of the so-called  {Kennicutt-Schmidt} (KS) relationship,
{\mbox{$\Sigma_{\rm SFR}$\,=\,$a$\,$\Sigma_{\rm H_2}^{N}$}}, where 
$\Sigma_{\rm SFR}$ and $\Sigma_{\rm H_2}$ are the SFR and molecular gas surface densities. {One obtains $N$$\approx$1.5 
assuming that a roughly constant fraction of the gas present in molecular clouds is subsequently 
converted into stars each free-fall time} \mbox{\citep[e.g.,][]{Madore1977,Elmegreen2002}}.

HCN has a high dipole moment (\mbox{$\mu_e$\,$=$\,2.99~D}),  30 times higher than that of CO. {The \mbox{HCN\,$J$\,$=$\,1--0}  line is commonly used as a tracer of dense gas because} of its high critical density ($n_\mathrm{cr}$), 
the density for which the net radiative decay from $J$=1 equals the rate of collisional 
(de-)excitations out of the upper level. This results in 
 \mbox{$n_\mathrm{cr}$(HCN $J$=1--0)\,$\simeq$\,3$\times$10$^5$~cm$^{-3}$} 
 for collisions with H$_2$ at 20~K ({see Table~\ref{tab:spec_n} for references on spectroscopy and collisional rate coefficients}).  
However, as lines become optically thick, radiative trapping becomes important, leading to 
lower effective critical densities {\citep[{$n_\mathrm{cr,\,eff}$;} e.g.,][]{Evans99,Shirley2015}}.

The end $^{14}$N atom has a large nuclear electric quadrupole  moment  \citep[][]{Okinski1968} and nuclear spin $I$=1. The large quadruple moment coupling with the molecular rotation induces a hyperfine splitting of each rotational level ($J$) of HCN, in three hyperfine levels $F$\,(=\,$I$\,$+$\,$J)$ that vary between $|I-J|$ and $I+J$, except for $J$=0 which {only has a single level.} 
The rotational transition \mbox{$J$\,=\,1--0} splits into three hyperfine transitions: \mbox{$F$=0--1}, \mbox{$F$=2--1}, and \mbox{$F$=1--1}, separated by -7.1~\kms and +4.9~\kms from the central component \mbox{$F$=2--1}, respectively \citep[{e.g.,}][]{Ahrens02,Goicoechea2022}.  The three {hyperfine structure} (HFS) lines of the \mbox{$J$=1--0} transition are usually well spectrally resolved by observations toward GMCs of the {G}alactic disk.  In principle, this is convenient since the relative \mbox{HCN~$J$=1--0} HFS line intensity ratios can provide the  line opacity  
and the excitation temperature ($T_\mathrm{ex}$), thus avoiding the need to observe isotopologues or  multiple-$J$ lines. 
However, only in the optically thin limit ($\tau\rightarrow0$) are
the relative  
HFS line intensity ratios 
equal to their relative line strengths (1:5:3), is
the linewidth  
the same for the three HFS lines, and is
$T_\mathrm{ex}$  
exactly the same for the three HFS transitions, with $T_\mathrm{ex}=T_\mathrm{k}$ if 
{local thermodynamic equilibrium} (LTE) prevails. For optically thick lines, the line intensity ratios approach unity. Overall, the expected \mbox{HCN~$J$=1--0} HFS line intensity ratio ranges are \mbox{$R_{02}$=$W$($F$=0--1)/$W$($F$=2--1)=[0.2, 1]} and \mbox{$R_{12}$=$W$($F$=1--1)/$W$($F$=2--1)=[0.6, 1]}, {where we define the integrated line intensity as \mbox{$W$\,=\,$\int T_{\rm mb}(v)\,dv$} (in K\,km\,s$^{-1}$)}. \mbox{Interestingly}, the observed interstellar line ratios  are usually outside these ranges. This is called anomalous HCN emission.

Early studies of the HCN~$J$=1--0 emission from warm GMCs revealed anomalous $R_{12}$\,<\,0.6 and $R_{02}\,\!\gtrsim$\,0.2 ratios \citep{Wannier1974,Clark1974,Gottlieb1975}. Cold dark clouds show  HFS anomalies toward embedded cores \citep{Walmsley1982} and 
{around them} \citep{Cernicharo1984HCN}. More modern  observations confirm the ubiquity of the HCN~$J$=1--0 HFS line intensity anomalies 
toward low- and high-mass star-forming cores \citep[e.g.,][]{Fuller91,Sohn07,Loughnane2012,Magalaes2018}.

Since the first detection of anomalous \mbox{HCN $J$=1--0} HFS  emission, several theoretical studies have tried to explain {its origin}. Proposed explanations are as follows: radiative trapping combined with efficient collisional excitation from $J$=0~to~2  \citep{Kwan1975};  HFS line overlap effects 
\citep{Guilloteau1981,Daniel2008,Keto2010}; resonant scattering by
low density halos \citep{Gonz-Alf1993}; and line overlaps together with electron-assisted weak collisional excitation  \citep{Goicoechea2022}. 
A proper treatment of the HCN excitation in GMCs 
{thus} requires (i) the radiative effects
induced by {high line opacities and}  HFS line overlaps to be modeled and (ii) the HFS-resolved inelastic collision rate coefficients to be known. Recent developments include 
collisions of HCN with $p$-H$_2$, $o$-H$_2$, and $e^-$ \citep{Faure2007,Faure2012,Vera2017,Magalaes2018,Goicoechea2022}.
\begin{figure*}[t]
    \centering
    \includegraphics[width=0.7\textwidth]{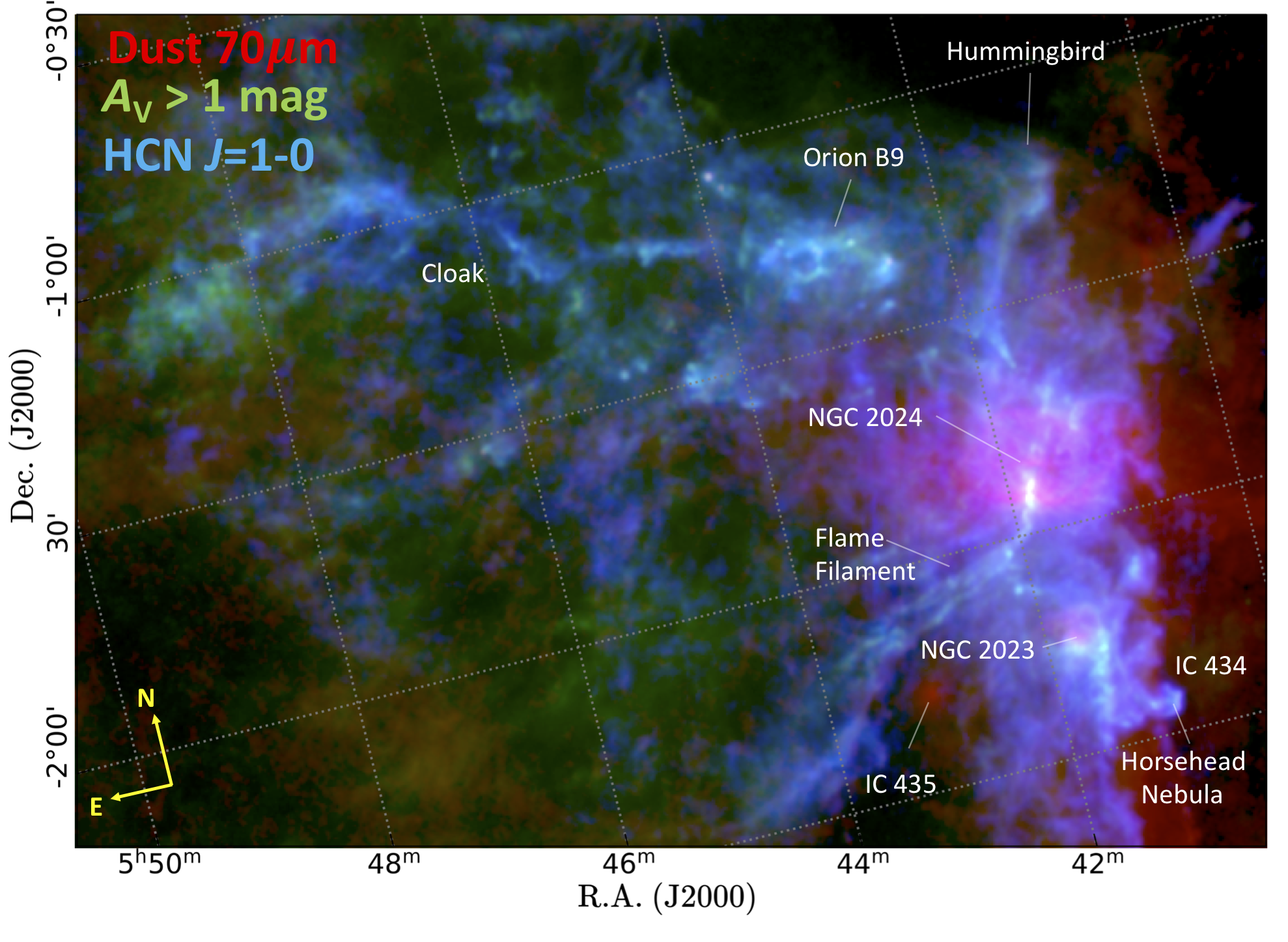} \vspace{-0.2cm}
    \caption{{Composite image of the $\sim$\,5\,deg$^2$ area  mapped in Orion~B. 
    Red color represents  the PACS~70~$\upmu$m  emission tracing FUV-illuminated extended warm dust. Green color represents the cloud depth in
     magnitudes of visual extinction, $A_\mathrm{V}$ $\propto$\,$N$(H$_2$).
     Blue color represents the HCN~$J$=1$-$0 line intensity.
    We note that outside the main filaments most of the 
    \mbox{HCN~$J$=1$-$0 emission} is at
    $A_\mathrm{V}$\,$<$\,4~mag.}}
    \label{fig:rgb_orionb}
\end{figure*}

\begin{table*}[h]
\centering

\caption{Spectroscopic parameters of  the  lines
{studied in this work} {\citep[from][and references therein]{Endres16}},
critical densities for collisions with $p$-H$_2$ and electrons at 20\,K (if LTE prevails, 99.82\,\% of H$_2$ is in $para$ form), and critical fractional abundance of electrons (see text).}
\label{tab:spec_n}
\vspace{-0.1cm}
\begin{threeparttable}
\resizebox{0.9\textwidth}{!}{%
\begin{tabular}{lcccccccccc@{\vrule height 7pt depth 2pt width 0pt}}
\toprule
Species        & \multicolumn{1}{c}{Transition}  & Frequency    & $E_\mathrm{u}$/$k_\mathrm{B}$ &  $A_\mathrm{ul}$ & $n^\mathrm{H_2}_{\mathrm{cr},u}$\,($T_\mathrm{k}$=20 K) & $n^e_{\mathrm{cr},u}$\,($T_\mathrm{k}$=20 K) & $\chi^*_{\mathrm{cr},e}$ & Ref. collisional rates\\
        &             & {[}GHz{]}    & {[}K{]}  & {[}s$^{-1}${]} & {[}cm$^{-3}${]}& {[}cm$^{-3}${]} & & $para$-H$_2$, e$^-$\\ \midrule
        
HCN & $J$=1--0  & \phantom{0}88.63185  & \phantom{0}4.25  &  2.41$\times$10$^{-5}$   &  2.7$\times$10$^{5}$ &  3.2 &  1.2$\times$10$^{-5}$ & a,b \\
HCN       & $J$=2--1  & 177.26122 & 12.76 & 2.31$\times$10$^{-4}$  &  2.3$\times$10$^{6}$ & 38 & 1.7$\times$10$^{-5}$ &\\
HCN       & $J$=3--2  & 265.88650 & 25.52 &  8.36$\times$10$^{-4}$  &  5.8$\times$10$^{6}$  & 168 & 2.9$\times$10$^{-5}$  & \\
HCN     & $J$=4--3  & 354.50548 & 42.53 &  2.05$\times$10$^{-3}$  & 1.6$\times$10$^{7}$  & 485 & 3.0$\times$10$^{-5}$ &  \\
\midrule
HNC     & $J$=1--0          & \phantom{0}90.66357  & \phantom{0}4.35  & 2.69$\times$10$^{-5}$   & 7.0$\times$10$^{4}$ & 3.6 & 5.2$\times$10$^{-5}$ & a,b \\
HNC     & $J$=3--2 & 271.98114 & 26.11 &  9.34$\times$10$^{-4}$   & 2.2$\times$10$^{6}$  & 191 & 8.6$\times$10$^{-5}$ & \\
\midrule
HCO$^+$ & $J$=1--0          & \phantom{0}89.18852  & \phantom{0}4.28  &  4.19$\times$10$^{-5}$    & 4.5$\times$10$^{4}$ & 2.0 & 4.4$\times$10$^{-5}$ & c,d\\
HCO$^+$ & $J$=2--1    & 178.37506 & 12.84 & 4.02$\times$10$^{-4}$   & 4.0$\times$10$^{5}$ & 21 & 5.3$\times$10$^{-5}$ & \\
HCO$^+$ & $J$=3--2  & 267.55763 & 25.68 &  1.45$\times$10$^{-3}$   & 1.5$\times$10$^{6}$  & 84 & 5.8$\times$10$^{-5}$ & \\
HCO$^+$ & $J$=4--3 & 356.73422 & 42.80 &  3.57$\times$10$^{-3}$  & 3.1$\times$10$^{6}$ & 223 &  7.1$\times$10$^{-5}$  & \\
\midrule
CO      & $J$=1--0        & 115.27120 & \phantom{0}5.53 &  7.20$\times$10$^{-8}$   & 5.7$\times$10$^{2}$ & ... & ... & e \\
\midrule

C$^0$ & $^3P_1$-$^3P_0$ & 492.16065 & 23.62 & 7.88$\times$10$^{-8}$ & 5.5$\times$10$^2$ & 410 & 7.0$\times$10$^{-1}$ & f,g \\

\bottomrule 
\end{tabular}}
\tablefoot{We define the critical density  as the H$_2$ (or $e^-$) density at which $A_{ul}$ equals the sum of all upward and downward collisional rates from the upper level. That is, $n_\mathrm{cr} = A_{ul}\,/\,\sum_{i\neq u}\gamma_{ui}$. For collisions with electrons, we consider only dipole-allowed transitions. We define the critical fractional abundance of electrons as $\chi^*_{\mathrm{cr},e}$ = $n^e_\mathrm{cr}$\,/\,$n^\mathrm{H_2}_\mathrm{cr}$. References: (a)~\cite{Vera2017}; (b)~\cite{Faure2007}; (c)~\cite{Denis2020}; (d)~\cite{FaureHCOp2007,FaureHCOp2009}; (e)~\cite{Yang2010}; (f)~\cite{Schroder1991}; (g)~\cite{Johnson1987}. \vspace{-0.2cm}}
\end{threeparttable}
\end{table*}

Mapping  large areas of nearby molecular clouds (a few hundred pc$^2$) in molecular rotational lines  different than CO, and at  the high spatial resolution ($<$\,0.1\,pc) needed to separate the emission from the different cloud component (cores, filaments, and ambient gas), has always been a difficult challenge. Recent surveys of GMCs, sensitive to the  line emission from star-forming clumps and their environment, suggest that a significant fraction of the \mbox{HCN $J$\,$=$\,1--0}  emission stems from  low visual extinctions 
\citep[$A_\mathrm{V}$, {i.e., from low density gas}; e.g.,][]{Pety17a,Shimajiri2017,Kauffmann2017,Evans2020,Barnes2020,Tafalla2021,Patra2022,Dame2023}. Even {the most} translucent \citep{Turner1997} and  diffuse molecular clouds ($A_\mathrm{V}$\,$<$\,1~mag) show \mbox{HCN\,$J$\,=1--0} 
emission {and} absorption lines \citep{Liszt2001,Godard2010} compatible with HCN abundances similar to those inferred in dense molecular clouds, 10$^{-8}$~--~10$^{-9}$ \citep[e.g.,][]{Blake1987}.

Giant molecular clouds are illuminated by UV  photons from nearby massive stars and by the interstellar radiation field. They are also bathed by cosmic ray particles. 
{Ultraviolet} 
radiation favors high electron abundances (the ionization fraction or $\chi_e$) in the first \mbox{$A_\mathrm{V}$\,$\approx$\,2--3~mag} 
 into the cloud \citep[e.g.,][]{Hollenbach1991}. In these cloud surface layers, most electrons arise from the photoionization of carbon atoms. Hence, \mbox{$\chi_e$\;$\simeq$\;$\chi$(C$^+$)\;$\simeq$ a few 10$^{-4}$} \citep{Sofia2004}. At intermediate cloud depths, from \mbox{$A_\mathrm{V}\!\approx$\,2-3} to \mbox{4-5~mag} depending on the gas density, cloud porosity to UV photons \citep{Boisse1990}, and  abundance of low ionization potential elements such as sulfur determine the ionization fraction  \citep[e.g., \mbox{$\chi_e$\;$\simeq$\;$\chi$(S$^+$)\;$\simeq$ a few 10$^{-5}$}
in Orion\,A;][]{Goicoechea2021}. At much larger $A_\mathrm{V}$, deeper inside the dense cores shielded from external UV radiation, $\chi_e$ is much lower, $\sim$\,10$^{-7}$--10$^{-8}$. These $\chi_e$ values apply to GMCs in the disk of the galaxy exposed to standard cosmic ray ionization rates, \mbox{$\zeta_{\rm CR}$\,=\,10$^{-17}$--10$^{-16}$\,s$^{-1}$}
\citep[][]{Guelin1982,Caselli1998,Goicoechea2009}.

 More than 45 years ago, \citet{Dickinson1977} suggested that electron collisions contribute to the rotational excitation of very polar neutral molecules \citep[see also][]{Liszt2012}. These molecules have large cross sections for collisions with electrons 
\citep{Faure2007}. This implies that the rate coefficients of inelastic collisions with electrons can be at least three orders of magnitude greater than those of collisions with H$_2$ and H. Hence, electron collisions contribute to, and even dominate, the excitation of these molecules when (i) $\chi_e$  is higher than the critical fractional abundance {of electrons}, \mbox{$\chi^*_\mathrm{cr}$(e$^-$)\,=\,$n_\mathrm{cr}$(e$^-$)\,/\,$n_\mathrm{cr}$(H$_2$)}, and (ii) the gas density $n$(H$_2$) is lower than the critical density for collisions with H$_2$, \mbox{$n$(H$_2$)\;$<$\;$n_\mathrm{cr}$(H$_2$)}. 
For \mbox{HCN~$J$\,=\,1--0}, this implies \mbox{$\chi_e$\,$\gtrsim$\,10$^{-5}$} and \mbox{$n$(H$_2$)$\;\lesssim$\;10$^5$~cm$^{-3}$} \citep{Dickinson1977,Liszt2012,Goldsmith17,Goicoechea2022}. 
Table~\ref{tab:spec_n} lists the frequency, upper level energy,  $n_\mathrm{cr}$, and critical fractional abundance $\chi^*_\mathrm{cr}$(e$^-$) of the lines
 relevant to this work.

Galactic and extragalactic studies  typically overlook the role of electron  excitation \citep[e.g.,][]{Yamada2007,Behrens2022}. However, the ionization fraction in the 
{interstellar medium} (ISM) of galaxies can be very high because of enhanced cosmic ray ionization rates and  \mbox{X-ray} fluxes driven by accretion processes in their nuclei  {\citep[][]{Lim2017}}. Mapping nearby GMCs in our Galaxy  offers a convenient template to spatially resolve and  quantify the amount of low surface brightness HCN emission (affected by electron excitation)  not directly associated with dense star-forming clumps. This emission component  is  usually not considered in extragalactic studies  \citep[e.g.,][]{Papadopoulos2014,Stephens2016}.

Here we carry out a detailed analysis of the extended \mbox{HCN~$J$\,$=$\,1--0} line emission, and that of related molecules, obtained in the framework of the 
 large program Outstanding Radio-Imaging of Orion\,B (ORION-B) over 5\,deg$^2$ {(see \mbox{Fig.~\ref{fig:rgb_orionb}} for an overview)}. These maps cover five times larger areas than those originally presented by \cite{Pety17a}. We revisit  the diagnostic power of the  \mbox{HCN~$J$\,$=$\,1--0} emission as a tracer of the dense molecular gas reservoir for star formation. This paper is organized as follows. In Sect.~\ref{sec:observations}, we introduce the most relevant  regions in Orion\,B as well as the observational dataset. In Sect.~\ref{sec:results}, we present and discuss the spatial distribution of different tracers. In Sect.~\ref{sec:analysis}, we analyze the extended HCN emission and derive gas physical conditions. In Sect.~\ref{sec:Meudon}, we reassess the chemistry of HCN and HNC in 
\mbox{FUV-illuminated gas}. In Sect.~\ref{sec:discussion}, we discuss the relevance and properties of the low-density extended cloud component, we determine the dense gas mass conversion factor $\alpha$(HCN), and discuss the {\mbox{$I_{\rm FIR}$--$W$}} scalings we find between different emission lines and FIR dust intensities. In Sect.~\ref{sec:conclusions}, we summarize our findings and give our conclusions.

\section{Observations}\label{sec:observations}
\subsection{The Orion\,B GMC}

\begin{figure*}[!hp]
       
    \begin{center}
    \includegraphics[width=\textwidth]{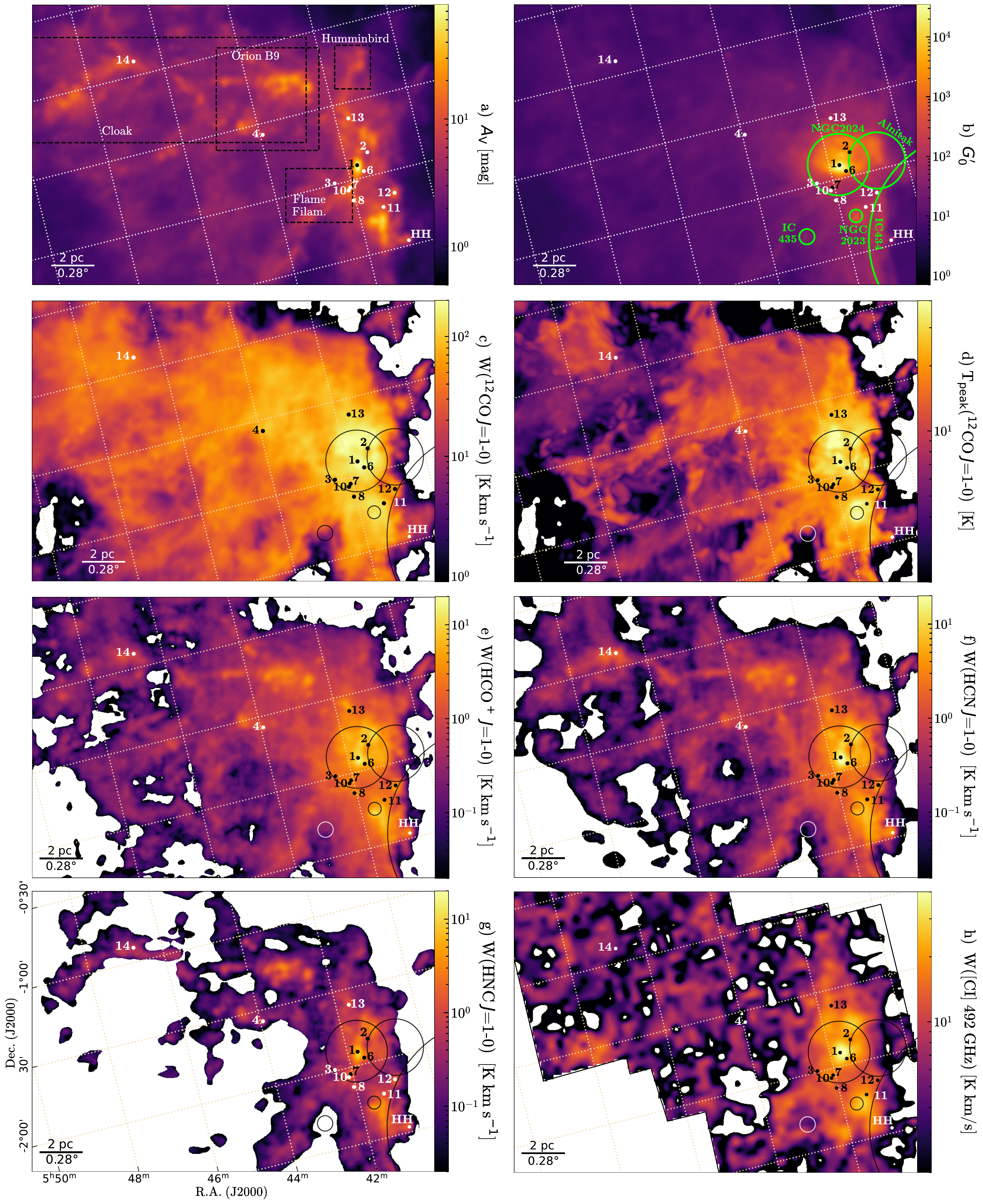}
    \end{center}
    \vspace{-0.5cm}
    \caption{Maps of Orion\, B in different tracers. (a) Visual extinction $A_\mathrm{V}$, 
    (b) Approximate FUV field, $G_0'$ (see text), (d) $^{12}$CO~\textit{J}=1--0 peak temperature (in K). 
    (c) and (e) to (h)~$^{12}$CO, HCO$^+$, HCN, 
    HNC~\textit{J}=1--0, and \CI~492~GHz \cite[from][]{Ikeda2002} integrated line intensity maps (in K~\kms)  spatially smoothed to an angular resolution of $\sim$2$'$. Dashed black boxes mark the Cloak, Orion B9, Hummingbird, and Flame filament. Circles mark the extension of the \HII\; regions in NGC~2024, NGC~2023, IC 434, IC 435, and around the star Alnitak. The {HH} dot marks the position of the Horsehead PDR, the projection center of the maps.}
    \label{fig:fullmaps120}
\end{figure*}

Orion\, B, in the Orion complex, east of the Orion Belt stars,  is one of the nearest GMCs \citep[e.g.,][]{Anthony1982}. 
{Here we adopt a distance\footnote{Interferometric observations of the $\sigma$~Ori system provides a distance of $\sim$\,388~pc \citep{Schaefer2016}. {Recent determinations using GAIA also
estimate $\sim$\,400\,pc \cite[e.g.,][]{Zucker19,Rezaei20}.}} of $d$\,=\,400\,pc}.
 Orion B is a  good template to study the star formation processes in the disk of a normal galaxy. This is an active but modest  star-forming region  {\citep[{with a} low SFR$\sim$1.6$\times$10$^{-4}$~\Ms~yr$^{-1}$ and low star-formation efficiency, SFE\,$\sim$\,1\%, e.g.,][]{Lada2010,Megeath2016,Orkisz2019}} that contains  thousands of  dense molecular cores: starless, prestellar, and protostellar cores \citep[e.g.,][]{Konyves2020}. Massive star formation  is highly concentrated in four main regions:  NGC~2071 and NGC~2068 in the northeast, and  NGC~2023 and NGC~2024 in the southwest.
 Table~\ref{tab:HII_reg} summarizes the properties of the massive stars that create \HII\;regions in the  field. Figure~\ref{fig:fullmaps120}b shows  the position and extent of
 these \HII\;regions (marked with circles). 
Orion~B hosts a complex network of  filaments. The main and longest filaments are the Flame and Hummingbird filaments, Orion~B9, and the Cloak \citep{Orkisz2019,Gaudel2022}.  {
Appendix~\ref{App:regions} outlines the main properties of these regions}.

\begin{table}[ht]
\centering
\caption{Properties of the massive stars creating \HII\;regions.}
\label{tab:HII_reg}
\vspace{-0.2cm}
\begin{threeparttable}

\resizebox{0.44\textwidth}{!}{%
\begin{tabular}{lcccc}
\toprule
\HII\;region & Spectral type   & Distance$^a$    & Radius$^b$  \\ 
                &                  & [pc] & [arcmin]  \\\midrule
 IC 434 & O9.5V B        & 388$\pm$1 & 42 \\
 NGC 2024    & O9.5V         & 415      & 12 \\
 Around Alnitak          & O9.7Ib+B0III C  & 290$\pm$21 & 11 \\
 IC 435   & B5V D          & 170$\pm$37 & 3 \\
NGC 2023 & B1.5V C         & 360$\pm$35 & 2.5 \\
\bottomrule 
\end{tabular}}
\vspace{-0.2cm}
\tablefoot{$^a$ Distance to the ionizing star  \citep[as in][]{Pety17a}. $^b$ Radius of the circles draw in \mbox{Fig.~\ref{fig:fullmaps120}b} 
\cite[as in][]{Gaudel2022}. \vspace{-0.4cm}}
\end{threeparttable}
\end{table}

\begin{table}[ht]

\caption{Representative environments of our pointed observations. We sort these
positions in decreasing order of $T_{\rm rot}$(HCN) (see also Table~\ref{tab:TTNN}).} 
\label{tab:Positions}

    \centering
    
    \begin{threeparttable}
    \resizebox{0.46\textwidth}{!}{%
    \begin{tabular}{lccc@{\vrule height 9pt depth 4pt width 0pt}}
    \toprule \vspace{-2pt} 
    Pos. &  \multicolumn{2}{c}{(R.A., Dec.)}        &  Environment  \\
        &   \multicolumn{2}{c}{[h:m:s, $\degree$:$\arcmin$:$\arcsec$]}  & \vspace{-2pt} \\
    \midrule

    \#1  &  \multicolumn{2}{c}{{5:41:44.4, -1:54:45.4}}      & \begin{tabular}[c]{@{}c@{}}Protostellar core \\ NGC~2024(FIR-4) \end{tabular}     \vspace{3pt} \\

   \#2   & \multicolumn{2}{c}{{5:41:24.1, -1:50:49.5}}  & \begin{tabular}[c]{@{}c@{}}Irradiated prestellar core \\ (NGC~2024 bubble) \end{tabular}   \vspace{3pt}  \\

    \#3   & \multicolumn{2}{c}{{5:42:25.3, -1:59:33.2}}   & Starless core   \vspace{3pt} \\

    \#4 &  \multicolumn{2}{c}{{5:43:55.9, -1:34:21.4}}  & Orion~B9 (south)   \vspace{3pt} \\
    
    HH PDR  &  \multicolumn{2}{c}{{5:40:53.9, -2:28:00.0}}  & \begin{tabular}[c]{@{}c@{}} Warm PDR \end{tabular}   \vspace{3pt} \\

    \#6  &  \multicolumn{2}{c}{{5:41:36.5, -1:57:38.5}}  & \begin{tabular}[c]{@{}c@{}}NGC 2024 south: \\ dust ridge\end{tabular}   \vspace{3pt} \\

     \#7    &  \multicolumn{2}{c}{{5:42:03.2, -2:02:35.8}}   & \begin{tabular}[c]{@{}c@{}} Prestellar core \\ in the Flame Filament\end{tabular}   \vspace{3pt} \\

    \#8   &  \multicolumn{2}{c}{{5:42:02.9, -2:07:45.2}}  & Protostellar core (YSO)  \vspace{3pt}  \\

     HH Core  &  \multicolumn{2}{c}{{5:40:55.6, -2:27:38.0}}  & \begin{tabular}[c]{@{}c@{}} DCO$^+$ emission peak \end{tabular}   \vspace{3pt} \\

    \#10    & \multicolumn{2}{c}{{5:42:06.8, -2:03:33.5}}   & \#7 surroundings   \vspace{4pt} \\

    \#11   &    \multicolumn{2}{c}{{5:41:20.2, -2:13:03.6}}   &  \begin{tabular}[c]{@{}c@{}}NGC~2023 north\\ surroundings\end{tabular}   \vspace{3pt} \\

    \#12  &  \multicolumn{2}{c}{{5:40:58.1, -2:08:41.9}} & Prestellar core   \vspace{3pt}  \\

     \#13  &  \multicolumn{2}{c}{{5:41:40.0, -1:36:13.6}}  & Filament surroundings   \vspace{3pt} \\

    \#14  &  \multicolumn{2}{c}{{5:46:43.7, -0:54:22.6}}   & Cloak  \vspace{-3pt} \\
    
    \bottomrule
    \end{tabular}}
    \end{threeparttable}
\end{table}

\begin{figure*}[!ht]
    
    \centering
    \includegraphics[height=14.cm]{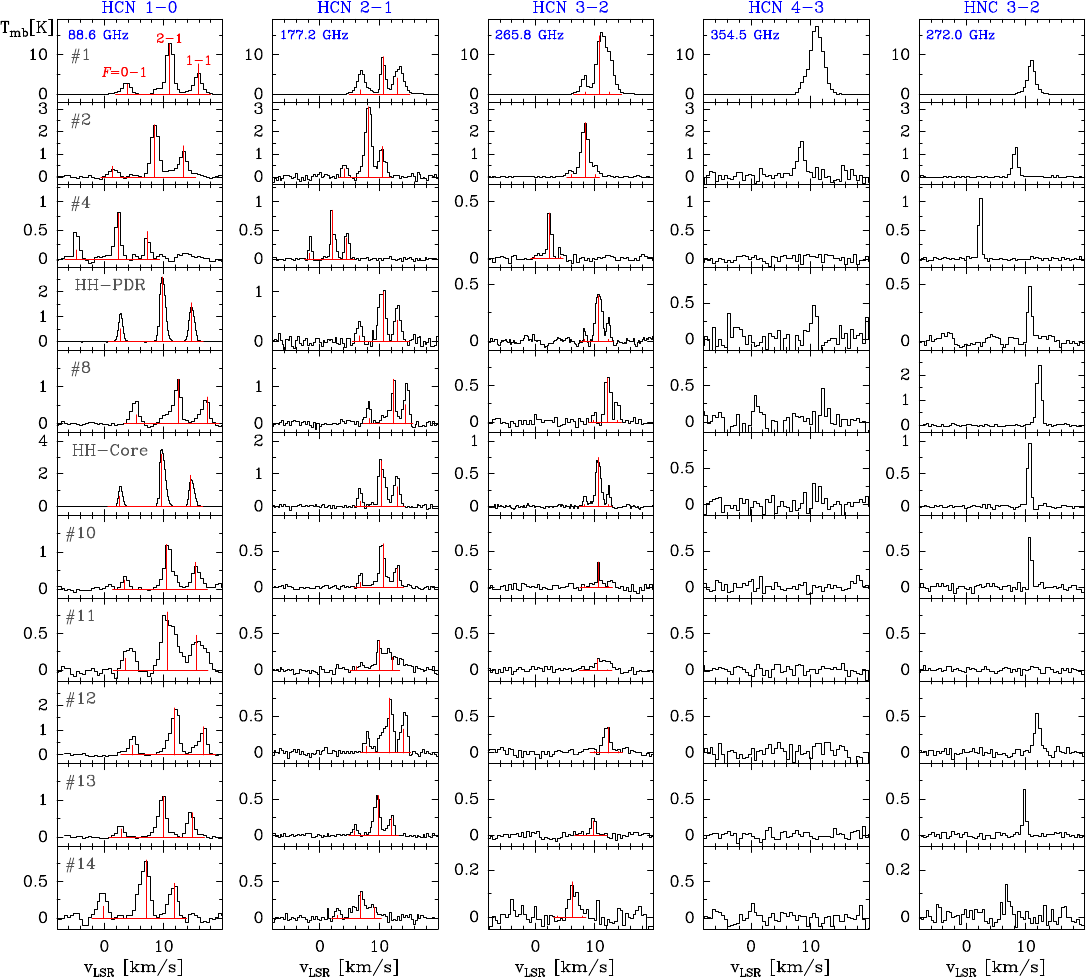}
    \caption{Selection of HCN $J$\,=\,1--0 to 4--3, and HNC~$J$=3--2 line detections toward representative cloud environments in Orion\,B. Red lines show the expected relative HFS line intensities in the LTE and optically thin limit.} 
    \label{fig:hcn_multiJ_resume} \vspace{-0.3cm}
\end{figure*}

\begin{table*}[ht]
\centering
\caption{Characteristics of the molecular line emission  over 5\,deg$^2$ of Orion\,B.}
\label{tab:lines_stats}
\vspace{-0.2cm}
\begin{threeparttable}
\resizebox{0.95\textwidth}{!}{
\begin{tabular}{@{}lcccccccc}
\toprule
(1) & (2) & (3) & (4) & (5)  & (6) & (7) & (8) &(9)   \\
   Line                & Intensity ($W_\mathrm{{average}}$)      & $T_\mathrm{mb}^\mathrm{peak}$           & $L$uminosity ($L'$)     & $L$uminosity ($L$)   &  $\alpha$\,(molecule)                  &  \multicolumn{2}{c}{$A_\mathrm{cov}$} & $A_\mathrm{cov}^{120''}$/$A_\mathrm{cov}^{30''}$    \\ 
                   & {[}K km/s{]}         & {[}K{]}                    &  {[}\Kkms pc$^2${]}      & [\Ls] & [\Ms/(\Kkms pc$^{2}$)] &  \multicolumn{2}{c}{{[}\%{]}}                                                                     &                                             \\ \midrule
                   &    \multicolumn{2}{c}{(average spectrum) }   &            &     &             & 30''          & 120''                         \\\midrule
                   
$^{12}$CO $J$=1--0  & 30.6\phantom{0}                       & 5.4\phantom{00}                                    & 7300    &  1.7$\times$10$^{0\phantom{-}}$  & \phantom{00}0.43                   & 78                                             & 92                                              & 1.2                                            \\
HCO$^+$ $J$=1--0    & \phantom{0}0.52                & 0.14\phantom{0}                                 &   \phantom{0}120  & 1.4$\times$10$^{-2}$   &       \phantom{0}25\phantom{.00}                           & 21                                             & 73                                              & 3.5                             \\
HCN $J$=1--0        & \phantom{0}0.46                 & 0.06\phantom{0}                                  &  \phantom{0}110 & 1.2$\times$10$^{-2}$         &     \phantom{0}29\phantom{.00}                           & 16                                             & 60                                              & 3.8                               \\

HNC $J$=1--0        & \phantom{0}0.15                      & 0.04\phantom{0}                                 & \phantom{00}40   & 4.1$\times$10$^{-3}$         &      \phantom{0}88\phantom{.00}                  & 9                                              & 33                                              & 3.8                               \\

\bottomrule
\end{tabular}%
}\vspace{-0.2cm}
\tablefoot{(2), (3): Line intensity and line peak temperature of the average spectrum in the mapped area.  (4) and (5): Line luminosity in  \Kkms\;pc$^2$ and \Ls\;units. 
(6) Dense gas mass $M_\mathrm{dg}$ to line luminosity ratio, where  $M_\mathrm{dg}$=3.1$\times$10$^3$~\Ms\
(mass at  $A_\mathrm{V}$>8~mag).
(7) and (8): Percentage of the mapped area with 3$\sigma$ line detections.
(9): Ratio of cloud areas with  3$\sigma$ line detection after spatial smoothing.}
\end{threeparttable}
\vspace{-0.2cm}
\end{table*}
\subsection{ORION-B molecular line maps in the {3\,mm band} {and spatial smoothing} }
The ORION-B project (PIs: J. Pety and M. Gerin) is a large program that uses the 
30m telescope of the Institut de Radioastronomie Millim\'etrique (IRAM) to map a large fraction of the Orion~B molecular cloud (5 square-degrees, 18.1$\times$13.7~pc$^2$). Observations were obtained using the EMIR090  receiver  at $\sim$21$''$$-$28$''$
resolution. The FTS backend provided a {channel spacing of 195~kHz (\mbox{0.5-0.7~\kms}\;depending on the line frequency).}
The typical 1$\sigma$ line sensitivity in these maps is $\sim$100~mK per velocity resolution channel.
The full field of view was covered in about 850~hours by (on-the-fly) mapping rectangular tiles  with a position angle of 14$\degree$ in the Equatorial J2000 frame that follows the global morphology of the  cloud.
Data reduction was carried out using \texttt{GILDAS}\footnote{See \url{https://www.iram.fr/IRAMFR/GILDAS} \label{footn:class}}\texttt{/CLASS} and \texttt{CUBE}. This  includes gridding of individual spectra to produce regularly sampled maps, at a common angular resolution of 30$''$, with pixels of 9$''$ size, about one third of the angular resolution of the telescope (half power beam width, HPBW). The projection center of the maps is located on the Horsehead {photodissociation region} (PDR) at 5$\mathrm{h}$40$\mathrm{m}$54.27$\mathrm{s}$, -02$\degree$28\arcmin00.0\arcsec. We rotated the maps  counter-clockwise by 14$\degree$ around this center.
\cite{Pety17a} presents a detailed description of the observing procedure and data reduction.  Here we focus on a global analysis of the \mbox{HCN $J$\,=\,1--0} emission, and its relation to that of \mbox{HNC}, \mbox{$^{12}$CO}, and \mbox{HCO$^+$}.
{Orion~B shows three main velocity  components at the  {local standard of rest} (LSR) velocities $\sim$2.5, $\sim$6, and $\sim$10~\kms~\citep[][]{Gaudel2022}. Here we obtained the 
line  intensity maps (zero-order moment maps) integrating each line spectrum
in the velocity ranges $[-5, +25]$~\kms~(for $^{12}$CO
and HCN~$J$=1$-$0 lines) and $[0, +18]$~\kms~(for HNC and HCO$^+$~$J$=1$-$0).}
We refer to \cite{Gaudel2022} for a thorough analysis of the $^{13}$CO and \mbox{C$^{18}$O~$J$\,=\,1--0} maps and gas kinematics.

To match the resolution of the \CI~492~GHz map (see next Section), and since  we are  interested in the faint and extended molecular emission, we spatially smoothed the original  line maps to an angular resolution of
$\sim$2$'$ ($\sim$0.2\,pc). This allows us to recover a significant fraction of low surface brightness line emission 
 at large spatial scales.
 Spatial smoothing improves the {root mean square} (rms) to {$\sim$25~mK} per velocity channel. 
Thus, it improves the detection limit and signal-to-noise ratio (S/N) of the faint and extended emission.
Figure~\ref{fig:fullmaps120} shows the spatially smoothed  maps.

\subsection{Wide-field \texorpdfstring{\CI}{[CI]}~492~GHz map}

 We complement our molecular line maps  with an existing wide-field map of the ground-state  fine structure line ($^3P_1$–$^3P_1$)  of neutral atomic carbon, the  \CI~492~GHz line, obtained with the Mount Fuji submillimeter-wave telescope. These observations reached a  rms noise of 
 $\sim$0.45~K per 1.0\,km\,s$^{-1}$ velocity channel
 \citep{Ikeda2002}. The angular resolution is $\sim$2$'$.
 {\mbox{Figure~\ref{fig:fullmaps120}h}} shows the \CI~492~GHz   line  integrated intensity map {in  the LSR velocity range $[+3, +14]$~\kms.}

\subsection{Pointed observations of rotationally excited lines}

In addition to the molecular \mbox{$J$=1--0}  line maps, we observed several cloud positions  (see \mbox{Table~\ref{tab:Positions}} for the exact coordinates and details) in rotationally excited lines (\mbox{$J$=2--1}, \mbox{3--2}, and \mbox{4--3}). We obtained these observations also with the IRAM-30m telescope.  We observed 14 positions representative of different cloud environments: cold and dense cores, filaments and their surroundings, PDRs adjacent to \HII~regions and cloud environment. \mbox{Figure~\ref{fig:fullmaps120}} shows the location of these positions. We used the EMIR receivers E150 (at 2~mm), E230 (at 1.3~mm), and E330 (at 0.9~mm) in combination with the FTS backend (195\,kHz spectral resolution). The HPBW varies as $\approx$2460/Freq[GHz]\footnote{See  \href{https://publicwiki.iram.es/Iram30mEfficiencies.}{https://publicwiki.iram.es/Iram30mEfficiencies}\label{foot:irameff}}. For the E1, E2, and E3 bands,
 the HPBW is $\sim$14$''$, $\sim$9$''$, and $\sim$7$''$, respectively. 
We carried out dualband observations combining the E1 and E3 bands, during December 2021 and 2022, under excellent winter conditions ($\sim$1--4~mm of precipitable water vapor, pwv). We obtained the E2 band observations during three different sessions: (i)  December 2021: pwv<4~mm (positions \#1, \#2, \#4,  \#10, and \#11), (ii) March 2022: pwv>6~mm (\#3 and \#7), (iii) May 2022: pwv$\sim$5~mm ( \#6, \#8, \#12, \#13, and \#14).

In order to compare line intensities at roughly the same 30$''$ resolution, we averaged
small raster maps centered around each target position  and  approximately covering the area of a 30$''$ diameter disk  (Fig.~\ref{fig:raster-map-post} in the Appendix shows our pointing strategy). The total integration time per raster-map was $\sim$1~h, including on and off integrations. The achieved rms noises of these observations, merging all observed positions of a given raster-map, are $\sim$22~mK ($J$=2--1), $\sim$20~mK ($J$=3--2), and $\sim$30~mK ($J$=4--3), per 0.5~\kms velocity channel. 
 Table~\ref{tab:obs_bands} in the Appendix summarizes the frequency ranges observed with each backend, the HPBW, and the number of pointings of each raster map.

We analyzed these pointed observations with \texttt{CLASS}.
We subtracted baselines fitting  line-free channels with first or second order polynomial functions. We converted the intensity scale from antenna temperature, $T_\mathrm{A}^*$, to main-beam temperature, $T_\mathrm{mb}$,  as \mbox{$T_\mathrm{mb}$ = $T_\mathrm{A}^* \times$ F$_\mathrm{eff}$/B$_\mathrm{eff}$,} where $F_\mathrm{eff}$ and  B$_\mathrm{eff}$ are the  forward  and beam efficiencies\cref{foot:irameff}. 
\mbox{Figure~\ref{fig:hcn_multiJ_resume}} shows a summary of the spectra.
\mbox{Figures~\ref{fig:HCN-multiJ}} and \ref{fig:HNC-multiJ} 
show the complete dataset.

\subsection{Herschel \texorpdfstring{T$_\mathrm{d}$}{Td}, \texorpdfstring{$A_\mathrm{V}$}{Av}, and \texorpdfstring{$G_0'$}{G0'} maps}\label{subsec-Herschel}

In addition to the molecular and atomic line maps, we also make use of
the dust temperature ($T_\mathrm{d}$) and 850~$\upmu$m dust opacity ($\tau_\mathrm{850\,\upmu m}$) 
maps fitted by \cite{Lombardi2014}  on a combination of Planck and Herschel
data from the {\textit{Herschel}} Gould Belt Survey {(HGBS)}  \citep{Andre2010}.
In {Appendix~\ref{Sect:App-SED}}  we provide additional details on these maps.
  We estimated the (line of sight) visual extinction
from the $\tau_{850\,\upmu\mathrm{m}}$  dust opacity  map 
following \cite{Pety17a}:
\begin{equation}
    A_\mathrm{V} = 2.7\times10^4\; \tau_{850\,\upmu\mathrm{m}}\; [\mathrm{mag}].
\end{equation}
{By taking the $\tau_{850\,\upmu\mathrm{m}}$ error map of \cite{Lombardi2014}, we determine that the mean 5$\sigma$ error of the  $A_\mathrm{V}$ map  is about 0.8\,mag. This value is slightly {above} our  molecular line detection threshold ({$ A_\mathrm{V}$}\,$\simeq$\,0.3-0.4~mag, see Fig.~\ref{fig:lines-Av_hist}). 
Thus, we caution that one can probably not trust any \mbox{$A_\mathrm{V}$--$W$} trend 
below {$ A_\mathrm{V}$}\,$\simeq$\,0.8\,mag}. 
For each line of sight, we determined the FIR surface brightness, $I_\mathrm{FIR}$, 
 from spectral energy distributions (SED) fits, by integrating: 
\begin{equation}
   I_\mathrm{FIR} =  \int I_\nu\;d\nu=\int B_\nu(T_\mathrm{d})[1-e^{-\tau_\nu}]~d\nu,
\end{equation}
from $\lambda$\,=\,40 to 500\,$\upmu\mathrm{m}$. In this expression, $B_\nu$($T_\mathrm{d}$) is the blackbody function, $\tau_\nu$=$\tau_{\nu_\mathrm{850\,\upmu\mathrm{m}}} ( \nu/\nu_{\mathrm{850\,\upmu\mathrm{m}}})^\beta$
is the {frequency-dependent} dust opacity  \cite[we adopt {the same emissivity exponent 
as}][]{Lombardi2014}, and  $T_{\rm d}$ is an effective dust temperature. 

We estimate the strength of the far-UV {(FUV)} radiation field (\mbox{6\,$<$\,$E$\,$<$\,13.6\,eV}), in Habing units ($G_0'$), from $I_\mathrm{FIR}$  using:
\begin{equation}
   G_{0}' =  \frac{1}{2} \frac{I_\mathrm{FIR}\,[\rm erg\,s^{-1}\,cm^{-2}\,sr^{-1}]} {1.3\times10^{-4}}.
   \label{eq:G0}
\end{equation}
In this expression we assume that the FIR continuum emission arises from  
dust grains heated by stellar FUV and visible photons  \mbox{\citep{Hollenbach1999}}.
We use the notation  $G_{0}'$  (meaning approximate $G_{0}$) because this expression is precise for a face-on PDR (e.g., it is valid for NGC\,2024 and NGC\,2023).
Because of their edge-on geometry, \mbox{Eq.~(\ref{eq:G0})} is less accurate for the Horsehead  PDR and IC\,434 front (although  it provides the expected $G_0$ within  factors of a few).
 In addition,
\mbox{Eq.~(\ref{eq:G0})} provides an upper limit to the actual $G_0$ toward embedded star-forming cores
(at high $A_\mathrm{V}$). {These cores emit}  significant non-PDR FIR dust continuum.
To directly compare our line emission maps with the $A_\mathrm{V}$, $G_{0}'$, and $I_\mathrm{FIR}$ maps, we also spatially
smoothed these SED-derived maps to an angular resolution of 120$''$ (Figs.~\ref{fig:fullmaps120}a and b).
 
\begin{table}[h]
\centering
\vspace{-0.2cm}
\caption{SED derived parameters from 5\,deg$^2$ maps of Orion~B.}
\label{tab:SED_values}
\begin{threeparttable}
\resizebox{0.48\textwidth}{!}{%
\begin{tabular}{@{}ccccc@{\vrule height 5pt depth 4pt width 0pt}}
\toprule
Parameter   & Median  & Mean value  & Std. dev.  & Units \\ \midrule
$T_\mathrm{d}$   & 18 & 20  & 4 & K  \\
$A_\mathrm{V}$   & 2.6 & 3.2   & 3 & mag \\
 $I_\mathrm{FIR}$ & {0.002} & {0.02} & {0.3} & erg s$^{-1}$ cm$^{-2}$ sr$^{-1}$  \\ 
 $G_0'$ & {9} & {70} & {1000} & Habing \\ \bottomrule 
\end{tabular}}  
\end{threeparttable}
\vspace{-0.6cm}
\end{table}

\section{Results}\label{sec:results}

\subsection{Spatial distribution of the HCN\,\textit{J}\,=\,1--0 emission,
relation to other chemical species, and \texorpdfstring{$A_\mathrm{V}$}{Av}  and \texorpdfstring{$G_0'$}{G0'} maps} \label{sec:HCNrelothers} 

{Figure~\ref{fig:rgb_orionb} shows a composite RGB image of the mapped area  
($\sim$\,5\,deg$^2$\,=\,250\,pc$^2$). This image shows  extended \mbox{HCN~$J$=1--0} emission far from the main dense gas filaments (where $A_\mathrm{V}$\,$>$\,8\,mag), as well as very extended 70\,$\upmu$m dust emission \mbox{\citep[e.g.,][]{Andre2010}} from FUV-illuminated
 warm grains.} 
\mbox{Figure~\ref{fig:fullmaps120}} shows the spatial distribution of the $^{12}$CO, HCO$^+$, HCN, and \mbox{HNC~$J$=1--0} integrated line intensities, $W$ (also dubbed line surface brightness) at a common  resolution of $\sim$2$'$ 
($\sim$0.2\,pc, thus matching the angular resolution of the \CI~492~GHz map in \mbox{Fig.~\ref{fig:fullmaps120}g}). 
\mbox{$W$(HCN~$J$=1--0)} refers the sum of the three  HFS components. 
The emission from all species peaks toward NGC~2024.
The last column in Table~\ref{tab:lines_stats} shows that spatial smoothing  (increasing the line sensitivity at the expense of lower spatial resolution) allows us
to detect HCN and \mbox{HNC~$J$=1--0} emission from a cloud area nearly four times
bigger than from maps at $\sim$30$''$ resolution. For \mbox{CO~$J$=1--0} (very extended emission) 
the recovered area is  smaller. 
 \mbox{CO~$J$=1--0} shows the most  widespread emission. It traces the most extended and translucent gas, arising from  90$\%$ of the mapped area.
 HCO$^+$ and \mbox{HCN~$J$=1--0} are the next molecular lines showing the most extended distribution,
 73\% and 60\% of the total observed area, respectively.
 On the other hand, \mbox{HNC~$J$=1--0} shows a similar  distribution {as}
 C$^{18}$O $J$=1--0 \citep{Gaudel2022}. 

{Table~\ref{tab:lines_stats} provides the total line luminosity ($L_{\rm line}$) from the mapped area  in \Ls~units. {$L_{\rm line}$ is} 
the power emitted through a given line. It also provides $L'_{\rm line}$ (in units of K\,km\,s$^{-1}$\,pc$^2$}), with:
 \begin{equation}\label{eq:L'}
{L'_{\mathrm{line}} =d^2\,\int T_{\mathrm{mb}}(v)\,dv\,d\Omega \,=\,W_{\mathrm{average}}\,\times\,A,}
 \end{equation}
{where $\Omega$ is the solid angle subtended by the source area and
$W_{\mathrm{average}}$ is the average spectrum over the mapped area $A$.
This last quantity
is commonly used to express mass conversion factors (see Sect.~\ref{sec:alphadense})
and {it} is also frequently used in the extragalactic context \citep[e.g.,][]{Gao2004a,Carilli13}}.
   The  \mbox{$^{12}$CO~$J$=1--0} luminosity over the mapped area, $\sim$1.7~\Ls, {is} more than a hundred times higher than $L_{\rm HCO^+\,1-0}$ and $L_{\mathrm{HCN\,1-0}}$.
 
 Figures~\ref{fig:fullmaps120}a and \ref{fig:fullmaps120}b  show maps of visual 
extinction  ($A_\mathrm{V}$) and the  approximate strength of the FUV radiation field 
$G_0'$ (see Sect.~\ref{subsec-Herschel}).
Table~\ref{tab:SED_values} summarizes the median, average, and standard deviation values of the SED-derived parameters.   
The star-forming cores at the center of the NGC~2024 have the highest  $A_\mathrm{V}$ values, with a secondary peak toward  NGC~2023 star-forming cores.
The highest values of  $G_0'$ correspond to cloud areas in the vicinity of the \HII\, regions
NGC\,2024 (\mbox{$G_0'$\,$\approx$\,10$^4$}), 
with a contribution from the neighbor \HII~region created by Alnitak star
(\mbox{$G_0'$\,$\approx$\,10$^3$}),
NGC\,2023 (\mbox{$G_0'$\,$\approx$\,10$^3$}), IC\,435 (\mbox{$G_0'$\,$\approx$\,10$^2$}), and the  ionization front IC\,434 (\mbox{$G_0'$\,$\approx$\,a few 10$^2$}) that includes the iconic Horsehead Nebula.
On the other hand, the easter part of the cloud shows  low surface brightness  $I_{\mathrm{FIR}}$ emission compatible with $G_0'$ of a few to $\simeq$\,10.
The {median} $G_0'$ in the mapped region is 9.

\begin{figure}[!t]
    \centering
    \includegraphics[width=0.48\textwidth]{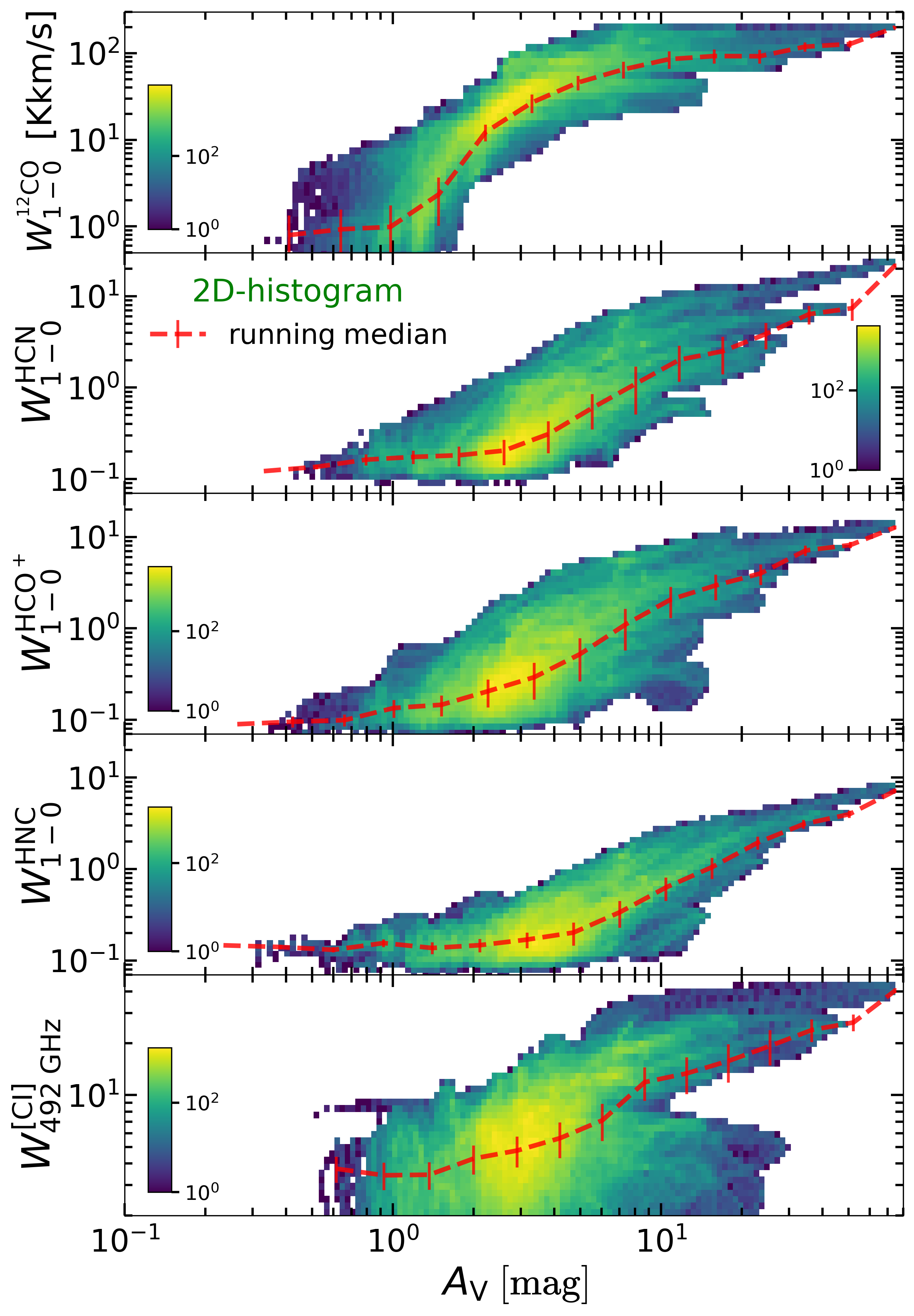}\vspace{-0.2cm}
    \caption{Distribution of $^{12}$CO, HCN, HCO$^+$, HNC~$J$=1--0, and \CI\,492\,GHz line
    intensities as a function of $A_\mathrm{V}$. The dashed red lines show the running median (median values of the line intensity within equally spaced log $A_\mathrm{V}$ bins). Error bars show the {line intensity dispersion. We note that the 5$\sigma$ error of $A_\mathrm{V}$ is 
    $\simeq$\,0.8\,mag. Thus, one cannot trust any trend below
    this threshold}.} 
    \label{fig:lines-Av_hist}
    \vspace{-0.4cm}
\end{figure}

\begin{figure}[!t]
    \centering
    \mbox{\hspace{0cm}\includegraphics[width=0.425\textwidth]{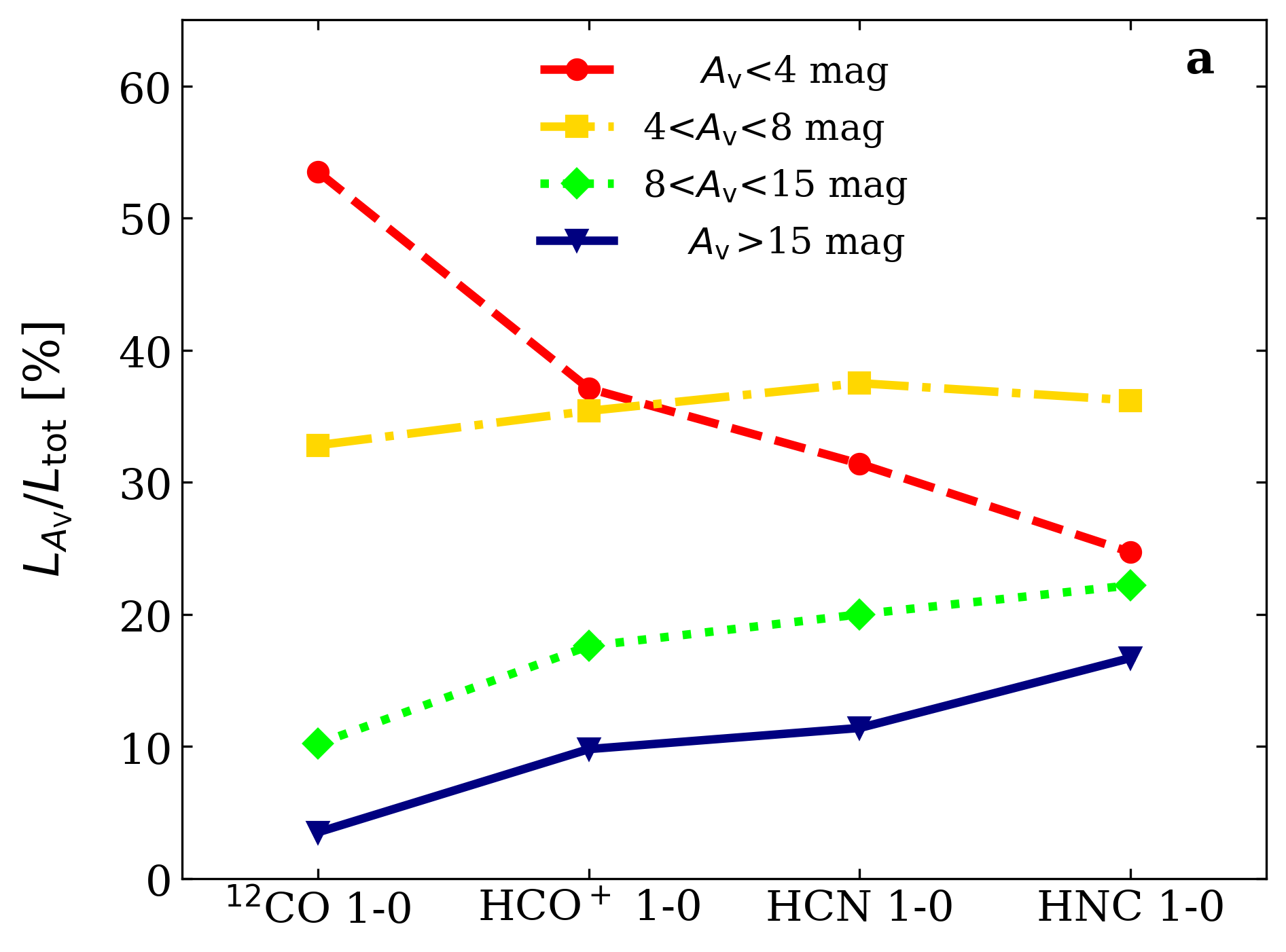}}\vspace{-0.05cm}
    \mbox{\hspace{0cm}\includegraphics[width=0.425\textwidth]{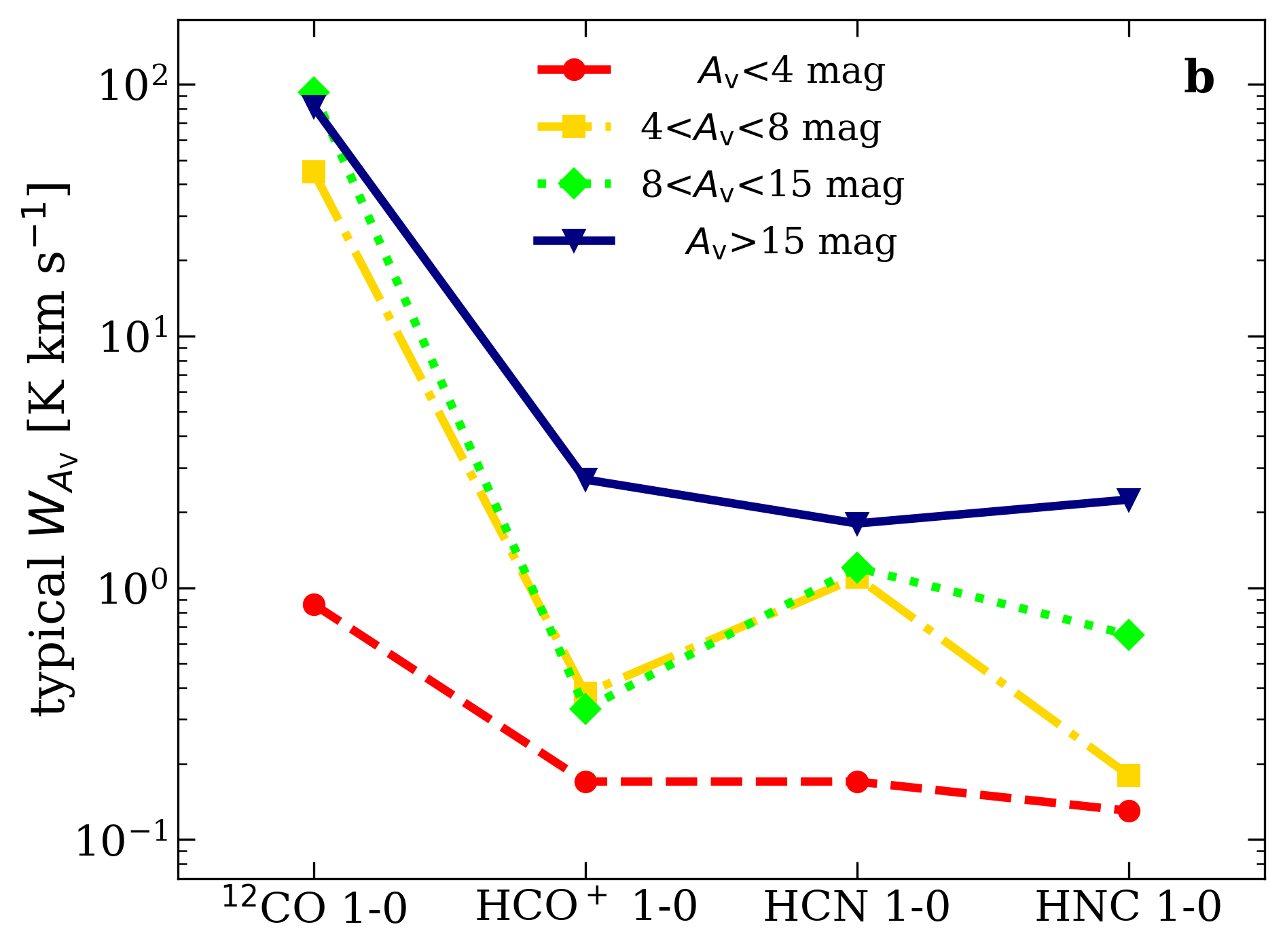}}\vspace{-0.05cm}
    \mbox{\hspace{0cm}\includegraphics[width=0.425\textwidth]{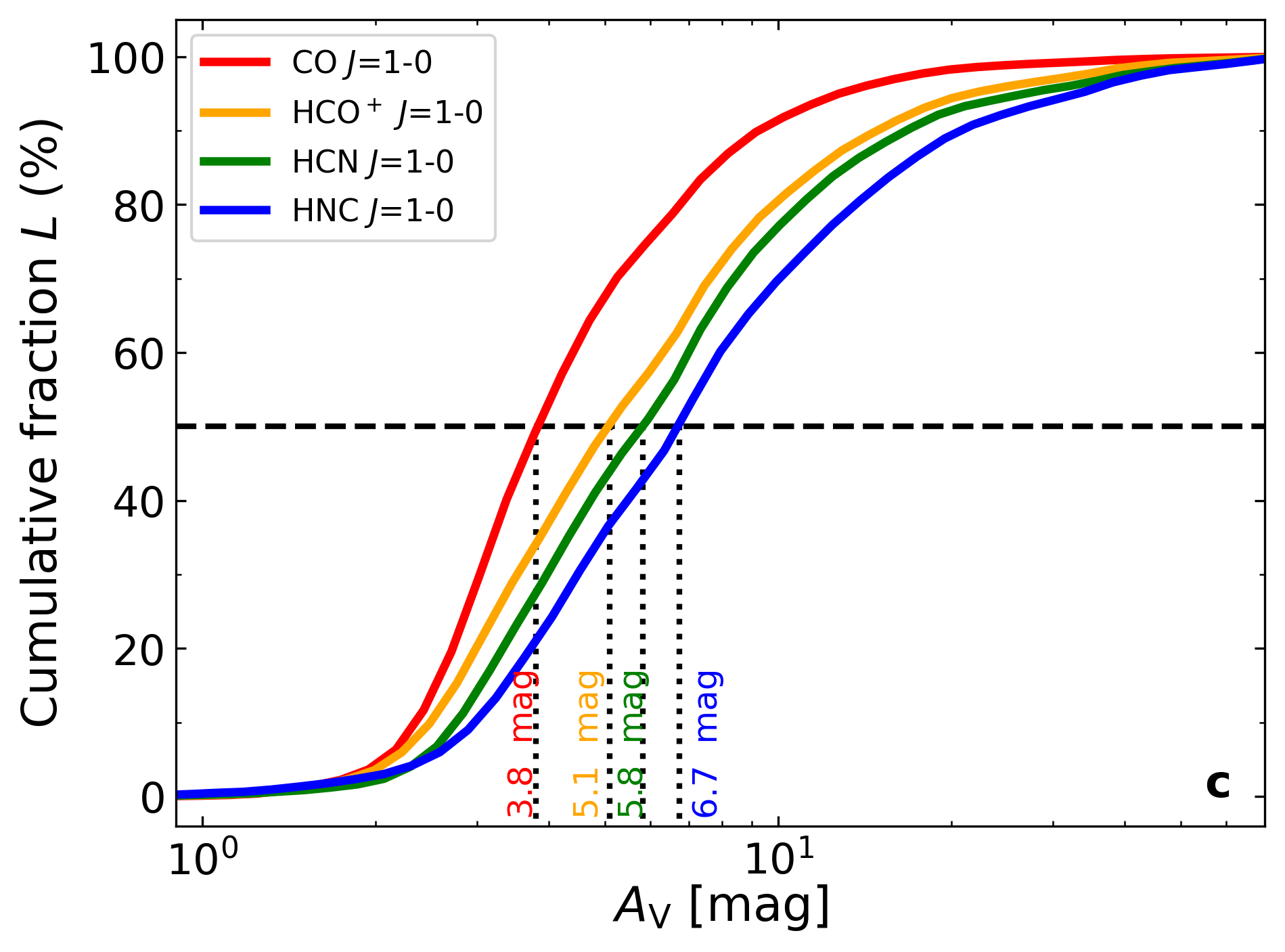}} \vspace{-0.3cm}
    \caption{{Line emission properties as a function of $A_\mathrm{V}$.} (a) Fractions (in \%) of  line {luminosities}  emitted in each  $A_\mathrm{V}$ mask.
    (b) Typical (the mode) line intensity  in each $A_\mathrm{V}$ mask.
    (c) Cumulative line {luminosity}.}
    \label{fig:lines-Av}
    \vspace{-0.1cm}
\end{figure}

Figure~\ref{fig:lines-Av_hist} shows 2D histograms of the CO, HCN, HCO$^+$, HNC~$J$=1--0, and \CI\,492\,GHz integrated line intensities   as function of the visual extinction into the cloud\footnote{{Section~\ref{sect:line-FIR} discusses the relation between
these  lines and $I_\mathrm{FIR}$.}}. 
The running median of the \mbox{HCN~$J$\,=\,1$-$0}  emission increases with extinction at $A_\mathrm{V}$\,>\,3~mag, whereas the running median of the \mbox{HNC~$J$\,=\,1--0} emission shows a similar change of tendency but at higher extinction depths $A_\mathrm{V}$\,>\,5~mag. 
As HCN, the largest number of \mbox{HCO$^+$~$J$\,=\,1$-$0} and \mbox{\CI\,492\,GHz} line  detections in the map occur at $A_\mathrm{V}$\,$\simeq$\,3~mag.
Atomic carbon, however, shows an approximate bimodal behavior with $A_\mathrm{V}$
(it shows both bright and faint emission at high $A_\mathrm{V}$). Indeed, while $A_\mathrm{V}$
is the total visual extinction along each line of sight, we expect that in many instances the
\CI\,492\,GHz emission mostly arises from cloud rims close to the
C$^0$/CO transition 
\citep[e.g.,][]{Hollenbach1991}. On the other hand, 
lines of sight of large $A_\mathrm{V}$ and very bright \CI\,492\,GHz emission
(such as NGC~2024) probably trace \mbox{FUV-illuminated}  surfaces of multiple dense cores and PDRs along the line of sight.

Following \cite{Pety17a}, \mbox{Fig.~\ref{fig:lines-Av}a} shows the fraction of total {$L_{\rm line}$}  within a set of four visual extinction masks. The mask with $A_\mathrm{V}$>15~mag ($\sim$1\% of the total mapped area) represents the highest density gas associated with dense cores. The mask within the $A_\mathrm{V}$ range \mbox{8 to 15~mag} ($\sim$3\% of the mapped area) is right above the extinction threshold above which the vast majority of prestellar cores are found in molecular clouds \citep[e.g.,][]{LadaE92,Lada2010,Wu2010,Evans2020}. Below this threshold, we create two masks to differentiate the emission associated with $A_\mathrm{V}$ below 4~mag (translucent and PDR gas; $\sim$80\% of the mapped area) and  \mbox{4\,$<$\,$A_\mathrm{V}$\,$<$\,8\,mag}  (intermediate cloud depths representing $\sim$16\% of the mapped area). We find that more than half  of the total 
\mbox{CO~$J$\,=\,1--0} intensity arises from the lowest extinction mask $A_\mathrm{V}$<4~mag. Interestingly, about a 30\% of the total \mbox{HCN~$J$\,=\,1--0} emission arises from gas also at  $A_\mathrm{V}$<4~mag. 
Most of the HCN and \mbox{HNC~$J$\,=\,1--0} emission arises from regions at visual extinctions between 4 and 8~mag, and only 10\% of the HCN emission arises from regions at very high visual extinctions, \mbox{$A_\mathrm{V}$\,$>$\,15~mag}. Likewise, the HCN and HNC 2D histograms peak at $A_\mathrm{V}$ lower than 3~mag (HCN) and 5~mag (HNC). 
This contrasts with the \mbox{N$_2$H$^+$~$J$=1--0} emission, {which}  arises from cold and dense gas {shielded from FUV radiation} at $A_\mathrm{V}$\,$>$\,\,15~mag  \citep[see][]{Pety17a}.
For each molecular line, Fig.~\ref{fig:lines-Av}b shows the typical (the statistical mode) intensity $W$  toward each of the four  extinction masks. 
The \mbox{CO~$J$\,=\,1--0} emission is bright ($\sim$1~\Kkms)  even at 
$A_\mathrm{V}$\,$<$\,4\,mag, and very bright ($>$\,10~\Kkms) toward all the other masks
(although  optically thick). 

The typical HCN, HNC, and HCO$^+$~$J$=1--0  line intensities are above  1~\Kkms\; 
for $A_\mathrm{V}$\,$>$\,15\,mag (dense gas). For lower $A_\mathrm{V}$, the lines are fainter but detectable. 
Since the translucent gas spans much larger areas  
than the dense gas
(96\% of the mapped cloud is $A_\mathrm{V}$\,<\,8~mag, 80\% at $A_\mathrm{V}$\,<\,3~mag), in many instances it is the  widespread and faint extended emission that dominates the total luminosity. 
We stress that  $\sim$70\% of the \mbox{HCN\,$J$\,=\,1--0} line  luminosity in  Orion\,B arises from gas at $A_\mathrm{V}$\,<\,8~mag { (and 50\%\, of the FIR dust luminosity)}.
Table~\ref{tab:lines_stats} 
summarizes the line intensities and line luminosities  over the mapped  area.

Figure~\ref{fig:lines-Av}c shows the cumulative fractions of the integrated intensities for CO, HCO$^+$, HCN, HNC~$J$=1--0 as a function of $A_\mathrm{V}$. The cumulative distributions are different for each species. We define the visual extinction that contains 50\% of the total integrated line intensity as the characteristic $A_\mathrm{V}$, such as \mbox{$W(A_\mathrm{V}$\,$<$\,$A_\mathrm{V}^\mathrm{char})$=50\%} \citep[e.g.,][]{Barnes2020}. We find that the characteristic $A_\mathrm{V}^\mathrm{char}$ for CO~$J$=1--0 is 3.8~mag, which implies that 50\% of the CO total intensity arises from {gas below $A_\mathrm{V}$\,=\,3.8~mag.} 
For  HCO$^+$, HCN, and HNC~$J$\,=\,1--0 lines, we find $A_\mathrm{V}^\mathrm{char}$ of  5.0, 5.8, and 6.7~mag, respectively. These values agree with recent studies of the  star-forming regions Orion~A and W49
\citep{Kauffmann2017,Barnes2020}.

\subsection{HCN/CO, HCN/HNC, \texorpdfstring{HCN/HCO$^+,$}{HCN/HCO+} and \texorpdfstring{\CI/CO}{[CI]/CO} line intensity ratio maps}
\label{subsec:ratios}

\begin{figure}[!t]
   \centering
    \includegraphics[width=0.4999\textwidth]{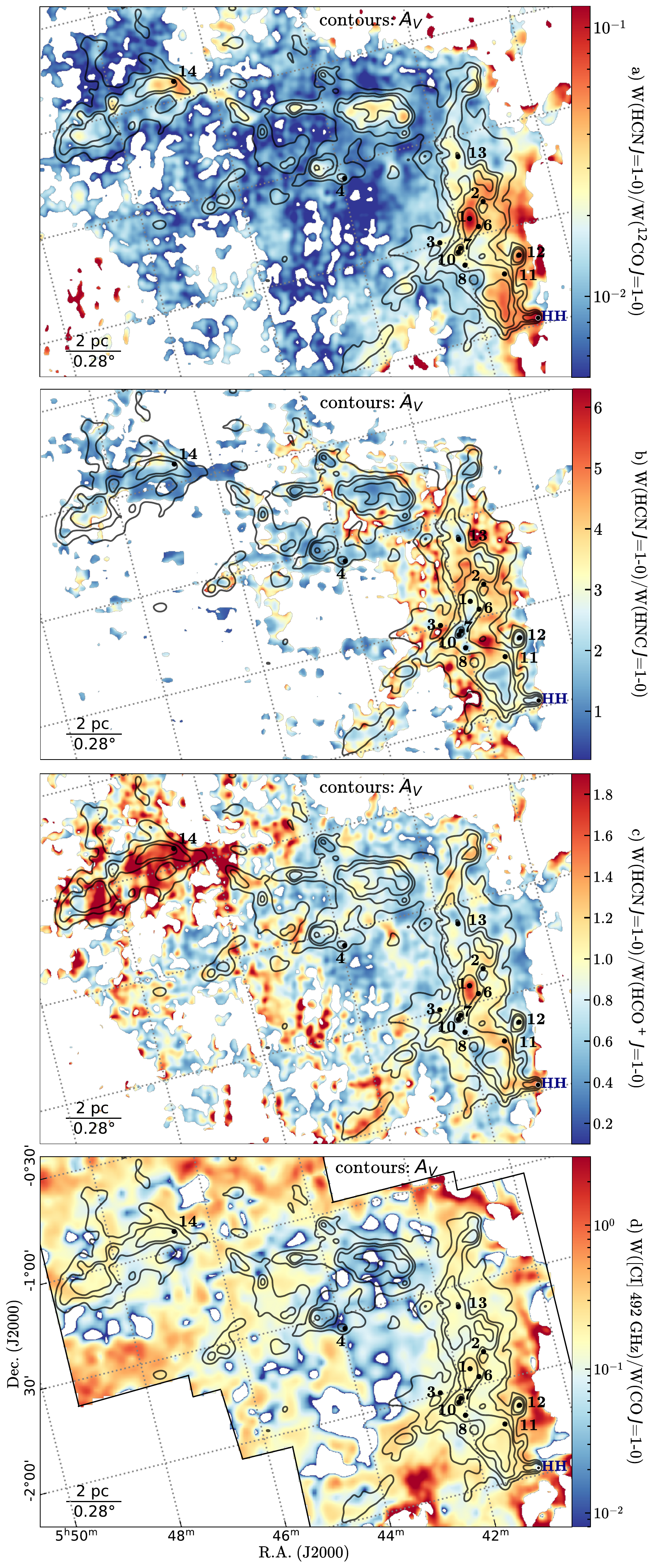}
    \caption{Line intensity ratio maps. 
    Contours show $A_\mathrm{V}$= 4, 6, 8, and 15~mag.} 
    \label{fig:hcnratios_120}
    \vspace{-0.5cm}
\end{figure}

The spatial distribution of the \mbox{HCN~$J$\,=\,1--0} line emission  compared to that of other molecules provides information about the origin and the physical conditions of the HCN-emitting gas. \mbox{Figure~\ref{fig:hcnratios_120}} shows the  \mbox{HCN\,/\,$^{12}$CO~$J$=1--0}, 
\mbox{HCN\,/\,HNC~$J$=1--0}, and \mbox{HCN\,/\,HCO$^+$~$J$=1--0} integrated line intensity ratios. We generated these maps by
taking only line signals above 3$\sigma$ for each species
(i.e., we show regions where the emission from both species spatially coexist along the line of sight). 
In addition, Fig.~\ref{fig:hcnratios_120}d shows a map of the
\CI~492~GHz/CO~$J$=1--0 integrated line intensity ratio\footnote{To obtain the \CI~492~GHz/CO~$J$=1--0 line ratio with line intensities in erg\,cm$^{-2}$\,s$^{-1}$\,sr$^{-1}$ one
has to multiply by (492\,GHz/115\,GHz)$^3$\,$\simeq$\,78.}).
Table~\ref{tab:linrat_values} summarizes the average and median line intensity ratios in the mapped region.

\begin{table}[ht]
\centering
\caption{\mbox{Statistics of  5 deg$^2$ line intensity ratio maps shown in Fig.~\ref{fig:hcnratios_120}.}}
\label{tab:linrat_values}
\vspace{-0.2cm}
\begin{threeparttable}
\resizebox{0.44\textwidth}{!}{%
\begin{tabular}{@{}lcccc@{\vrule height 5pt depth 1pt width 0pt}}
\toprule

Line Intensity Ratio  & Median &  Average   & std & $\rho$\,($G_0'$)$^{\dagger}$ \\ \midrule
HCN\,/\,CO $J$\,=\,1--0   & 0.011 & 0.015  & 0.023 & \phantom{-}{0.4} \\
HCN\,/\,HNC $J$\,=\,1--0 & 2.7 & 3.1  & 1.2 & \phantom{-}0.6 \\
HCN\,/\,HCO$^+$ $J$\,=\,1--0 & 0.9 & 0.88   & 0.4 & \phantom{-}{0.2} \\ 
\CI\,/\,CO $J$\,=\,1--0 & 0.15 & 0.20 & 0.26 & {-0.3} \\\bottomrule
\end{tabular}}

\tablefoot{$^{\dagger}$$\rho$($G_0'$) is the Spearman correlation rank with the $G_0'$ map. See also Fig.~\ref{fig:ratiosFIR} in the Appendix.}
\end{threeparttable}

\end{table}


--- HCN/CO~$J$\,=\,1--0: 
The average line intensity ratio is {0.015}, with a standard deviation of {0.023}. In the inner regions of the cloud, close to the Cloak, Orion B9, Hummingbird, and Flame filaments, the ratio increases  with $A_\mathrm{V}$ (shown in contours). We also find high line intensity ratios ($\sim$\,0.1) in the FUV-illuminated cloud edges  (see discussion in Sect.~\ref{sec:HCN-CO_Av}).

--- HCN/HNC~$J$\,=\,1--0: The average line intensity ratio is 3.1, with a standard deviation of 1.2. The lowest ratios, \mbox{$\sim$0.5-0.9}, appear in cold and low $I_{\rm FIR}$ regions such as the Cloak and Orion~B9. 

--- HCN/HCO$^+$~$J$\,=\,1--0: The average line intensity ratio is {0.9}, with a standard deviation of 0.4. 
In general, this line ratio displays small variations across the cloud.
The  Cloak, the center of NGC~2024, the Flame Filament, and the Horsehead show a line intensity ratio above one (reddish  areas in Fig.~\ref{fig:hcnratios_120}c). All these regions host starless and prestellar cores \citep{Konyves2020}.
 
--- \CI~492~GHz/CO~$J$=1--0: The average line intensity ratio is 0.20, with a standard deviation of  0.26.
In PDR gas, this ratio is roughly inversely proportional to the gas density \cite[see e.g.,][]{Kaufman1999}. 
We find the highest ratios, above one, toward the FUV-illuminated edges of {the} cloud.

We also investigate the possible spatial correlations of the above line intensity ratios with the
SED derived parameters $G_0'$, $T_{\rm d}$, $T_{\rm peak}$(CO), and $A_{\rm V}$. Only the \mbox{HCN\,/\,HNC $J$\,=\,1--0}
line intensity map shows a (weak) monotonic correlation with $G_0'$ (\mbox{Spearman} correlation rank of 0.6;
see \mbox{Table}~\ref{tab:linrat_values}).
This spatial correlation is not linear (the Pearson correlation rank is 0.5 in log-log scale and 0.008
in linear scale) but suggests a connection between the HCN/HNC abundance ratio and the FUV radiation field. \mbox{Figures~\ref{fig:ratiosAv}} and \ref{fig:ratiosFIR} 
show 2D histograms of the studied line intensity ratios as function a of 
$A_\mathrm{V}$ and $I_{\rm FIR}$, respectively.

\section{HCN excitation, radiative transfer models, and gas  physical conditions} \label{sec:analysis}

In this section we analyze the large scale HCN \mbox{$J$\,=\,1--0}  emission in detail. We, 
(i) derive excitation temperatures ($T_{\rm ex}$) and HCN column densities, $N$(HCN), 
using the LTE-HFS fitting method, (ii) analyze the anomalous HCN~$J$=1--0 HFS emission, (iii)  determine the physical conditions of the widespread and extended  \mbox{HCN $J$\,=\,1--0} emitting gas, and (iv) derive rotational temperatures, $T_\mathrm{rot}$, and $N$(HCN) in a sample of representative positions observed in rotationally excited HCN and H$^{13}$CN lines.  
In order to determine all these parameters at the highest possible spatial resolution, {throughout} all this section we make use of  maps and pointed
observations at an effective 30$''$  resolution ($\sim$0.06\,pc).

\subsection{HCN column density and \texorpdfstring{$T_{\rm ex}$}{Tex} using the LTE-HFS method}\label{sect:res-HFSfitting}

{Firstly,} we determine \mbox{$T_\mathrm{ex}$\,($J$\,=\,1--0)} and  the opacity-corrected column density $N^\mathrm{\tau,corr}$(HCN) by applying the LTE-HFS fitting method  in \texttt{CLASS}\cref{footn:class}
 (Appendix~\ref{App:HFS fit}).
 This method uses as input the  line separations  and  1:5:3 intrinsic line strengths of the  \mbox{$J$\,=\,1--0}  HFS components. 
 The {LTE-HFS} fitting method assumes that: {(i)  all HFS lines  have the same $T_\mathrm{ex}$ and linewidth $\Delta v$, and
 (ii)~the velocity-dependent line opacities have Gaussian profiles. Thus, one can express the continuum-substracted main beam temperature at 
 a given velocity $v$ of the \mbox{$J$\,=\,1--0} line profile as:}
 \begin{equation}
{T_{\mathrm{mb}}(v) = [J(T_{\rm ex}) - J(T_{\rm bg})]\, [1\,-\, e^{-\tau(v)}],}
 \end{equation}
{where $\tau(v)$ is the sum of the HFS line opacities:}
 \begin{equation} 
{\tau(v) = \sum_{i}^{\rm HFS} \tau_i\,(v-v_{i}^{0}), 
\,\,\,\, {\rm{with}}\,\,\,\,  \tau_i\,(v-v_{i}^{0})= \tau_{i}^{0}\, \phi(v-v_{i}^0).}
 \end{equation}
{In the above expressions, $\tau_{i}^{0}$ is the
opacity of each HFS \mbox{component $i$} at (each) line center
($v_{i}^{0}$), 
and $\phi(v-v_{i}^{0})$ is
a Gaussian profile centered at $v_{i}^{0}$. In the LTE-FTS fitting method,  one
fixes the sum of all \mbox{HCN $J$\,=\,1--0} HFS  line center opacities, $\tau^0$, following their
intrinsic line strengths:}
\begin{equation}
{\tau^0 = 
\tau_{F=1-0}^{0} + \tau_{F=2-1}^{0} + \tau_{F=1-1}^{0} =
\frac{3}{9}\,\tau^0 + \frac{5}{9}\,\tau^0 + \frac{1}{9}\,\tau^0.
}
\end{equation}

{In the Rayleigh-Jeans regime,  $J(T_{\rm ex})\rightarrow T_{\rm ex}$. Thus,  the \mbox{LTE-HFS} fitting method returns  $T_\mathrm{ex}$ and {$\tau^0$}  as outputs. {We} use these parameters} to derive \mbox{$N^\mathrm{\tau,corr}$\,(HCN)}.
In order to obtain satisfactory fits,
we  applied this  method to the brightest regions, those \mbox{$T_{\rm mb}^{\rm peak}$(HCN $J$\,=\,1--0 $F$=2--1, $\geq$ 0.5 K = 5$\sigma$}),  associated with the
{main} cloud velocity component at \mbox{$v_{\mathrm{LSR}}\simeq$10~\kms}.

\begin{figure*}[!h]
    \centering
    \includegraphics[height=5.57cm]{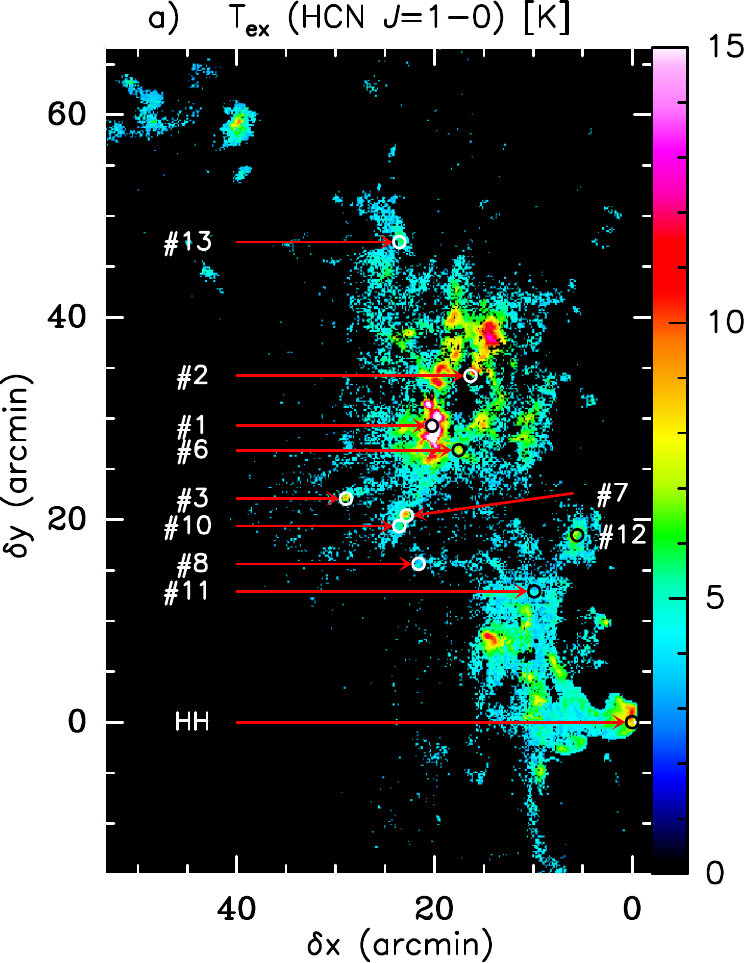}\hspace{0.1cm}
    \includegraphics[height=5.63cm]{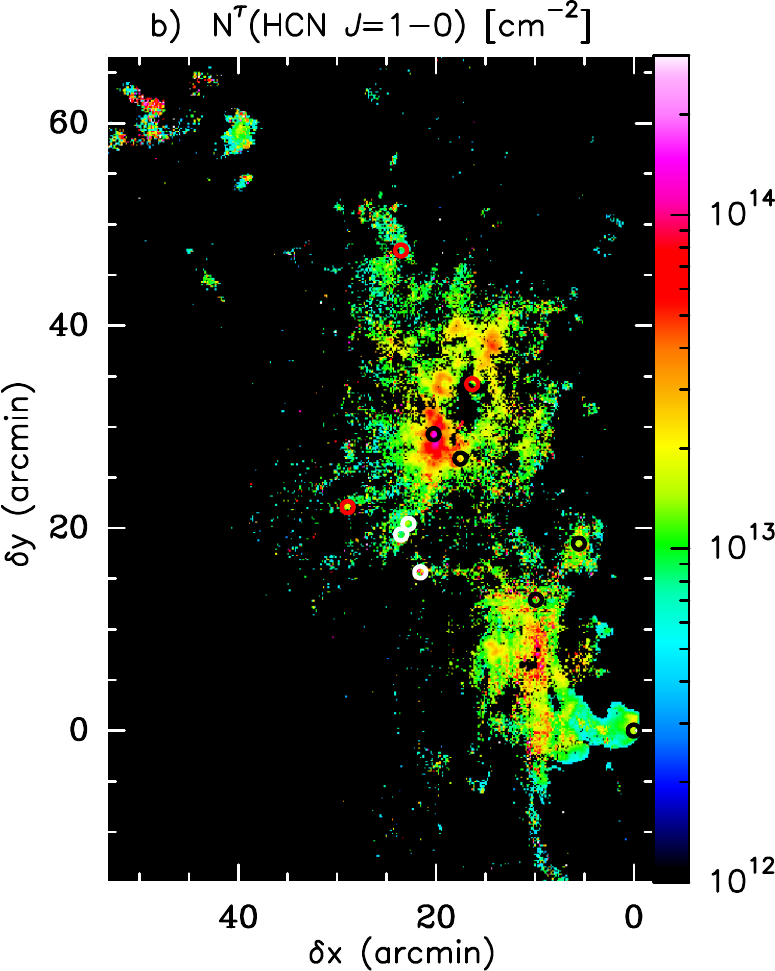}\hspace{0.1cm}
    \includegraphics[height=5.63cm]{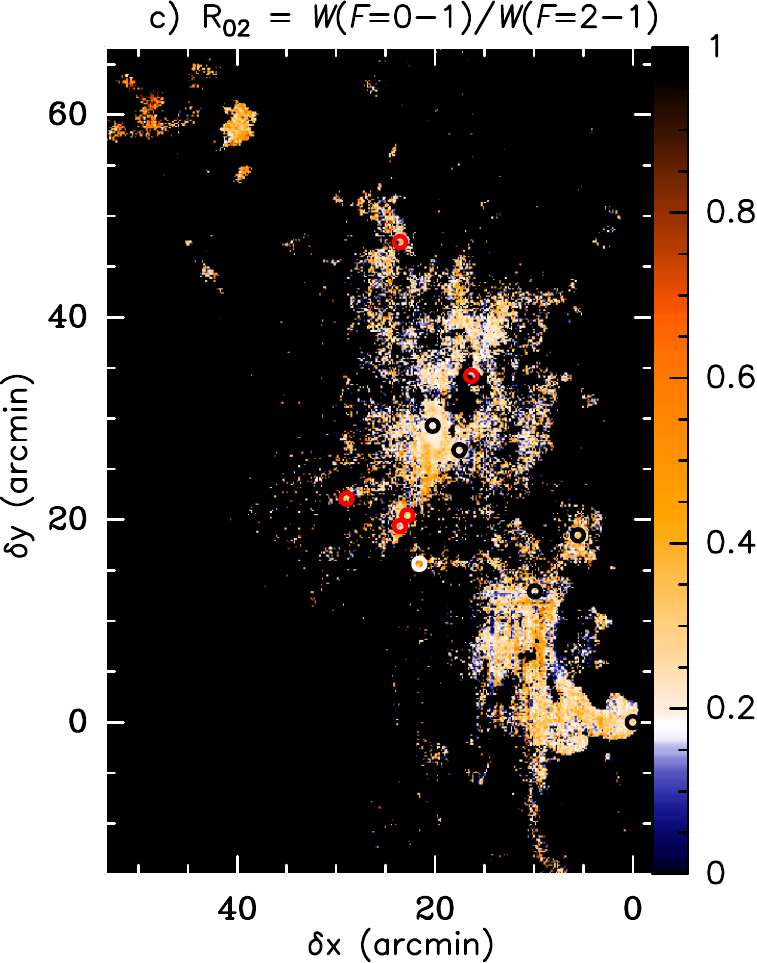}\hspace{0.1cm}
    \includegraphics[height=5.63cm]{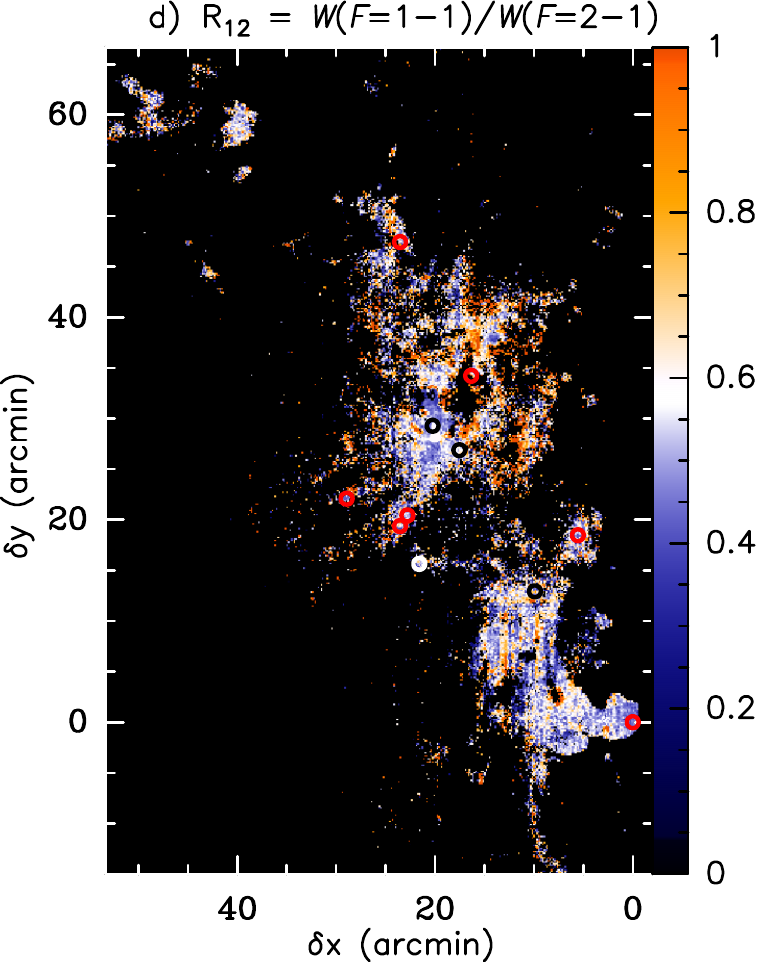}
    \caption{{Spatial distribution of \mbox{$T_{\mathrm{ex}}$(HCN $J$=1--0)}
     and $N$(HCN) estimated from  LTE-HFS fits, and maps of  \mbox{HCN~$J$=1--0} HFS intensity ratios.} (a)\,\mbox{$T_{\mathrm{ex}}$(HCN $J$=1--0)}. (b)\,Opacity-corrected  column densities $N$(HCN). (c)\,and\,(d)\,$R_{02}$ and $R_{12}$ (white color corresponds to non-anomalous  ratios).}
    \label{fig:hcn_hfs_Tex-tau}
\end{figure*}

\begin{figure*}[!th]
    \centering
    \includegraphics[width=0.465\textwidth]{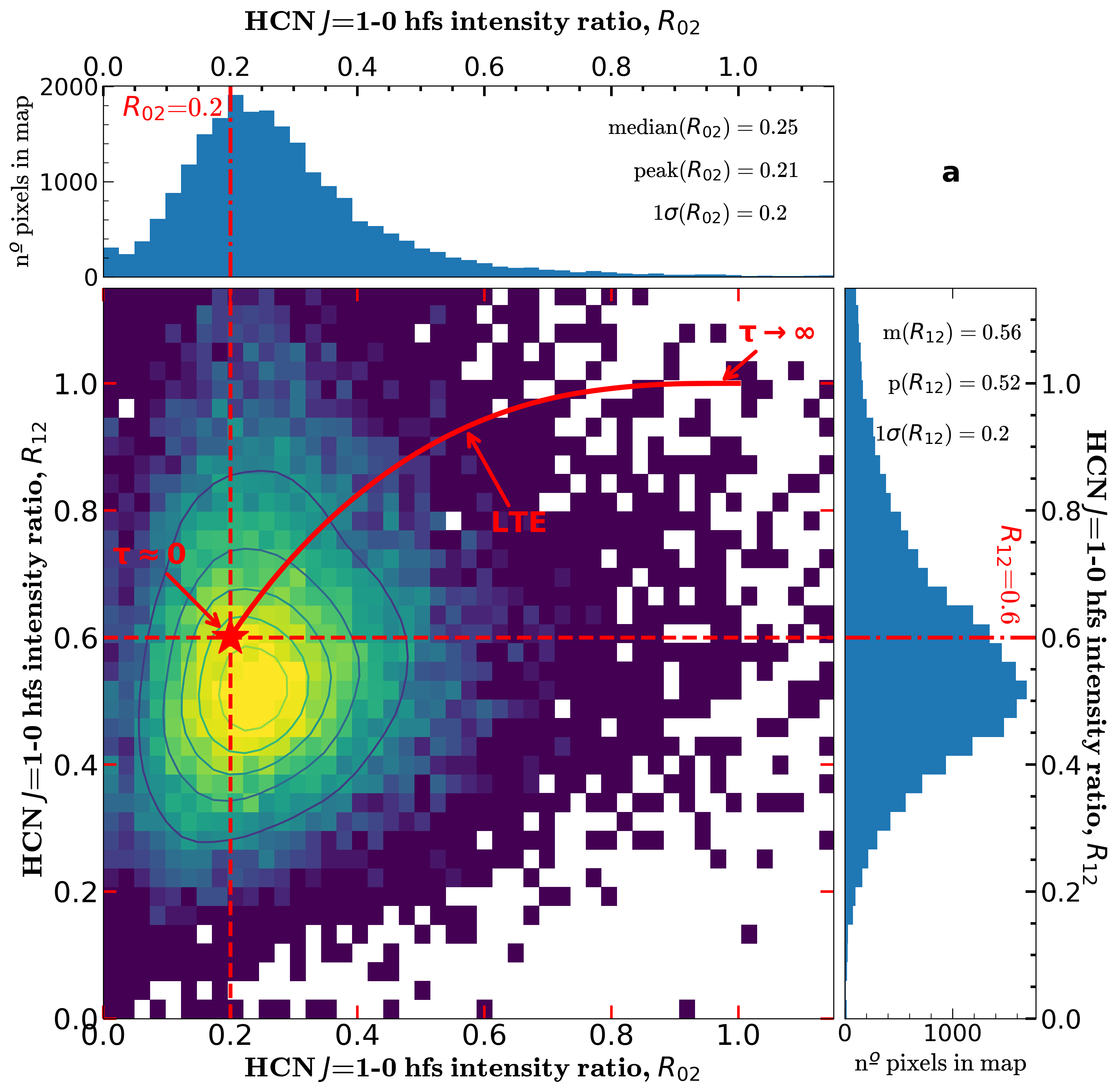}\hspace{0.3cm}
    \includegraphics[width=0.465\textwidth]{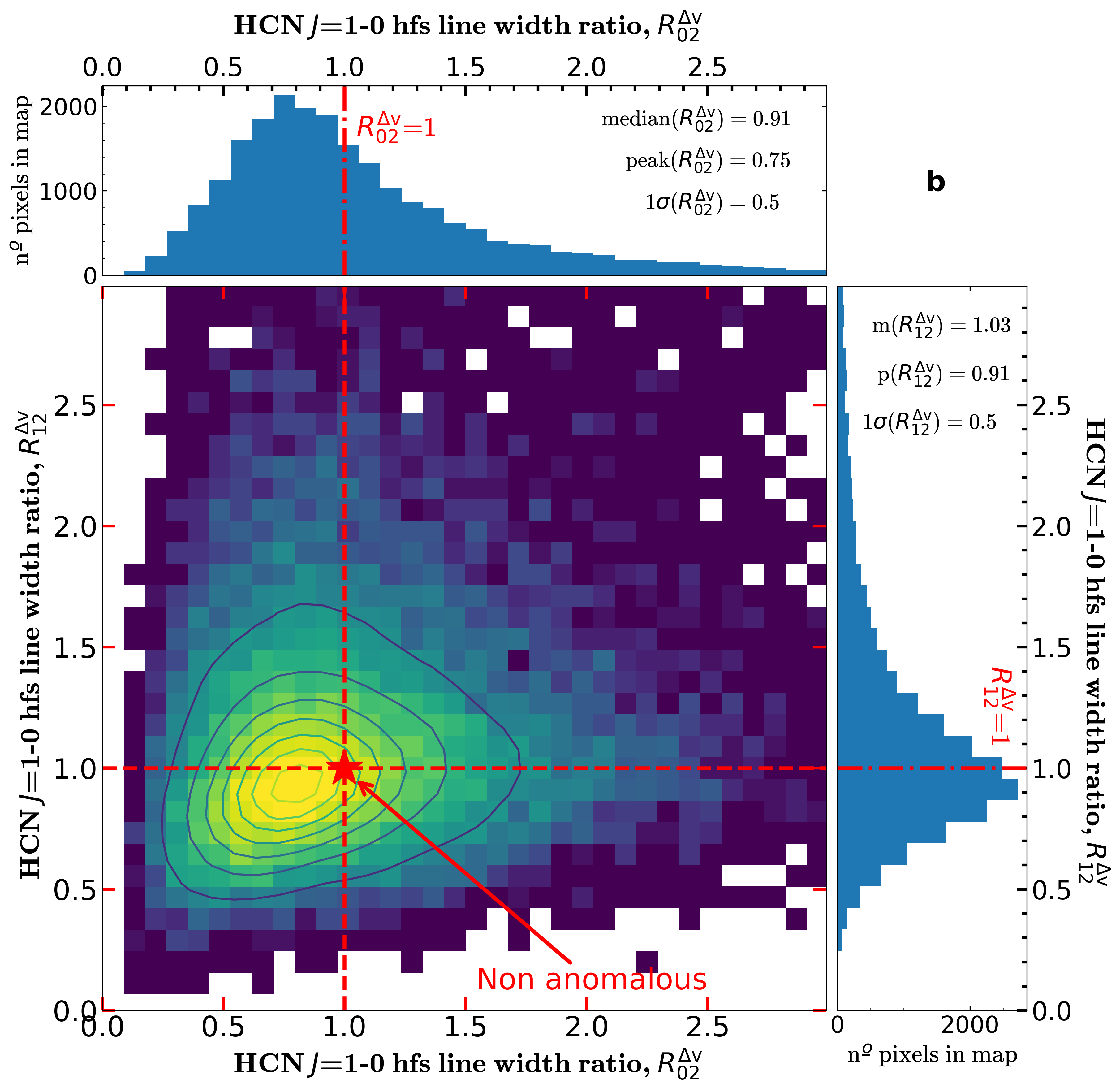}
    \caption{Histograms of HCN $J$=1--0 HFS (a)  Line intensity ratios, and (b) Line-width ratios observed in Orion\,B. $R_{02}$ stands for \mbox{$W$($F$=0--1)/$W$($F$=2--1)} and $R_{12}$ stands for \mbox{$W$($F$=1--1)/$W$($F$=2--1)}. {The red curve in panel (a) shows the expected LTE ratios as line opacities increase.}
    {The red star marks the non-anomalous 
    ratios in the optically thin limit  $\tau \rightarrow 0$
    ({1$\sigma$} is the standard deviation relative to the mean line ratios)}.} 
    \label{fig:hcn_hfs_anomalies}
\end{figure*}

Figure \ref{fig:hcn_hfs_Tex-tau} shows the resulting HCN column density  and $T_\mathrm{ex}$ maps in a smaller but high S/N submap.  
The average (median) \mbox{$T_\mathrm{ex}$\,($J$\,=\,1--0)} in the region is 5~K (4.5~K), implying subthermal emission{, that is,} \mbox{$T_\mathrm{ex}$\,$\ll$\,$T_\mathrm{k}$}
and \mbox{$n$(H$_2$)\,$<$\,$n_{\rm cr,\,eff}$}. NGC\,2024 shows the highest values, with \mbox{$T_\mathrm{ex}$ > 15 K}. 
 $N^\mathrm{\tau,corr}$(HCN) ranges between 10$^{13}$ and a few 10$^{14}$~cm$^{-2}$. The average (median) column is 3.4$\cdot$10$^{13}$~cm$^{-2}$ (1.3$\cdot$10$^{13}$~cm$^{-2}$).

\begin{figure*}[!ht]
    \centering
    \includegraphics[width=0.494\textwidth]{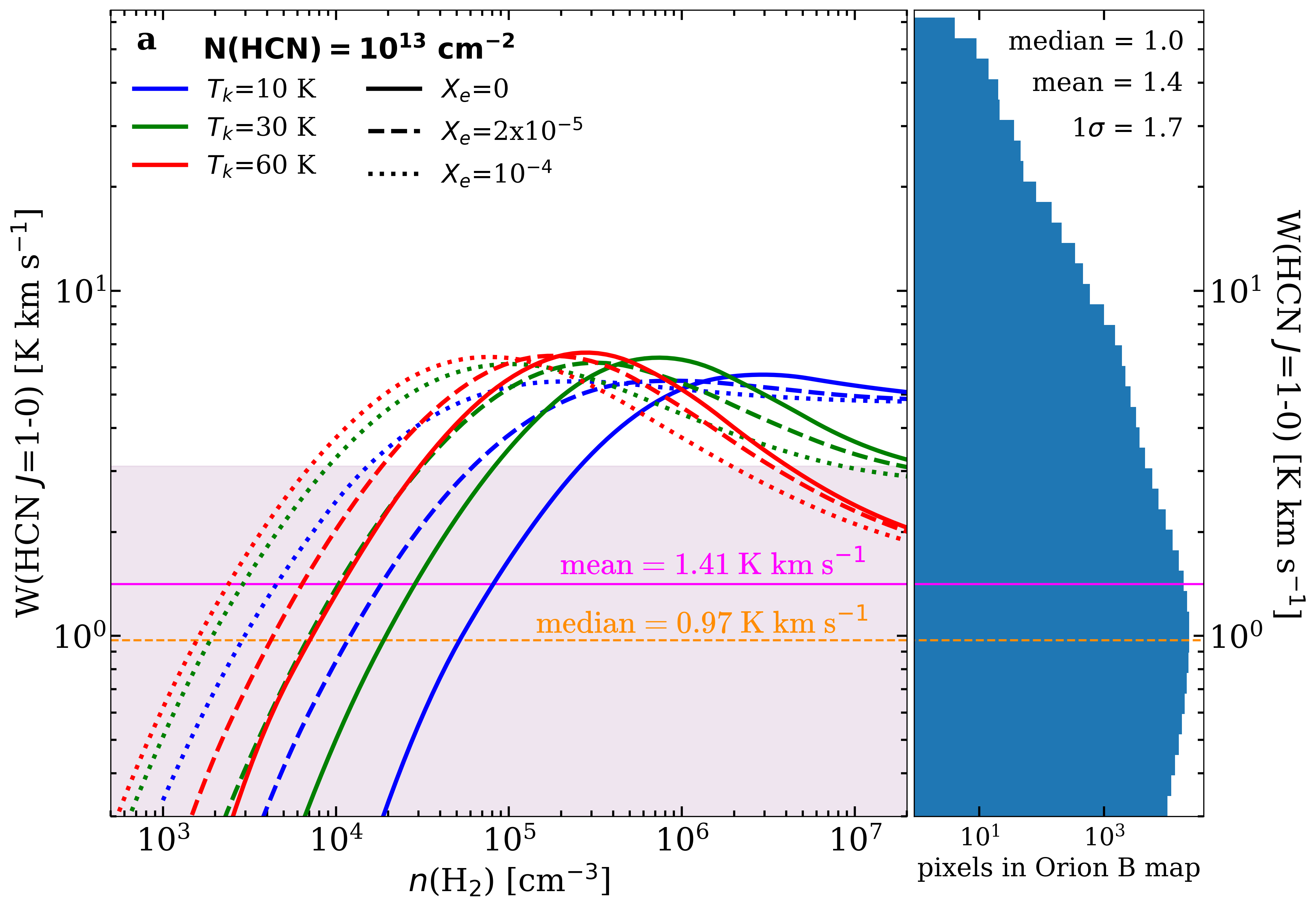} \hspace{-0.15cm}
    \includegraphics[width=0.494\textwidth]{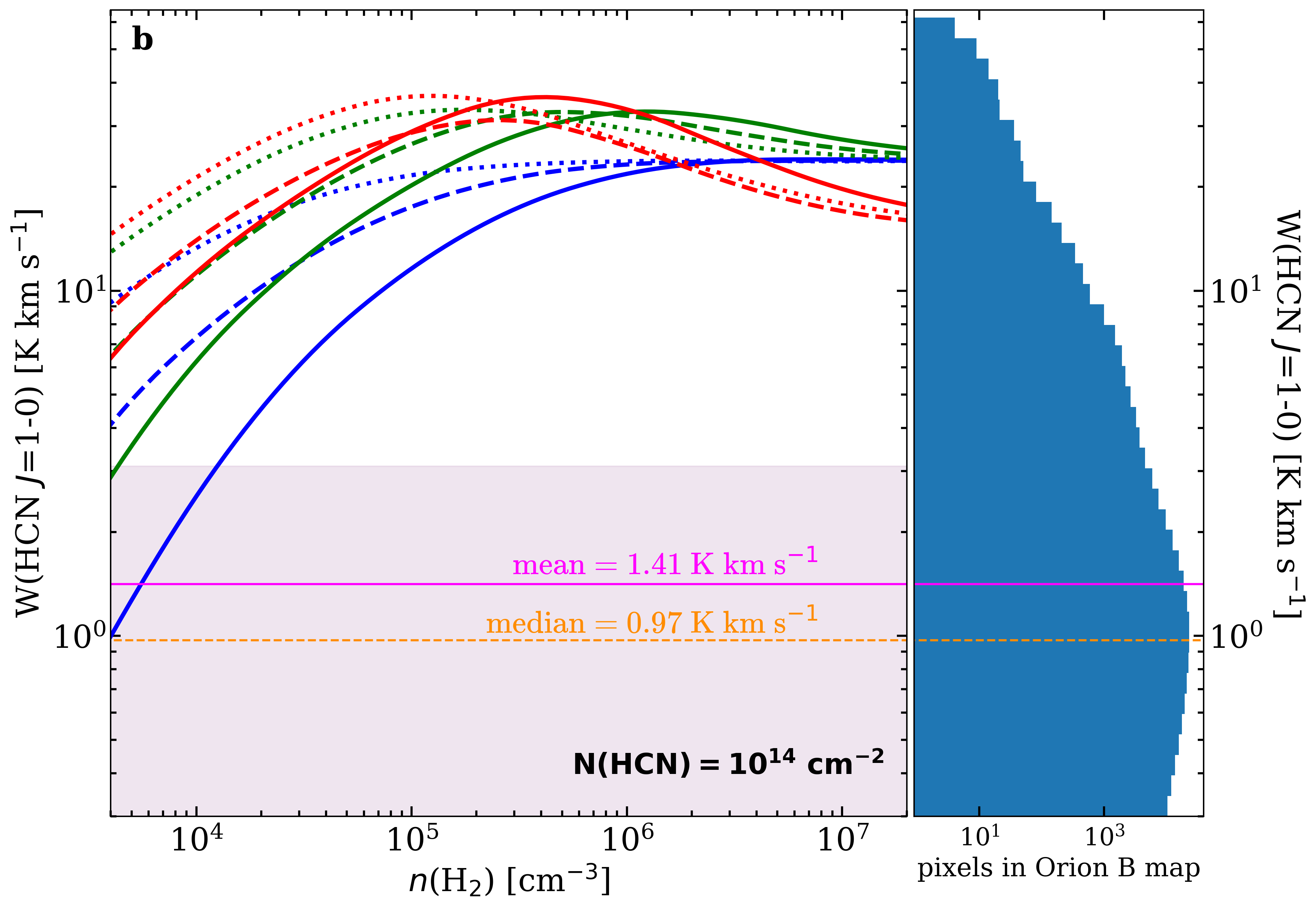}
    \caption{Comparison of observed $W$(HCN~$J$=1--0) intensities in Orion\,B and predictions from  nonlocal and non-LTE radiative transfer models including line overlap for
    (a)~$N$(HCN)=10$^{13}$~cm$^{-2}$ and (b)~$N$(HCN)=10$^{14}$~cm$^{-2}$. The continuous curves show model results for $T_\mathrm{k}$=60, 30, and 10~K (red, green, and blue curves, respectively),  different ionization fractions: \mbox{$\chi_\mathrm{e}$\,=\,0} (continuous curves), \mbox{$\chi_\mathrm{e}$\,=\,2$\cdot$10$^{-5}$} (dashed curves), and \mbox{$\chi_\mathrm{e}$\,=\,10$^{-4}$} (dotted curves). The pink and orange horizontal line mark the mean {and} median  values of \textit{W}(HCN $J$=1--0). The pink shaded area represents the standard {deviation (1$\sigma$)}  {relative to} the mean
    {detected} \textit{W}(\mbox{HCN $J$=1--0}) {intensities} in Orion\,B (at 30$''$). 
    {Positions in the pink area}  account for $\sim$70\,\% of the total
    $L_{\rm HCN\,1-0}$ in the map. The right panels  show an histogram with the distribution of 
     \textit{W}(HCN $J$=1--0) detections in individual map pixels.}
    \label{fig:hcn_WMCmod}
\end{figure*}

\subsection{Large-scale anomalous HCN \textit{J}\,=\,1--0 HFS emission} \label{sect:HCNHFS}

To study the \mbox{HCN $J$\,=\,1--0} emission in more detail we extracted
the intensity and linewidth of each HFS component individually (by fitting Gaussians). {Figures~\ref{fig:hcn_hfs_Tex-tau}c and \ref{fig:hcn_hfs_Tex-tau}d  show the spatial distribution
of the HFS line intensity ratios,
{\mbox{$R_{02}$=$W$($F$=0--1)/$W$($F$=2--1)} and \mbox{$R_{12}$=$W$($F$=1--1)/$W$($F$=2--1)}}, respectively, and Fig.~\ref{fig:hcn_hfs_anomalies}a shows their histograms.}
{In addition, Fig.~\ref{fig:hcn_hfs_anomalies}b shows the histograms of the  HFS linewidth ratios, {\mbox{$R^{\Delta v}_{02}$=$\Delta v$($F$=0--1)/$\Delta v$($F$=2--1)} and \mbox{$R^{\Delta v}_{12}$=$\Delta v$($F$=1--1)/$\Delta v$($F$=2--1)}}}. {The red curve in Fig.~\ref{fig:hcn_hfs_anomalies}a shows the expected $R_{02}$ and $R_{12}$ ratios in LTE as line opacities increase. We note that the majority of observed {ratios} in the map are  far from the LTE curve. Indeed,}
the histogram of the  intensity ratio $R_{02}$  peaks at 0.21, with a median value of 0.25 whereas the histogram of the  intensity ratio $R_{12}$ peaks at 0.52, with a median value of 0.56. Therefore, the intensity ratio $R_{12}$ is typically anomalous\footnote{We recall that 
the range of possible  HFS line intensity ratios in LTE are $R_{02}$=$[$0.2, 1$]$ and $R_{12}$=$[$0.6, 1$]$. {Outside these ranges, HFS ratios are called} anomalous
({and very anomalous if \mbox{$R_{02}$\,$<$\,0.2} and \mbox{$R_{12}$\,$<$\,0.6}}). 
Only in LTE, HFS lines have the same $T_\mathrm{ex}$
and  same  linewidths, thus  $R^{\Delta v}$=1.} over large cloud scales.

\mbox{Non-LTE} radiative transfer models including line overlap
{effects} show that these anomalous intensity ratios imply 
that {lines are optically thick} and that a single $T_\mathrm{ex}$  does not represent the excitation of these HFS levels \citep[][]{Gonz-Alf1993,Goicoechea2022}.
This questions the precision of the  parameters obtained from the
LTE-HFS fitting method. To illustrate this,
\mbox{Fig.~\ref{fig:HFS-gauss}}  shows the (poor) 
best LTE-HFS fit to the \mbox{HCN $J$=1--0} HFS  lines toward the Horsehead PDR.

The linewidth of the  faintest \mbox{$F$=0--1} HFS component in the map ranges from $\sim$1 to $\sim$2\,km\,s$^{-1}$ (see also \mbox{Table~\ref{tab:HCNJfit}}). These linewidths are broader than the  narrow  linewidths,  $\sim$0.5\,km\,s$^{-1}$, typically observed in Orion\,B toward dense and \mbox{FUV-shielded} cold cores in molecules such as  H$^{13}$CO$^+$ \citep[e.g.,][]{Gerin2009}. 
{Thus, \mbox{HCN $J$=1--0} traces a different cloud component}. \mbox{Figure~\ref{fig:hcn_hfs_anomalies}b} shows the histogram of the HFS linewidth ratios $R_{02}^{\Delta v}$ and $R_{12}^{\Delta v}$. They peak at 0.75 and 0.91 respectively, with median values of 0.91 and 1.03. That is, the linewidths of the different HFS components are not the same and line opacity broadening matters. 
 Non-LTE models including line overlap predict these anomalous linewidth ratios,  $R^{\Delta v}$$\neq$1, when HFS lines become optically thick  \citep[e.g., see \mbox{Fig.~3} of][]{Goicoechea2022}.

\subsection{Physical conditions of the extended  low surface brightness HCN~\textit{J}=1--0 emitting  gas }\label{sect:comp-models}

{Here} we compare the observed line integrated intensities \mbox{$W$(HCN~$J$\,=\,1--0}) with a grid of non{-}local and non-LTE radiative transfer models
{calculated by \cite{Goicoechea2022}}. These  models include HFS line overlaps and use new HFS-resolved collisional rate coefficients for inelastic collisions of HCN with $para$-H$_2$, $ortho$-H$_2$,
and electrons in warm gas.

The grid of single-component ($T_\mathrm{k}$=60, 30, and 10~K)  static-cloud  (no velocity field) models encompass the HCN  column densities 
{predicted by our chemical models (Sect.~\ref{sec:Meudon})} and
typically  observed in Orion~B {(Fig.~\ref{fig:hcn_hfs_Tex-tau}b)}: $N$(HCN)=10$^{13}$~cm$^{-2}$,  representative of optically thin or marginally optically thick HCN $J$=1--0 HFS lines, and $N$(HCN)=10$^{14}$~cm$^{-2}$, representative of bright optically thick lines.
The range in gas densities $n$(H$_2$) goes from $\sim$10$^7$~cm$^{-3}$, only relevant to hot cores and protostellar envelopes, to nearly  10$^2$~cm$^{-3}$, relevant to the most extended {and FUV-illuminated} component of GMCs. As we are mostly interested in this component, these models
compute the HCN excitation for three different electron abundances: \mbox{$\chi_\mathrm{e}$\,=\,10$^{-4}$}, \mbox{2$\times$10$^{-5}$}, and 0.
 Figure~\ref{fig:hcn_WMCmod} shows model results (continuous curves) in the form
 of predicted line intensities \mbox{$W$(HCN~$J$\,=\,1--0}) as a function of  $n$(H$_2$).

The right panels in Fig.~\ref{fig:hcn_WMCmod}  show histograms with the distribution of  \mbox{$W$(HCN~$J$\,=\,1--0)} {detections ($>$\,3$\sigma$)} in individual pixels of the map. {The mean (median) intensity\footnote{{This value is higher than the line intensity of the average \mbox{HCN~$J$\,=\,1--0} spectrum
over the  full mapped
 area (see Table~\ref{tab:lines_stats}),  thus including emission free pixels. The mean and
 median \mbox{$W$(HCN~$J$\,=\,1--0)} intensity values computed
 considering pixels with CO~$J$\,=\,1--0
 detections above 3$\sigma$ are 1.0 and 0.7~\Kkms, respectively.}\label{foot:HCN-CO_W}} in these  pixels  is 1.4~\Kkms~
 (0.97~\Kkms). The pink shaded area in Fig.~\ref{fig:hcn_WMCmod} represents the 1$\sigma$ dispersion 
relative to the mean \mbox{$W$(HCN~$J$\,=\,1--0)} value. 
 However, while about 70\% of the observed intensities have a value below the mean, less than 1\% of the observed intensities have a value above 10~\Kkms\, (very bright HCN emission).}
In the following 
{we take} \mbox{$W$(HCN~$J$\,=\,1--0)}\,=\,1~\Kkms~
 as the reference\footref{foot:HCN-CO_W} for the extended cloud emission.  
Models with \mbox{$N$(HCN)=10$^{13}$~cm$^{-2}$} (Fig.~\ref{fig:hcn_WMCmod}a) {encompass this \mbox{$W$(HCN~$J$\,=\,1--0)} intensity level}. The gas 
temperature in this cloud component is $T_\mathrm{k}$\,$\simeq$\,30 to 60\,K (translucent gas and \mbox{UV-illuminated} cloud edges; see specific PDR models in Sect.~\ref{sec:Meudon}). 
Using $N$(HCN)\,=\,10$^{13}$\,cm$^{-2}$ 
{and neglecting electron collisional excitation} ($\chi_e$\,$=$\,0) we determine an upper
limit to the gas density of
\mbox{$n$(H$_2$)\,$\simeq$\,(1--3)\,$\times$\,10$^{4}$\,cm$^{-3}$}.

\begin{figure}[h]
    \centering
    \includegraphics[height=0.40\textwidth]{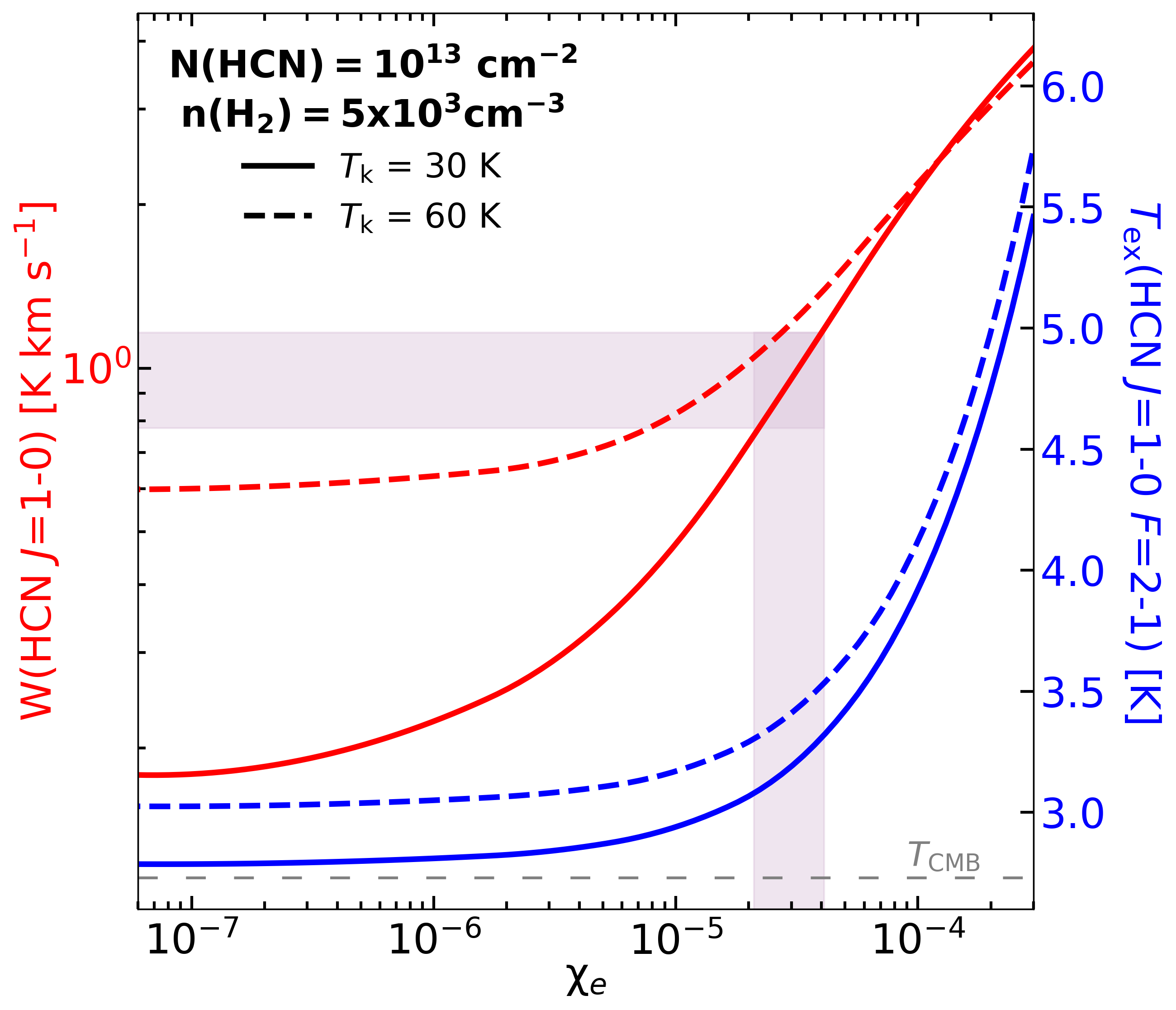} 
    \caption{\mbox{$W$(HCN~$J$\,=\,1--0)}
    (red curves) and  \mbox{$T_\mathrm{ex}$\,(HCN $J$=1--0 $F$=2--1)}
     (blue curves)  predicted by \mbox{non-LTE} radiative transfer models, appropriate to  extended and translucent gas, as a function of {\mbox{electron abundance}}. 
      The vertical pink shaded area intersects 
      {the typical \mbox{$W$(HCN~$J$\,=\,1--0)\,$=$\,1\,\Kkms} 
       intensity level ($\pm$\,20\%)}.}\vspace{-0.3cm}
    \label{fig:hcn_WTexXeMCmod}
\end{figure}

{\mbox{Figure~\ref{fig:hcn_WTexXeMCmod}} shows the effect of 
electron excitation predicted by radiative 
transfer models appropriate to this extended and translucent gas 
(adopting \mbox{$n$(H$_2$)\,=\,5\,$\times$10$^3$\,cm$^{-3}$})}.
{The plot shows how
\mbox{$W$(HCN~$J$\,=\,1--0)} (red curves) and  \mbox{$T_\mathrm{ex}$(HCN~$J$\,=\,1--0 $F$\,=\,2--1)} (blue curves) increase as the electron abundance $\chi_e$\, increases.
The reference intensity value, \mbox{$W$(HCN~$J$\,=\,1--0)\,=\,1\,\Kkms},  intersects
the model curves at an electron abundance of a few 10$^{-5}$
and \mbox{$T_\mathrm{ex}$\,$\simeq$\,3.2--3.5\,K}. These low excitation
temperatures imply weak collisional excitation, but still
$T_\mathrm{ex}$\,$>$\,$T_{\rm CMB}$}.

{Interestingly}, the extended \mbox{HCN~$J$\,=\,1--0} emission observationally correlates well
with the \CI~492\,GHz emission (see Fig.~\ref{fig:fullmaps120} and Sect.~\ref{subsec:porosity}). In addition, our photochemical models show 
that $\chi_e$ reaches $\gtrsim$\,10$^{-5}$  in the \CI~492\,GHz emitting cloud layers  (Figs.~\ref{fig:hcn_Meudon}a,b).
{For such high $\chi_e$ values,  electron excitation enhances the \mbox{HCN~$J$\,=\,1--0} emission at low gas densities 
\citep[see Fig.~\ref{fig:hcn_WTexXeMCmod} and ][]{Goldsmith17,Goicoechea2022}}.
Hence,  we estimate that the median gas density in the extended cloud component is \mbox{$n$(H$_2$)\,$\simeq$\,(4--7)\,$\times$\,10$^{3}$\,cm$^{-3}$}
{if $\chi_e$\,$\simeq$\,2$\times$10$^{-5}$}
({or} $\simeq$\,10$^3$\,cm$^{-3}$ if $\chi_e$\,$\simeq$\,10$^{-4}$).

{On the other hand,}
the strongest HCN-emitting regions in Orion\,B, those with $W$(HCN $J$\,=\,1--0)\,>\,6~\Kkms,
only represent  $\sim$15$\%$ of the total \mbox{HCN~$J$\,=\,1--0} luminosity in the map.
This bright HCN emission can only be reproduced by models
with $N$(HCN)=10$^{14}$~cm$^{-2}$ and higher gas densities, 
\mbox{$n$(H$_2$)\,$>$\,10$^{5}$\,cm$^{-3}$}.

\begin{table*}[!ht]
\caption{HCN $J$\,=\,1--0 HFS line intensity ratios,  
$W$(HCN $J$\,=\,2--1)\,/\,$W$(HCN $J$\,=\,1--0) and
$W$(HCN $J$\,=\,3--2)\,/\,$W$(HCN $J$\,=\,1--0)
line intensity ratios, and
 parameters derived from rotational population diagrams,
 {computed in Appendix~\ref{app:rot-diag}, toward}
 a sample of representative cloud positions.} 
\label{tab:TTNN}
    \centering
    \begin{threeparttable}
    
    \begin{tabular}{lccccccccccc@{\vrule height 7pt depth 3pt width 0pt}}
    \toprule
    Pos. &  \multicolumn{4}{c}{HCN \textit{J}=1--0}        & \multicolumn{2}{c}{Excited HCN}        &\multicolumn{4}{c}{Rotational diagrams}    \\\vspace{0.1cm}
        &   $R_{02}$ & $R_{12}$    & $R_{02}^{\Delta v}$ & $R_{12}^{\Delta v}$ & $R\frac{J=2-1}{J=1-0}$    & $R\frac{J=3-2}{J=1-0}$   & $T_{\mathrm{rot}}^\mathrm{thin}$ & $N^\mathrm{thin}$ & $T_{\mathrm{rot}}^\mathrm{\tau, corr}$ & $N^{\mathrm{\tau,\mathrm{corr}}}$        \\
          &  & & & & &  & [K]  & 10$^{13}$ [cm$^{-2}$] & [K]  & 10$^{13}$ [cm$^{-2}$]   \\
    \midrule 
    \#1  & 0.3 & 0.4 & 1.1  & 1.1  & 0.9$\pm$0.4 & 1.2$\pm$0.5 & 18$\pm$5 & 5$\pm$3 & 38$\pm$10 & 86$\pm$20          \\ 
    \#2   & 0.2 & 0.7 & 1.2  & 1.9 & 1.0$\pm$0.4 & 0.7$\pm$0.3  & 11$\pm$2 & 0.8$\pm$0.3 & 10$\pm$1           & 1.5$\pm$0.3            \\
    \#3   & 0.3 & 0.4 & 0.9 & 1.1 & 1.0$\pm$0.4  & 0.5$\pm$0.2  & 10$\pm$1 & 0.9$\pm$0.4 & 7$\pm$1  \\
    \#4   & 0.6 & 0.5 & 1.0  & 1.0 & 0.7$\pm$0.3 & 0.2$\pm$0.1 & 9$\pm$2 & 0.2$\pm$0.1 & 7$\pm$1 & 0.5$\pm$0.3  \\
    HH PDR  & 0.3 & 0.5 & 0.7 & 0.9 & 0.5$\pm$0.2 & 0.13$\pm$0.04  & 8$\pm$1 & 0.5$\pm$0.3  & 5$\pm$1   & 1.6$\pm$1.4  \\
    \#6  & 0.2 & 0.6 & 0.8 & 0.9 & 0.6$\pm$0.2  & 0.4$\pm$0.2 & 7$\pm$1 & 1.1$\pm$0.7 &  &  \\
    \#7   & 0.3 & 0.5 & 0.8 & 1.0 & 0.8$\pm$0.3  & 0.3$\pm$0.1 & 6.6$\pm$0.8 & 0.8$\pm$0.3 & 5.2$\pm$0.3 & 2.4$\pm$0.4  \\  
    \#8   & 0.5 & 0.6  & 0.8 & 1.0 & 0.7$\pm$0.3 & 0.3$\pm$0.1 & 6.1$\pm$0.6 & 0.5$\pm$0.1 & 4.2$\pm$0.2 & 3.0$\pm$0.7  \\
    HH  Core & 0.3 & 0.5 & 0.7  & 1.0 & 0.5$\pm$0.2 & 0.16$\pm$0.08 & 5.3$\pm$0.7 & 0.7$\pm$0.4 & 4.3$\pm$0.5   & 2.2$\pm$1.1  \\
    \#10    & 0.2 & 0.5 & 0.7 & 1.1 & 0.4$\pm$0.2 & 0.12$\pm$0.04 & 5.0$\pm$0.9 & 0.4$\pm$0.2 &  &  \\
    \#11   &  0.3   & 0.6  & 0.8 & 1.0 & 0.4$\pm$0.2 & 0.10$\pm$0.04 & 6.8$\pm$0.9           & 0.5$\pm$0.3    & &         \\
    \#12  & 0.4 &  0.5 & 0.8 & 0.8 & 0.3$\pm$0.1 & 0.10$\pm$0.04 & 4.7$\pm$0.9 & 0.7$\pm$0.5 & &  \\
     \#13  & 0.2 &  0.6 & 1.0 & 1.1 & 0.3$\pm$0.1 & 0.09$\pm$0.04 & 4.7$\pm$0.8 & 0.4$\pm$0.2 & &  \\
    \#14  & 0.4 &  0.5 & 1.0 & 1.1 & 0.3$\pm$0.1 & 0.06$\pm$0.02  & 4.4$\pm$0.5 & 0.4$\pm$0.2 & &  \\
    \bottomrule
    \end{tabular}
    \tablefoot{
    The uncertainty in R$_{02}$, R$_{12}$, R$^{\Delta v}_{02}$, and R$^{\Delta v}_{12}$ is between 10 and 20\%.} 
    \end{threeparttable}
\end{table*}

\begin{figure}[h]
    \centering
    \includegraphics[height=0.45\textwidth]{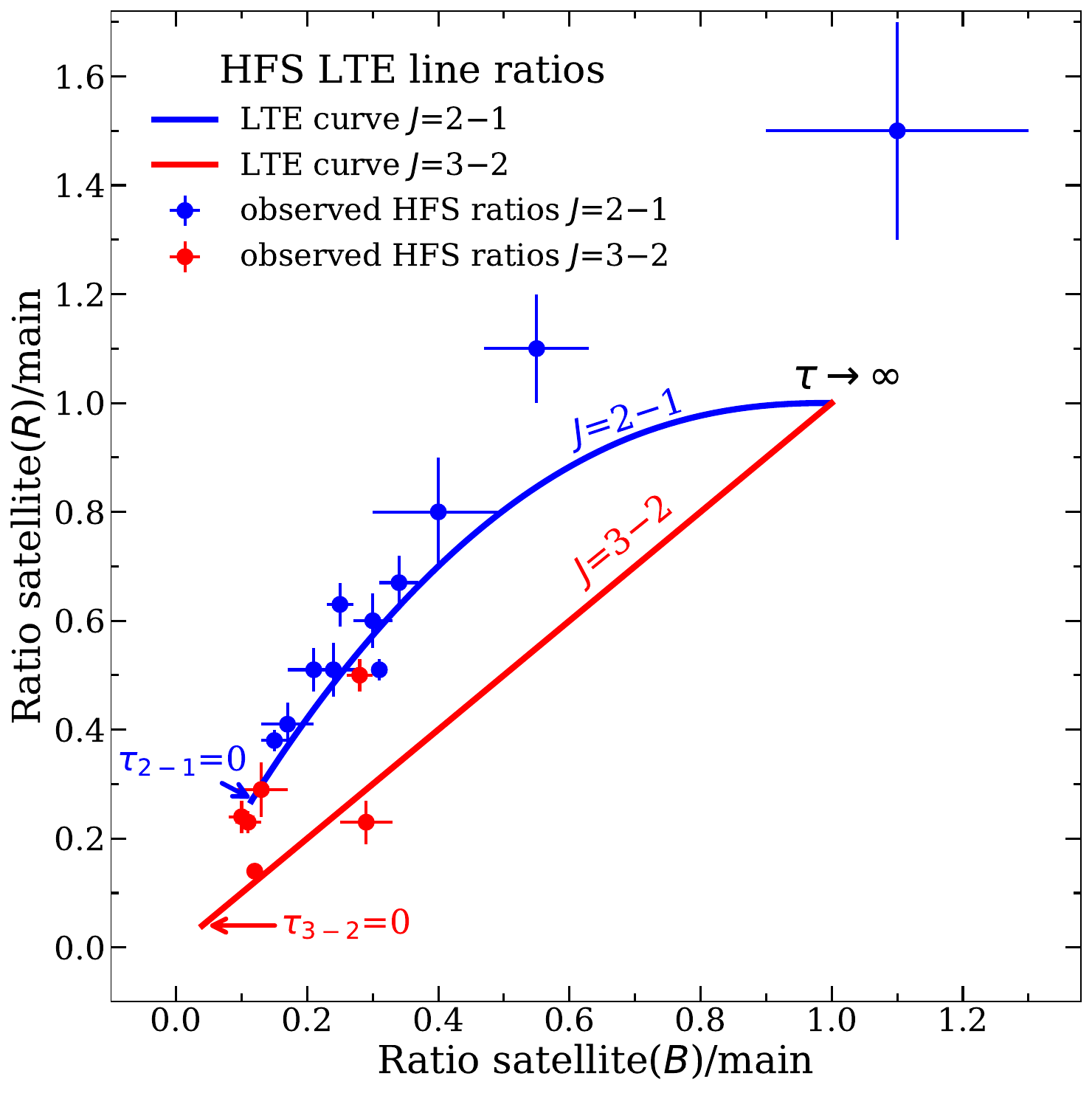} 
    \caption{{HCN~$J$\,=\,2--1 and $J$\,=\,3--2 HFS  intensity ratios \mbox{satellite($R$)/main} versus
    \mbox{satellite($B$)/main} (see \mbox{Sect.~\ref{sect:taupopdia}}  for their definition) in LTE and as line opacities increase. 
    Blue and red dots show the observed HFS line ratios toward the representative positions (see Table~\ref{tab:HCN-HFS_multiJ}).}}  
    \label{fig:hcn_hfs_21_32}
\end{figure}

\subsection{Rotationally excited HCN and \texorpdfstring{H$^{13}$CN}{H13CN} toward representative cloud environments in Orion B}\label{sect:taupopdia}

To complement our analysis of the \mbox{HCN~$J$\,=\,1--0} emission at large spatial scales, and to determine more accurate HCN column densities, here we analyze our multiple-$J$ HCN and H$^{13}$CN line observations 
toward 14 positions in Orion\,B (see Table~\ref{tab:Positions} for a brief explanation).
Figure~\ref{fig:hcn_multiJ_resume} shows a selection of the spectra.
We detect \mbox{HCN~$J$\,=\,2--1} and \mbox{$J$=3--2} toward  all positions, and 
\mbox{HCN $J$\,=\,4--3} toward five of the 14 observed positions
(Fig.~\ref{fig:HCN-multiJ} in the \mbox{Appendix} shows  the spectra of all observed positions).

The HCN~$J$=2--1 transition has six HFS lines that, for the narrow line widths
in Orion\,B, blend 
into three lines with apparent relative intensity ratios  \mbox{$\sim$1:9:2} in the LTE and optically thin limit (red vertical lines in Fig.~\ref{fig:hcn_multiJ_resume}).
The HCN~$J$=3--2 transition also has six HFS lines. Only the central ones are blended and cannot be spectrally resolved. This gives the impression of three lines with relative intensity ratios \mbox{1:25:1} in the LTE and optically thin limit {\citep[e.g.,][]{Ahrens02,Loughnane2012}}. 
 {We term these  three apparent components (blueshifted, central, and redshifted) of 
 the \mbox{$J$=2--1}  and \mbox{$J$=3--2} rotational lines 
 as ``satellite ($B$),'' ``main,'' and ``satellite ($R$),''  respectively.}
We recall that these overlapping lines in the \mbox{HCN~$J$\,=\,2--1} and 3--2 transitions are responsible of the observed anomalous \mbox{HCN~$J$\,=\,1--0} HFS line intensity ratios
 \citep[][and references therein]{Goicoechea2022}.

 {Blue and red curves in Fig.~\ref{fig:hcn_hfs_21_32} show the expected \mbox{HCN~$J$\,=\,2--1} and \mbox{$J$\,=\,3--2} HFS  intensity ratios  \mbox{satellite ($R$)/main} versus \mbox{satellite ($B$)/main} in  LTE
 as line opacities increase. Only when \mbox{$\tau_{2-1}$\,$\rightarrow$\,0} 
 and  \mbox{$\tau_{3-2}$\,$\rightarrow$\,0}, one should detect the 
\mbox{$\sim$1:9:2} and \mbox{$\sim$1:25:1} HFS ratios. The filled dots in \mbox{Fig.~\ref{fig:hcn_hfs_21_32}} show the observed ratios (summarized in {Table~\ref{tab:HCN-HFS_multiJ}} of the Appendix) toward the sample of representative positions that could be fitted with three Gaussian lines. This plot shows that 
several HCN~$J$\,=\,2--1,  and specially $J$\,=\,3--2,  HFS line intensity ratios do not lie on the LTE curves even for elevated line opacities. That is, the emission of rotationally excited HCN lines  can also be anomalous. }

\subsubsection{{HCN and HNC rotational diagrams}}
\label{subsec:DRs_HCN_HNC}

{Here we estimate the degree of excitation (by determining \mbox{rotational} temperatures, $T_\mathrm{rot}$) and column densities of HCN and HNC toward the sample of representative positions.}
We  analyze the detected  rotationally excited HCN (up to \mbox{$J$\,=\,4--3}) and HNC (up to \mbox{$J$\,=\,3--2}) lines   by constructing  \mbox{rotational} population diagrams {in Appendix~\ref{app:rot-diag}} {\citep[][]{Goldsmith_1999}}. 
{We derive $N$(HCN), and
$N$(HNC) ignoring their HFS structure (i.e., only the total line intensity of each rotational transition matters). This is a valid approximation 
to obtain $T_\mathrm{rot}$ from observations of multiple-$J$ lines.}
We  derive  the HCN
column density and rotational temperature under the assumption of optically thin emission ($N^\mathrm{thin}$ and $T_\mathrm{rot}^\mathrm{thin}$). We also determine  
their opacity-corrected
values ($N^{\tau,\mathrm{corr}}$ and $T_\mathrm{rot}^{\tau,\mathrm{corr}}$) 
by using the  H$^{13}$CN line {intensities}
(see \mbox{Fig.~\ref{fig:HCN-multiJ}}) and assuming that  {the emission from} HCN
and  H$^{13}$CN lines arise from the same gas.
Except  for the brightest position \#1, the derived HCN rotational temperatures range from 4 to 10~K {(i.e. subthermal excitation)}, and  $N^\mathrm{\tau, corr}$(HCN) ranges from 5$\times$10$^{12}$ to  3.4$\times$10$^{13}$~cm$^{-2}$. 
Table~\ref{tab:TTNN} summarizes the derived values and
\mbox{Appendix~\ref{app:rot-diag}}  shows  the resulting rotational diagram {plots}. 
{We employ the same methodology for HNC and HN$^{13}$C. Rotational temperatures
are also low, from 5 to 11~K. HNC column densities range from
$\sim$10$^{12}$ to 1.6$\times$10$^{13}$~cm$^{-2}$ (see Table~\ref{tab:HNC_TexN}).}

\subsubsection{{Comparison with single-component non-LTE models}}

{Most  of the observed representative positions likely have  velocity, temperature, and density gradients (specially prestellar cores and protostars). However, carrying {out a} complete, source-by-source, radiative transfer analysis  is beyond the scope of this study (more focused on the extended cloud component). 
Here we just used the outputs of the grid of single-component and static models computed by \cite{Goicoechea2022}, and presented in Sect.~\ref{sect:comp-models}, to estimate the  physical conditions (gas temperature and densities) compatible by the detected  rotationally excited HCN line emission.} 
{We compared} the observed line intensity ratios \mbox{$R^J_{21}$=$W$(HCN~$J$=2--1)/$W$(HCN~$J$=1--0)} and \mbox{$R^J_{31}$=$W$(HCN~$J$=3--2)/$W$(HCN~$J$=1--0)} summarized in \mbox{Table~\ref{tab:TTNN}}
with the models {shown in \mbox{Fig.~10} of} \cite{Goicoechea2022}. 
As an example, the observed line ratios 
toward the Horsehead PDR  are \mbox{$R^J_{21}$\,=\,0.5} and \mbox{$R^J_{31}$\,=\,0.13}. These ratios can be explained by models with  \mbox{$T_\mathrm{k}$\,=\,30-60~K} and $n$(H$_2$) of a few 10$^{4}$~cm$^{-3}$.
\mbox{Position \#14} shows the lowest line ratios of the sample, \mbox{$R^J_{21}$\,=\,0.3} and  \mbox{$R^J_{31}$\,=\,0.06}, which is consistent with \mbox{$n$(H$_2$) of a few
10$^4$\,cm$^{-3}$}. 
On the other hand, the observed line intensity ratios toward position \#1 
(center of NGC~2024)
are \mbox{$R^J_{21}$\,=\,0.9} and \mbox{$R^J_{31}$\,=\,1.2}. {In this position we derive}
 the highest HCN rotational temperature  ($\sim$38$\pm$10~K).
The observed \mbox{$R^J_{31}$\,>\,$R^J_{21}$} intensity ratios 
{are consistent with the presence of} dense gas, $n$(H$_2$)\,$\geq$\,10$^6$~cm$^{-3}$.

\subsection {HCN/HNC intensity and column density ratios}

\mbox{Table~\ref{tab:Trot_HCN-HNC_Nratios}} summarizes the resulting  $N$(HCN)/$N$(HNC) column density {ratios} {obtained from rotational population
diagrams} (Sect.~\ref{subsec:DRs_HCN_HNC})  as well as
{$W$(HCN)/$W$(HNC) \mbox{$J$\,=\,1--0} and \mbox{$J$\,=\,3--2}} line intensity ratios.
The positions that host  lower excitation conditions (e.g., low $T_{\rm rot}$(HNC)) tend to have lower  $N$(HCN)/$N$(HNC) column and $W$(HCN)/$W$(HNC) intensity ratios.   
Mildly \mbox{FUV-illuminated} environments such as the \mbox{Horsehead} nebula show  $W$(HCN)/$W$(HNC)  line intensity ratios of about two, whereas the most \mbox{FUV-shielded} cold cores and their surroundings (positions \#7, \#8, \#10, \#13) display ratios of about one. On the other hand, the most FUV-irradiated and densest cloud environments (those with \mbox{HCN $J$\,=\,4--3} detections in NGC\,2024) show ratios of at least four (positions \#1 and \#2). 

\begin{figure}[!t]

    \mbox{\includegraphics[width=0.47\textwidth]{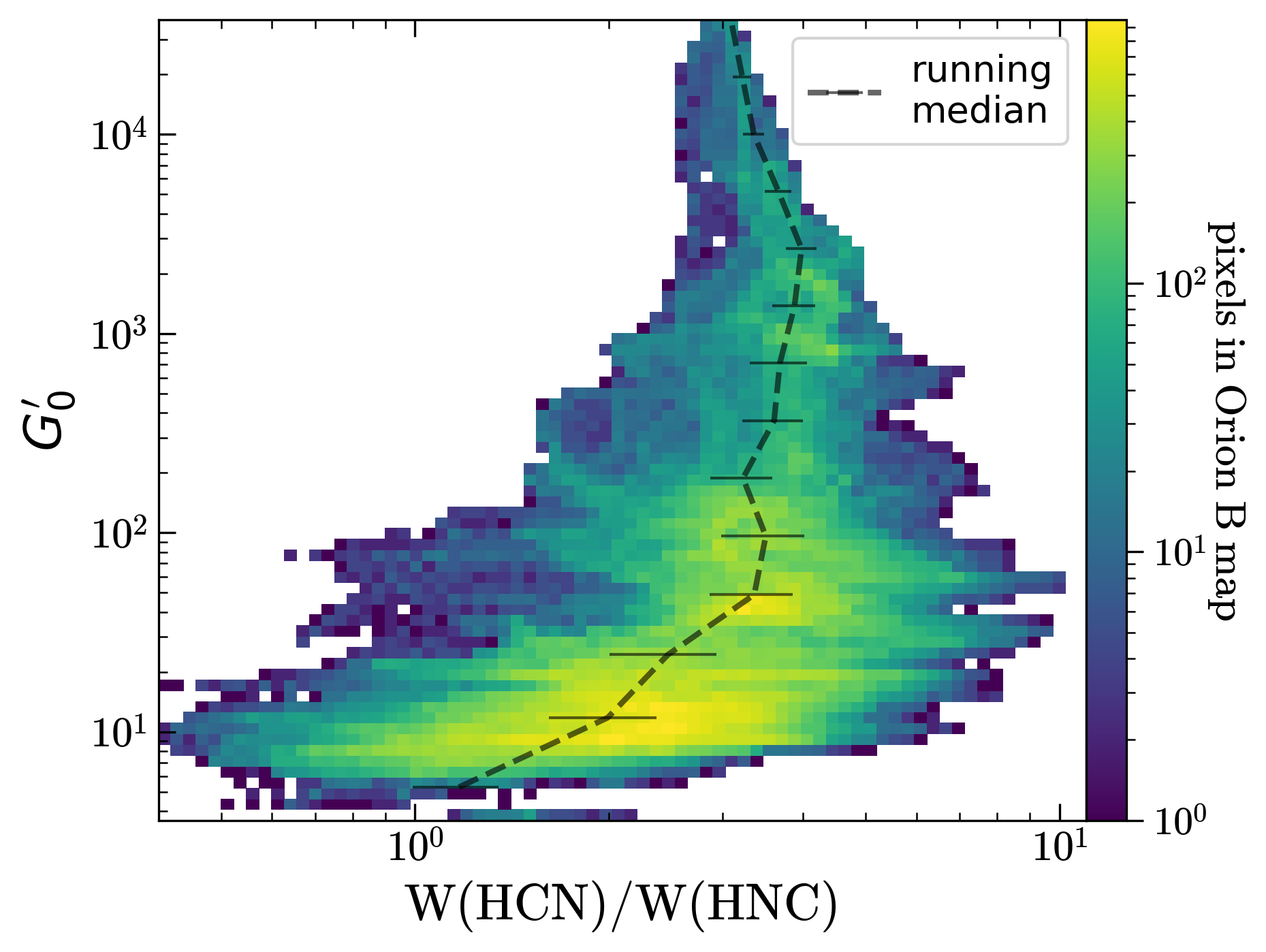}}
    
    \caption{2D histogram of the observed HCN/HNC $J$=1--0 line  intensity ratio  as  a function of $G_0'$ in the Orion\,B map. The dashed black curve shows the running median. The error bars show the standard deviation.} 
    \label{fig:HCN-HNC_Ifir}
\end{figure}

These results  roughly agree with the spatial {correlation}
between the \mbox{HCN/HNC $J$\,=\,1--0} integrated line intensity ratio
and $G_0'$ in the entire region. 
Figure~\ref{fig:HCN-HNC_Ifir} shows 
a 2D histogram of the observed \mbox{$W$(HCN)/$W$(HNC) $J$\,=\,1--0}  intensity ratio as 
a function of $G_0'$ in the mapped region. The running median \mbox{HCN/HNC $J$\,=\,1--0} intensity ratio increases
from $\simeq$\,1 at $G_0'$\,$\simeq$10 to  $\simeq$\,3 at $G_0'$\,$\simeq$\,200.
For higher values of $G_0'$, the running median intensity ratio stays roughly constant at 
\mbox{$\simeq$\,3-4}. 
We estimate that the higher  \mbox{HCN $J$\,=\,1--0} line opacity 
toward these bright positions ({at least 
{$\tau$}\,$\simeq$\,5--10}{, see Table~\ref{tab:HCN-HFS_multiJ}})
compared to that of \mbox{HNC $J$\,=\,1--0} (on the order of 
{$\tau$}\,$\simeq$\,1--3,\mbox{Table~\ref{tab:HNC_TexN}}) 
contributes to the observed constancy of the line intensity ratio.

\begin{table}[b]
\caption{HNC rotational temperatures, HCN/HNC column density, and line intensity ratios toward selected positions in Orion B.}
\label{tab:Trot_HCN-HNC_Nratios}  
\centering
\begin{threeparttable}
 \resizebox{0.4\textwidth}{!}{%
\begin{tabular}{lcccc@{\vrule height 7.5pt depth 5pt width 0pt} }
\toprule
 Pos.    &  $T_\mathrm{rot}$(HNC) &        $\frac{N(\mathrm{HCN})}{N(\mathrm{HNC})}^\dagger$  &  $\frac{W(\mathrm{HCN})}{W(\mathrm{HNC})}$ & $\frac{W(\mathrm{HCN})}{W(\mathrm{HNC})}$ \\
    &      &  & $J$=1--0 &  $J$=3--2 \\
  \midrule
\#1 & 11\,$\pm$\,2 &  3 -- 54 & 4 & 4 \\
\#2 & 9\,$\pm$\,2 &   4 -- 7.5 & 4 &  3  \\
\#3 & 7\,$\pm$\,1 &  3 -- 14 &   4 &  5  \\
\#4 & 6\,$\pm$\,1 &   0.7 -- 1.7  & 0.8 & 0.6 \\
HH PDR  & 5\,$\pm$\,1 & 1.7 -- 5 &  2 & 2  \\
\#6 & 6\,$\pm$\,1 &    3  &  3 & 5  \\
\#7 & 8\,$\pm$\,1 &    1.2 --  4 & 1 &  0.6 \\
\#8 & 8\,$\pm$\,1  &    0.8 -- 5  & 1 & 0.5 \\
HH Core & 6\,$\pm$\,1 &  2 --  5  & 2 & 1.6 \\
\#10 & 6\,$\pm$\,1 &   1.3  & 1  & 0.8 \\
\#11 & --  &  --  & 3  & -- \\
\#12 & 6\,$\pm$\,1 &   1.7  &  2 & 1.2 \\
\#13& 5\,$\pm$\,1 &    1.3  &  1  & 0.8 \\
\#14& 5\,$\pm$\,1 &   4  & 3  & 2 \\
\bottomrule \vspace{-0.6cm}      
\end{tabular}} 
\tablefoot{{$^\dagger$}{The lower value of the ratio adopts column densities obtained from
rotational diagrams in the optically thin limit. The higher value of the ratio
implements a line opacity correction (see Tables~\ref{tab:TTNN} and \ref{tab:HNC_TexN})} .}
\end{threeparttable}
\end{table}

In  FUV--illuminated environments, the strength of the radiation field influences the gas chemistry and determines much of the gas temperature and electron abundance (see Sect.~\ref{sec:Meudon}). 
{At a given abundance, HNC responds more weakly to electron excitation than HCN.}
In particular, {the \mbox{HCN $J$\,=\,1--0} critical fractional abundance of electrons
(\mbox{$\chi_{\rm cr,e}^{*}$} in \mbox{Table~\ref{tab:spec_n}})
is a factor of $\sim$\,4 lower than that of \mbox{HNC $J$\,=\,1--0}}. Moreover, HNC
 is typically less abundant in \mbox{FUV-illuminated} gas (\mbox{Sect.~\ref{sec:Meudon}}). 

{In addition,} as HCN and HNC rotational lines become optically thick, HFS line overlap effects become important for both species.
{However,} their relative effect as a function of $J$ are different \citep{Daniel2008}.
These aspects ultimately drive their excitation and contribute to the 
slightly different rotational temperatures we infer for the two species. Still,
modeling the HFS resolved excitation of HNC is beyond the scope of our study.
{Future determinations of \mbox{HFS-resolved} \mbox{HNC--H$_2$} inelastic collision rate
coefficients will make such detailed studies feasible}.

\begin{figure*}[h]
    \centering
    \includegraphics[width=0.8\textwidth]%
    {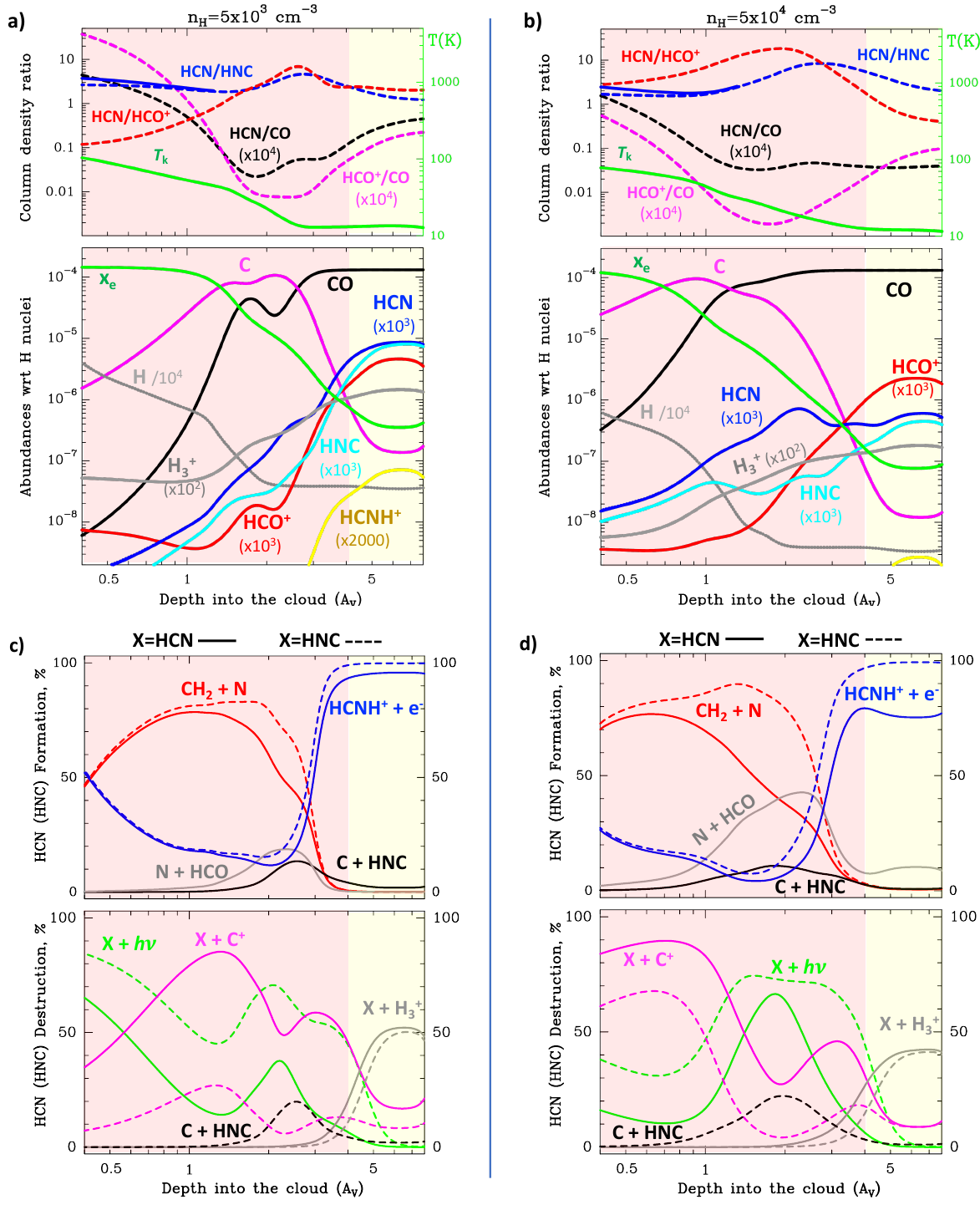}
    \caption{Constant density gas-phase PDR models with  $G_0$=100 and $n_\mathrm{H}$=5$\times$10$^3$~cm$^{-3}$ (\textit{left}) and 5$\times$10$^4$~cm$^{-3}$ (\textit{right}). 
    {These models adopt $E_b$\,=\,1200\,K for Reaction~(\ref{reac:isom_H})}.
    \textit{Upper panels} in (a) and (b):  {Dashed curves show the depth-dependent column
    density ratios of selected species (left y-axis)}.
    {The blue continuous curves in the upper panels of (a) and (b) show the HCN/HNC column density ratio
    adopting  $E_b$\,=\,200\,K}.
    Green continuous curves  show the temperature structure as a function of $A_\mathrm{V}$ (right y-axis). 
    \textit{Lower panels in (a) and (b):} Abundance profiles with respect to H nuclei. $(c)$ and $(d)$:
    Contribution (in percent) of the main formation and destruction reactions for HCN (continuous curves) and HNC (dashed curves).\vspace{-0.3cm}}  
    \label{fig:hcn_Meudon}
\end{figure*}

\section{HCN and HNC chemistry in FUV-illuminated gas}\label{sec:Meudon}

To {guide} our interpretation of the extended \mbox{HCN $J$\,=\,1--0} emission, here  we reassess  the chemistry of HCN, HNC, and related species in FUV-illuminated gas. The presence of FUV photons, C$^+$ ions, C atoms, and high electron abundances
triggers a distinctive nitrogen chemistry, different to that prevailing in
   cold and dense cores \mbox{\citep[e.g.,][]{Hily-Blant2010}} shielded from FUV radiation. 
\mbox{\citet{Sternberg1995}}, \citet{Young_Owl2000} and \citet{Boger2005} previously studied the formation
and destruction of HCN in \mbox{FUV-irradiated gas}. Here we used an updated version of the Meudon PDR code \citep{LePetit2006} that implements 
a detailed treatment of the penetration of FUV photons \mbox{\citep{Goicoechea2007}}
and includes $v$-state-dependent reactions of \mbox{FUV-pumped} \mbox{H$_2$($v$)} (hereafter H$_{2}^{*}$) with neutral N atoms
leading to \mbox{NH\,$+$\,H} \mbox{\citep{Goicoechea2022b}}, as well as reactions of \mbox{$o$-H$_2$} and \mbox{$p$-H$_2$} with N$^+$ ions, leading to \mbox{NH$^+$\,$+$\,H~
\citep{Zymak2013}}.

We also included the  isomerization reaction:
\begin{equation}
\rm{HNC\,+H\,\rightarrow \,HCN\,+\,H},
\label{reac:isom_H}
\end{equation}
with a rate coefficient \mbox{$k$($T$)\,=\,10$^{-10}$\,exp\,($-E_b/T$)},  where 
$E_b$ is the reaction energy barrier. Theoretical calculations agree on
the presence of a barrier, however, different methods provide slightly different barrier
heights: $\sim$2130\,K \citep{Talbi1996}, $\sim$1670\,K \citep{Sumathi1998},
and $\sim$960\,K \citep{Petrie2002}. 
In our models we initially adopt  \mbox{$E_b$\,=\,1200\,K} \citep[see][]{Graninger2014}.
We also included the isomerization reaction:
\begin{equation}\label{reac:c_iso}
\rm{HNC\,+C\,\rightarrow \,HCN\,+\,C},
\end{equation}
which is generally not included in {dark cloud} chemical models but plays a role in FUV-illuminated gas. 
We adopt a rate coefficient \mbox{$k$($T$)\,=\, 1.6$\times$10$^{-10}$\,cm$^3$\,s$^{-1}$} and no energy barrier \citep[from calculations by][]{Loison14,Loison2015}.

In order to {accurately} treat the photochemistry of HCN and HNC, our models explicitly integrate their photodissociation
and photoionization cross sections at each cloud depth. We use the wavelength-dependent  cross sections
tabulated in \citet{Heays2017}, which include a theoretical calculation of the HNC photodissociation cross section by \citet{Aguado2017}. For the interstellar radiation field, this cross section implies that  HNC is photodissociated
about two times faster than HCN.

We adopted a H$_2$ cosmic ray ionization rate 
$\zeta_{\rm CR}$ of \mbox{10$^{-16}$\,s$^{-1}$}, typical of translucent gas and cloud edges
in the disk of the galaxy \citep[e.g.,][]{Indriolo2015}. We assumed standard interstellar dust grain properties
and extinction laws.  We ran photochemical models adapted to the illumination conditions and gas densities
  at large  scales in Orion\,B. {In particular, we adopted a representative FUV  field of \mbox{$G_0$\,=\,100}, typical of the  Horsehead edge, the IC\,434 ionization front, and close to the mean $G_0$
  in the mapped area (see \mbox{Table~\ref{tab:SED_values})}}. Nonetheless, we note that adopting lower $G_0$ values basically shifts the abundance profiles to lower cloud depths but the following chemical discussion remains very similar.
 \mbox{Figure~\ref{fig:hcn_Meudon}} shows the predictions of constant density models, with
  \mbox{$n_{\rm H}$\,=\,5$\times$10$^{3}$\,cm$^{-3}$} (left panels) and 
   \mbox{$n_{\rm H}$\,=\,5$\times$10$^{4}$\,cm$^{-3}$} (right panels).
  Figures~\ref{fig:hcn_Meudon}a and \ref{fig:hcn_Meudon}b
  show the predicted
 {column density ratios} (upper panels) and abundance\footnote{Because at low $A_\mathrm{V}$ the abundance of H atoms can be significant,
 in this Section we provide the abundance of a given species~($x$) with respect to H nuclei.
 That is, \mbox{$x$(species)\,=\,$n$(species)\,/\,$n_{\rm H}$}, where 
 \mbox{$n_{\rm H}$\,=\,$n$(H)\,+\,2$n$(H$_2$)}. If the abundance of H atoms is negligible, then
 \mbox{$x$(species)\,=\,0.5\,$\chi$(species)}.} 
 profiles (lower panels) as a function of cloud depth, in mag of visual extinction\footnote{In these 1D PDR models, the cloud depth or shielding ($A_\mathrm{V}$ in mag of visual extinction)  refers to the extinction normal to the cloud surface and parallel to the FUV illumination direction. In general, this extinction is different from $A_\mathrm{V}$ determined from observations and the dust SED along a given  line of sight. Only for a face-on cloud (with the illuminating stars in the observed line of sight) both magnitudes are equivalent.}.

{We determine molecular column densities at a given cloud depth $A_\mathrm{V}$
(or cloud path length $l$)
by  integrating the predicted depth-dependent abundance profile, \mbox{$x$(species)}, 
from \mbox{0 to $A_\mathrm{V}$}:}
\begin{equation}
{ N(l) = \int_{0}^{{\rm{A_\mathrm{V}}}}  x(l) \, n_{\rm H} \, dl,}
\label{eq:PDR_column}
\end{equation}
{where \mbox{$x$($l$)} is the species abundance, with respect to H nuclei$^7$, at a cloud path length $l$}.

\subsection{Chemistry at cloud edges, \texorpdfstring{$A_\mathrm{V}<4$\,mag}{Av<4~mag}}\label{sec:Meudon_0-4}

\begin{figure}[!t]
    \centering
    \includegraphics[width=0.36\textwidth]{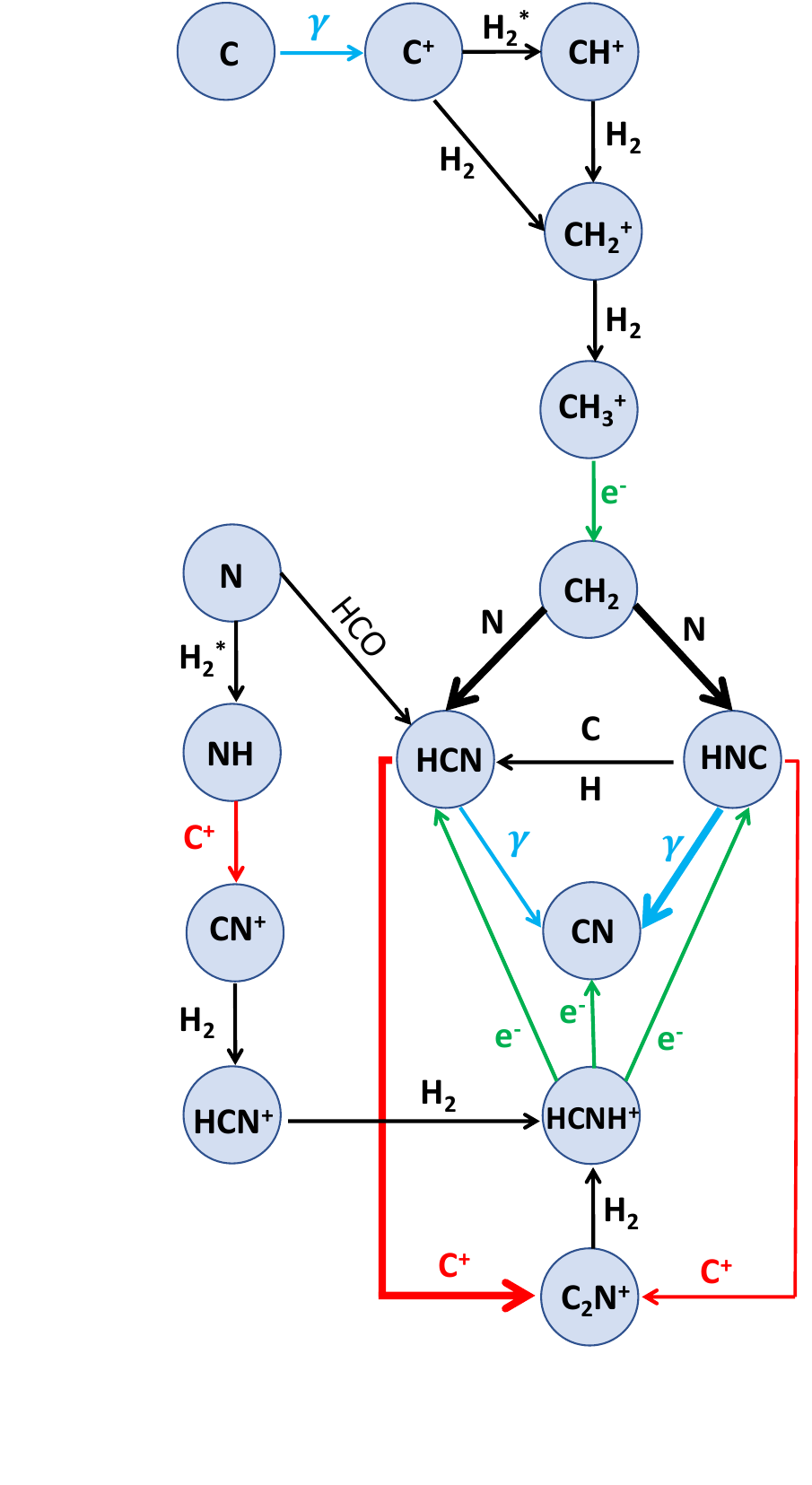} \vspace{-1.cm}
    \caption{Dominant chemical reactions in FUV-illuminated gas.} 
    \label{fig:app_PDRchem}
\end{figure}

The red shaded areas in \mbox{Fig.~\ref{fig:hcn_Meudon}} show model results
for \mbox{$A_\mathrm{V}<4$\,mag} {typical of FUV-illuminated cloud edges}.
FUV photons drive the chemistry in these translucent layers that host the C$^+$ to C transition and have high electron abundances: from $x_{\rm e}$\,$\simeq$\,$x$(C$^+$)\,$\simeq$\,10$^{-4}$ 
to  $x_{\rm e}$\,$\simeq$\,10$^{-6}$ depending on $A_\mathrm{V}$ and $G_0$/$n_{\rm H}$.
{To simplify our chemical discussion,} Fig.~\ref{fig:app_PDRchem} 
{summarizes} the  network of dominant chemical reactions at $A_\mathrm{V}<4$\,mag. {Wherever} C$^+$ is abundant, reactions of  CH$_2$ with N atoms dominate the formation of HCN and HNC, {as shown by the red curves in Fig.~\ref{fig:hcn_Meudon}c and \ref{fig:hcn_Meudon}d}.
{These two figures show the contribution (in percent)
of the main HCN and HNC formation and destruction reactions as a function of 
cloud depth. The second most important path for HCN and HNC formation
at \mbox{$A_\mathrm{V}<4$\,mag} is HCNH$^+$ dissociative recombination.  HCN and HNC \mbox{destruction} is governed by photodissociation and  by reactions  with C$^+$}. Their exact contribution depends on the gas density and $G_0$. {Our model assumes that} the  rate coefficient of reactions \mbox{C$^+$\,$+$\,HCN} and \mbox{C$^+$\,$+$\,HNC}, as well as the branching ratios of dissociative recombination 
\mbox{HCNH$^+$\,$+$\,e\,$\rightarrow$\,HCN/HNC\,$+$\,H}, are  {identical} for both isomers \citep[e.g.,][]{Semaniak2001}. Therefore, the \mbox{$N$(HCN)/$N$(HNC)} column density ratio at  \mbox{$A_\mathrm{V}<4$\,mag}
basically depends on the differences between  HCN and HNC photodissociation cross sections. 
{Wherever photodissociation dominates (e.g., green curves
in \mbox{Figs.~\ref{fig:hcn_Meudon}c} and \ref{fig:hcn_Meudon}d), we predict
\mbox{$N$(HCN)/$N$(HNC)\,$\simeq$\,1.5--2.5}. These values are consistent with
the ratio inferred toward the rim of the  Horsehead, a nearly edge-on PDR 
(see Table~\ref{tab:Trot_HCN-HNC_Nratios})}.

Neutral atomic carbon reaches its abundance peak at \mbox{$A_\mathrm{V}$\,$\simeq$\,1--3\,mag} (depending on $n_{\rm H}$), {which is relevant to understand the nature
of the extended \CI\,492\,GHz emission in Orion~B (\mbox{Fig.~\ref{fig:fullmaps120}h})}. The isomerization reaction \mbox{C\,+\,HNC\,$\rightarrow$\,HCN\,+\,C} 
as well as reaction \mbox{N\,+\,HCO\,$\rightarrow$\,HCN\,+\,O}
provide additional formation paths for HCN at 
\mbox{$A_\mathrm{V}$\,$<$\,4\,mag} (black and gray curves in 
\mbox{Figs.~\ref{fig:hcn_Meudon}c} and \ref{fig:hcn_Meudon}d). These two reactions enhance the HCN/HNC 
column density ratio  to $\sim$5-15 at \mbox{$A_\mathrm{V}$\,$\simeq$\,3\,mag}. These  ratios  {agree with the high $N$(HCN)/$N$(HNC)  ratios we
infer toward  NGC\,2024 (e.g., positions \#1 and \#2 in  Table~\ref{tab:Trot_HCN-HNC_Nratios})}. 

{We end this subsection by giving the HCN and HNC column densities
predicted by the \mbox{$n_{\rm H}$\,$=$\,5$\times$10$^3$\,cm$^{-3}$}
(5$\times$10$^4$\,cm$^{-3}$) models 
at \mbox{$A_\mathrm{V}$\,$=$\,4\,mag}:}
 \mbox{$N$(HCN)\,=\,4.5\,$\times$\,10$^{12}$\,cm$^{-2}$} 
 \mbox{(2.4\,$\times$\,10$^{12}$\,cm$^{-2}$)}
 and 
\mbox{$N$(HNC)\,=\,1.8\,$\times$\,10$^{12}$\,cm$^{-2}$}
 \mbox{(4.5\,$\times$\,10$^{11}$\,cm$^{-2}$)}.
{These column densities are representative of  extended and translucent gas}.

\subsection{Intermediate depths, \texorpdfstring{4\,mag\,$<\!A_\mathrm{V}\!<$\, 8\,mag}{4~mag<Av<8~mag}}\label{sec:Meudon_4-8}
 
\begin{figure}[!t]
    \centering
    \includegraphics[width=0.36\textwidth]{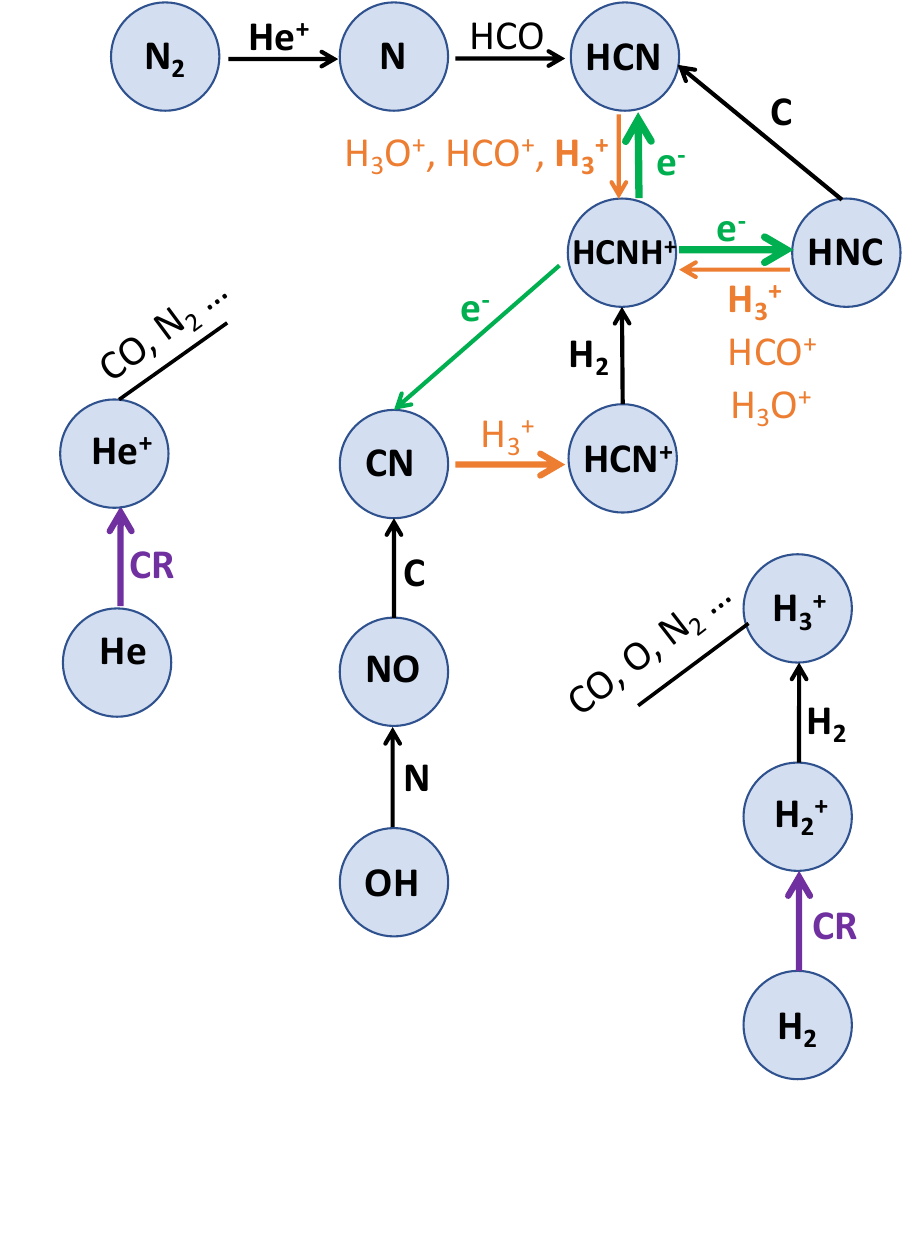}
    \vspace{-1.cm}
    \caption{Dominant chemical reactions in FUV-shielded gas.} 
    \label{fig:app_deepchem}
\end{figure}

The yellow shaded areas in \mbox{Fig.~\ref{fig:hcn_Meudon}} show model results for
\mbox{4\,mag\,$<\!A_\mathrm{V}\!<$\, 8\,mag}.  In these intermediate-depth cloud layers, the FUV flux diminishes and most  carbon becomes locked in CO.  
 \mbox{Figure~\ref{fig:app_deepchem}}  summarizes the dominant chemical reactions 
in these molecular cloud layers.
{As shown in Figs.~\ref{fig:hcn_Meudon}c and \ref{fig:hcn_Meudon}d}, HCN and HNC are now predominantly destroyed by reactions with abundant molecular and atomic ions (H$_{3}^{+}$ and C$^+$ at low densities, H$_{3}^{+}$, HCO$^+$, and H$_3$O$^+$ at higher densities). 
The main formation route for HCN and HNC switches to HCNH$^+$ dissociative recombination \mbox{(blue curves in Figs.~\ref{fig:hcn_Meudon}c and \ref{fig:hcn_Meudon}d)}. For equal branching ratios \citep[][]{Semaniak2001}, the predicted {$N$(HCN)/$N$(HNC)} column density ratio is \mbox{$\simeq$\,1--2}. Indeed, our observations of Orion\,B reveal  {$N$(HCN)/$N$(HNC)} 
ratios and 
\mbox{$W$(HCN $J$=1--0)/$W$(HNC $J$=1--0)} line intensity ratios 
of \mbox{$\simeq$\,1--2}  toward positions with low $G_0'$ values  {(see \mbox{Table~\ref{tab:Trot_HCN-HNC_Nratios}} and  Fig.~\ref{fig:hcnratios_120})}.

The abundance of HCNH$^+$, {the precursor of HCN and HNC at
large $A_\mathrm{V}$ (Figs.~\ref{fig:hcn_Meudon}c and \ref{fig:hcn_Meudon}d)}, depends on the H$_{3}^{+}$ abundance, {which is sensitive to the penetration of FUV radiation 
and to the cosmic ray ionization rate}.
The H$_{3}^{+}$ abundance is higher  at lower $n_{\rm H}$ because
{the higher penetration of}
FUV radiation 
reduces the abundances of the neutral species (CO, O, N$_2$, and S) that destroy H$_{3}^{+}$. 
In addition, the H$_{3}^{+}$ abundance  scales with $\zeta_{\rm CR}$.
{We run a few} models with $\zeta_{\rm CR}$ rates significantly lower than assumed in Fig.~\ref{fig:hcn_Meudon} {and indeed they produce} lower HCNH$^+$ abundances {than those shown in Figs.~\ref{fig:hcn_Meudon}a
and \ref{fig:hcn_Meudon}b}. This leads to higher HCN/HNC abundance ratios
\citep[see also ][]{Behrens2022} because reaction:
\begin{equation}\label{reac:n_HCO}
\rm{N\,+HCO\,\rightarrow \,HCN\,+\,O},
\end{equation}
becomes more important than HCNH$^+$ dissociative recombination. 
Reaction~(\ref{reac:n_HCO}) is often quoted in chemical networks \citep{Mitchel84,Young_Owl2000} but no detailed study seems to exist. 

{We end this subsection by  {providing   
HCN and HNC column densities predicted at}  \mbox{$A_\mathrm{V}$\,$=$\,8\,mag}}.
The PDR model with \mbox{$n_{\rm H}$\,$=$\,5$\times$10$^3$\,cm$^{-3}$} 
{(\mbox{5$\times$10$^4$\,cm$^{-3}$})}
predicts \mbox{$N$(HCN)\,=\,6.2\,$\times$\,10$^{13}$\,cm$^{-2}$}
{(\mbox{6.4\,$\times$\,10$^{12}$\,cm$^{-2}$})}
and 
\mbox{$N$(HNC)\,=\,5.2\,$\times$\,10$^{13}$\,cm$^{-2}$}  
{(\mbox{3.2\,$\times$\,10$^{12}$\,cm$^{-2}$})}. 
{These column densities encompass the
range of HCN  (see Table~\ref{tab:TTNN}) and
HNC (see Table~\ref{tab:HNC_TexN}) column densities we infer toward the 
observed sample
of representative positions in Orion~B.}
\subsection{On HNC destruction reactions}\label{sec:isomerization}
Previous studies invoked that the isomerization reaction
\mbox{H\,$+$\,HNC\,$\rightarrow$\,HCN\,$+$\,H} determines a temperature dependence of
the  \mbox{$N$(HCN)/$N$(HNC)} ratio in warm molecular gas \citep{Schilke1992,Herbst2000,Graninger2014,Hacar2020}. 
In our PDR  models, the gas temperature is \mbox{$T$\,$\simeq$\,50\,K} at \mbox{$A_\mathrm{V}$\,$\simeq$\,1\,mag},
and \mbox{$T$\,$\simeq$\,15\,K} at \mbox{$A_\mathrm{V}$\,$\simeq$\,4\,mag} (upper panels of \mbox{Fig.~\ref{fig:hcn_Meudon}a} and \mbox{\ref{fig:hcn_Meudon}b}).
{We run the same two  models adopting $E_b$\,=\,200~K
for \mbox{reaction~(\ref{reac:isom_H})}} and found that
 reducing $E_b$  has little effect on the predicted \mbox{$N$(HCN)/$N$(HNC)} ratio
{(blue continuous curves in the upper panels of 
Figs.~\ref{fig:hcn_Meudon}a and \ref{fig:hcn_Meudon}b).
Even at $A_\mathrm{V}$\,$<2$\,mag, where the abundance of H atoms and $T$ are moderately high,
the \mbox{$N$(HCN)/$N$(HNC)} ratio increases by less than 30\,\%
(i.e., the effects are very small)}.
We note that in all these models,  HCN and HNC photodissociation, as well as
\mbox{C\,$+$\,HNC\,$\rightarrow$\,HCN\,$+$\,C} reactions, are faster than the isomerization reaction \mbox{H\,$+$\,HNC} (see Figs.~\ref{fig:hcn_Meudon}c and \ref{fig:hcn_Meudon}d).

{We run a more extreme model adopting $E_b$\,=\,0\,K. That is to say, as if
\mbox{reaction~(\ref{reac:isom_H})} was barrierless. Only in this case,
the isomerization reaction \mbox{H\,$+$\,HNC\,$\rightarrow$\,HCN\,$+$\,H}  would dominate HNC destruction 
(specially at large $A_\mathrm{V}$), increasing the 
\mbox{$N$(HCN)/$N$(HNC)} ratio. However, this choice of $E_b$ results in very low HNC column densities, \mbox{2$\times$10$^{12}$\,cm$^{-2}$} and
\mbox{3$\times$10$^{11}$\,cm$^{-2}$} at $A_\mathrm{V}$\,=\,8\,mag for
\mbox{$n_{\rm H}$\,=\,5$\times$10$^3$\,cm$^{-3}$} and
\mbox{5$\times$10$^4$\,cm$^{-3}$}, respectively. These $N$(HNC) values are
much lower than the $N$(HNC) column {densitie}s we infer from observations (\mbox{Table~\ref{tab:HNC_TexN})}.
In addition, models with \mbox{$E_b$\,=\,0\,K} would imply very high 
\mbox{$N$(HCN)/$N$(HNC)\,=\,30--75}  ratios, something  not seen in our observations (\mbox{Table~\ref{tab:Trot_HCN-HNC_Nratios}})}.

Some studies also suggest that in cold molecular gas, reaction 
\mbox{HNC\,$+$\,O\,$\rightarrow$\,CO\,$+$\,NH} dominates HNC destruction, and thus it
controls the HCN/HNC abundance ratio if
the energy barrier {of this particular reaction} is low, $E_b$\,$\simeq$\,20--50\,K \citep{Schilke1992,Hacar2020}. 
However, these values are much lower than the expected theoretical barrier 
\citep[A. Zanchet, priv.comm. and ][]{Lin1992}.
Overall, our observational results are more consistent, at least for the extended cloud emission, with a greater dependence of the $N$(HCN)/$N$(HNC)  ratio on the FUV radiation field \citep[as suggested in planetary nebulae,][]{Bublitz2019,Bublitz2022}.
\section{Discussion}\label{sec:discussion}
In this section we discuss the nature of the extended
\mbox{HCN $J$\,=\,1--0} emission observed in Orion\,B and its relation to  other species. We conclude by comparing 
the {observed} {line intensity vs. FIR dust continuum intensity} scalings
with the line luminosity vs. SFR scaling laws typically inferred in extragalactic studies.
\begin{figure*}[!ht]
    \centering
    \includegraphics[width=0.85\textwidth]{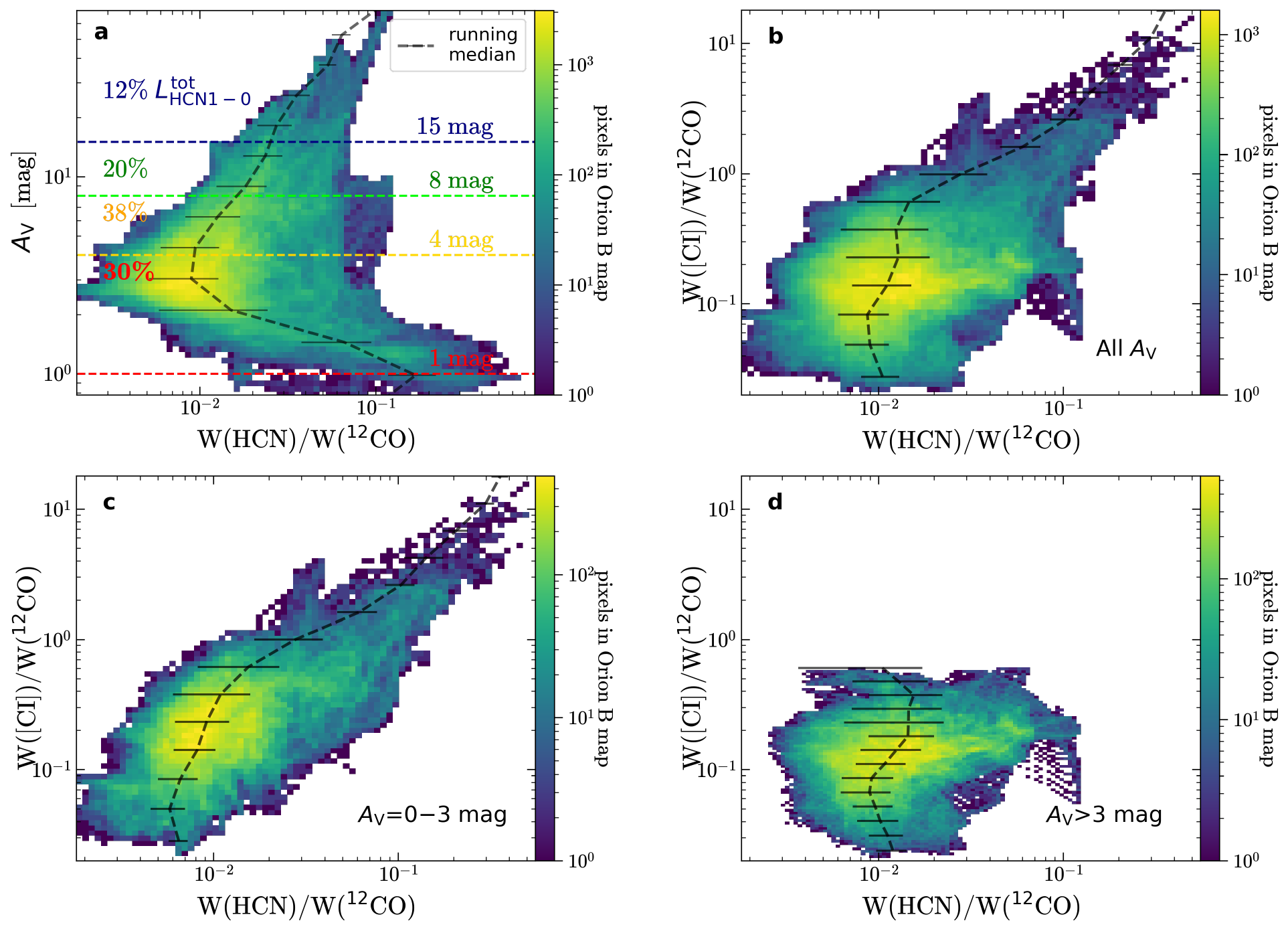} 
    \vspace{-0.1cm}
    \caption{2D histograms. (a)  Visual extinction $A_\mathrm{V}$ as a function of the HCN/CO $J$=1--0 integrated intensity ratio (from  maps at 120$''$ resolution).  Dashed red, yellow, green, and blue horizontal  lines are the visual extinction values 1, 4, 8, and 15~mag, respectively. Above each line, we show the percentage of the total 
    HCN $J$\,=\,1--0 {luminosity} that comes from the different \AV\;ranges. \mbox{(b), (c), and (d)}~2D histogram of the observed \CI~492\,GHz/CO~$J$=1$-$0 line intensity ratio (in units of \Kkms) as function of the observed HCN/CO~$J$=1--0 line  ratio for all $A_\mathrm{V}$, for $A_\mathrm{V}$\,<\,3~mag, and for  $A_\mathrm{V}$\,>\,3~mag. The dashed black curve shows the running median. Error bars show the standard deviation in the x-axis.} 
    \label{fig:ratios-AV_Ifir}
\end{figure*}
\subsection{{The origin of the extended HCN \textit{J}=1--0 emission: weak collisional excitation vs. scattering}}\label{sec:HCNelectrons}
The existence of a widespread \mbox{HCN $J$\,=\,1--0} emission component in low density gas{,} weakly collisionally excited, but enhanced by electron collisions (see Sect.~\ref{sect:comp-models}), may affect the interpretation of the extragalactic relationship HCN luminosity versus SFR.
Alternatively, the extended \mbox{HCN $J$\,=\,1--0} emission
we observe in Orion\,B might arise from photons emitted in  dense  star-forming cores that become resonantly scattered by halos of low density gas. This seems to be the case, albeit at much smaller spatial scales, in dense cores inside cold dark clouds shielded from stellar FUV radiation \citep[e.g.,][]{Langer1978,Walmsley1982,Cernicharo1984,Gonz-Alf1993}.

The above two scenarios lead to different \mbox{HCN $J$\,=\,1–0} HFS  line intensity ratios \citep[see predictions by][]{Goicoechea2022}, which can be tested on the basis of HFS resolved observations of the extended gas emission in GMCs.
In particular, if the observed \mbox{HCN $J$\,=\,1–0 HFS} photons arise from dense gas and become resonantly scattered by interacting with a low density halo, then both the  $R_{02}$ and $R_{12}$ HFS line intensity ratios should be very anomalous. That is, \mbox{$R_{02}$\,<\,0.2} and 
\mbox{$R_{12}$\,<\,0.6}.
On the other hand, if the  \mbox{HCN $J$\,=\,1–0} emission intrinsically arises from low density gas, far from dense cores, models predict that  weak collisional excitation drives the HFS intensity ratios to $R_{02}\!\gtrsim$\,0.2 and $R_{12}$\,$\lesssim$\,0.6.

\mbox{Figure~\ref{fig:hcn_hfs_anomalies}a} shows that the most common 
\mbox{HCN $J$\,=\,1–0 HFS} line intensity ratios in Orion\,B are \mbox{$R_{02}$\,$\gtrsim$\,0.2} and \mbox{$R_{12}$\,<\,0.6} (Fig.~\ref{fig:hcn_hfs_Tex-tau}). Hence, the very anomalous ratios predicted by the scattering halo scenario are rarely encountered {at large scales}. Therefore, we conclude that the extended \mbox{HCN~$J$\,=\,1--0} emission in Orion\,B is weakly collisionally excited, and it mostly arises from low density gas. 
In particular, we determined $n$(H$_2$) of several 10$^3$\,cm$^{-3}$
{to 10$^4$\,cm$^{-3}$} (see Sect.~\ref{sect:comp-models}).
This result contrasts with the prevailing view of \mbox{HCN\,$J$\,=\,1--0} emission as a tracer of dense gas \citep[{e.g.,}][]{Gao2004a,Gao2004b,Rosolowsky2011,JimenezDonaire2017tauneff,JimenezDonaire2019,SanchezGarcia2022HCN,Rybak2022}.
\subsection{Bimodal behavior of the HCN/CO \textit{J}=1--0 line intensity ratio as a function of \texorpdfstring{$A_\mathrm{V}$}{Av}} \label{sec:HCN-CO_Av}

Extragalactic studies frequently interpret  the \mbox{HCN/CO~$J$\,=\,1--0} line luminosity ratio as a tracer of the dense gas fraction \citep[{e.g.,} ][]{LadaE92,Gao2004b,Gao2004a,Usero2015,Gallagher2018,JimenezDonaire2019,Neumann2023}.
This {interpretation} assumes that CO{~$J$\,=\,1--0 line emission}  is a tracer of the bulk molecular gas, whereas 
\mbox{HCN~$J$\,=\,1--0} traces dense gas in star-forming cores (at high $A_\mathrm{V}$). Normal galaxies have low luminosity ratios
\mbox{$L_{\rm HCN}$/$L_{\rm CO}$\,=\,0.02–0.06} while luminous and ultraluminous galaxies have \mbox{$L_{\rm HCN}$/$L_{\rm CO}$\,$>$\,0.06}. {By contrast, \cite{Helfer1997} {argue} that the HCN/CO intensity ratio could  measure the total hydrostatic gas pressure.}

Figure~\ref{fig:ratios-AV_Ifir}a shows a 2D histogram of the
\mbox{HCN/CO~\,$J$\,=1--0} line intensity ratios  in Orion\,B as a function of 
$A_\mathrm{V}$. The 2D histogram shows a bimodal behavior. There is a first branch at \mbox{$A_\mathrm{V}$\,$>$\,3\,mag} where \mbox{$W$(HCN)/$W$(CO)~\,$J$\,=1--0} increases with extinction  (the assumed behavior in extragalactic studies).  The running median \mbox{$W$(HCN)/$W$(CO)~\,$J$\,=1--0} ratio increases from $\gtrsim$\,0.02 (at \mbox{$A_\mathrm{V}$\,$\simeq$\,8\,mag}) to $\sim$\,0.1 
(dense cores at larger $A_\mathrm{V}$).
In addition, there is a second branch at \mbox{$A_\mathrm{V}$\,$<$\,3\,mag} where \mbox{$W$(HCN)/$W$(CO)~\,$J$\,=1--0} increases with decreasing extinction.  This is somehow unexpected, because the running median intensity ratio  reaches high values, $\gtrsim$\,0.1,
in diffuse gas at \mbox{$A_\mathrm{V}$\,$\simeq$\,1\,mag}.

Fig.~\ref{fig:hcnratios_120}a shows the spatial distribution 
of the \mbox{HCN/CO~$J$\,=\,1--0} intensity ratios in Orion\,B. The ratio is indeed
high toward the dense gas in filaments and cores. In addition, \mbox{$W$(HCN)/$W$(CO)~\,$J$\,=1--0} also increases toward the east rim of the cloud
that borders the ionization front IC~434.
Owing to the roughly edge-on geometry with respect to the 
illuminating stars,  we can easily spatially resolve
these \mbox{FUV-illuminated} cloud edges (high $\chi_e$) from
the more shielded cloud interior. 
 This picture agrees with \mbox{HCN~$J$\,=\,1--0} emission {arising from} extended and
 relatively low density gas, \mbox{$n$(H$_2$)\,$\leq$\,10$^4$\,cm$^{-3}$},  in GMCs illuminated by
  FUV radiation, and {being} boosted by electron excitation. This {extended} cloud component must be common in GMCs that host young massive stars, or have massive stars in their vicinity. 
\subsubsection{{Cloud porosity to FUV radiation:} HCN 1--0 emission from  high electron abundance gas traced by \texorpdfstring{\CII}{[CII]}\,158\,\texorpdfstring{$\upmu$}{u}m and  extended \texorpdfstring{\CI}{[CI]}\,492\,GHz emission}
\label{subsec:porosity}
 The ionization fraction in  cloud edges and in gas translucent {to FUV--radiation} is high. It starts at \mbox{$\chi_e$\,$\simeq$\,a few 10$^{-4}$},  where the electron abundance is controlled by the photoionization of carbon atoms, thus leading to
\mbox{$\chi_e$\,$\simeq$\,$\chi$(C$^+$)}  at  \mbox{$A_\mathrm{V}$\,$\lesssim$\,2\,mag}. These cloud layers emit bright FIR \CII\,158~$\upmu$m  fine-structure line emission. 
Slightly deeper inside the molecular cloud, at \mbox{$A_\mathrm{V}$\,$\lesssim$\,3~mag}, the flux of FUV photons decreases to the point  where the gas becomes fully molecular, and neutral atomic carbon (C$^0$) becomes more abundant than C$^+$ (see PDR models in Figs.~\ref{fig:hcn_Meudon}a and \ref{fig:hcn_Meudon}b). 
{Our} models predict  \mbox{$\chi_e$\,$\gtrsim$\,10$^{-5}$}
at the C$^0$ abundance peak, {where the \CI~492\,GHz line emission 
 reaches its intensity peak}.

\begin{figure*}[!ht]
    \centering
    \includegraphics[width=0.64\textwidth]{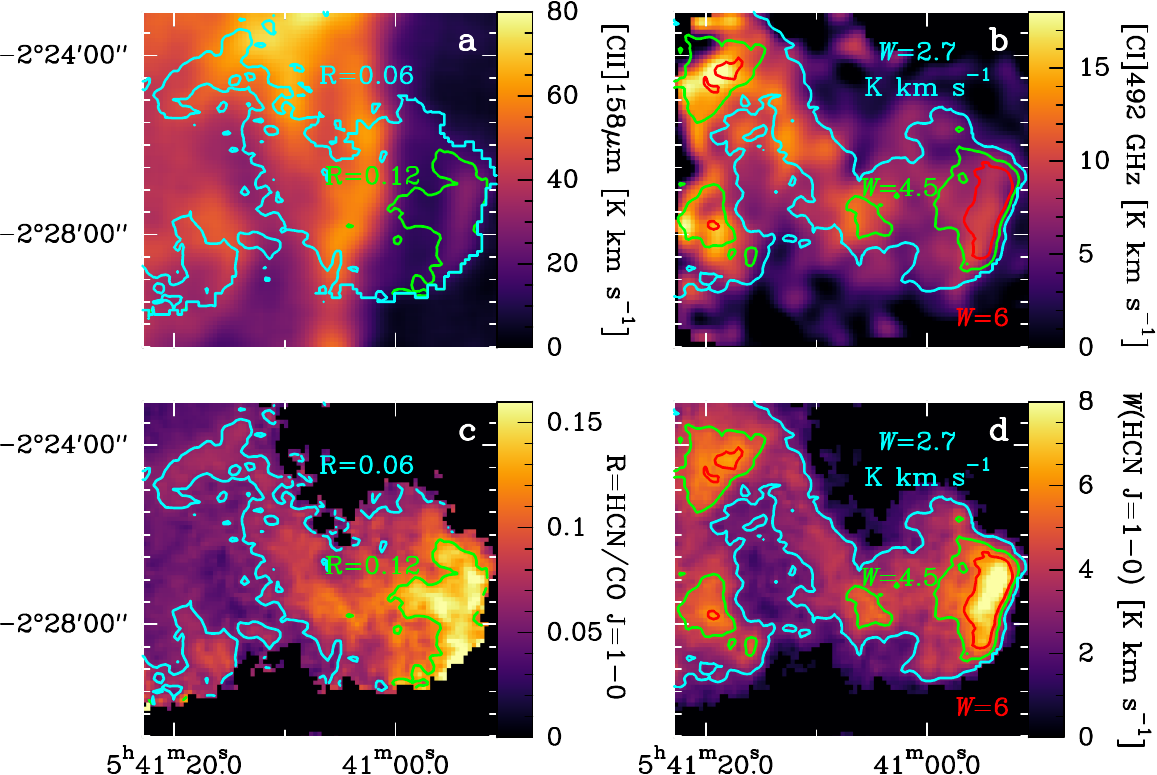}
     \caption{30$''$-resolution view of the Horsehead. (a)~\CII~158~$\upmu$m \citep{Pabst2017} and (b)~\CI~492~GHz \citep{Philipp2006} integrated line intensity maps (from 7 to 18 \kms). (c)~\textit{R}=HCN/CO  intensity ratio and (d)~HCN~$J$=1--0 line intensity maps.   Contours on \CII~158~$\upmu$m~map: HCN/CO ratio (\textit{R}=0.06 and 0.12). Contours on \CI~492~GHz map: HCN~$J$=1--0 surface brightness (\textit{W}=2.7, 4.5, and 6 \Kkms). \vspace{-0.2cm}} 
    \label{fig:CII-HCN-CO}
\end{figure*}

Figure~\ref{fig:CII-HCN-CO}a shows a \CII~158~$\upmu$m  line emission map of the Horsehead nebula and the ionization front IC\,434 observed with SOFIA \citep{Pabst2017} and convolved to the 30$''$ resolution of the \mbox{ORION-B} maps. The  C$^+$ map shows faint \CII~158~$\upmu$m emission from the (nearly edge-on) molecular PDR at the rim of the Horsehead. It also shows bright  \CII\,158\,$\upmu$m  emission from the neutral atomic PDR, \mbox{$\chi$(H)\,$\gg$\,$\chi$(H$_2$)}  at \mbox{$A_\mathrm{V}$\,$<$\,1\,mag}, that delineates  the edge of IC\,434 and {shows very little} CO emission \citep[e.g.,][]{Bally2018}.
Fig.~\ref{fig:CII-HCN-CO}c shows a closer look to the \mbox{HCN/CO\,$J$\,=\,1--0} line intensity ratio.
The ratio is particularly high, $\geq$\,0.12 (green contours), toward the 
\mbox{FUV-illuminated} cloud edge. This area matches the \CII~158~$\upmu$m emission from the  rim of the Horsehead. As $n$(H$_2$) is a few 10$^4$\,cm$^{-3}$ \mbox{\citep[e.g.,][]{Pabst2017}} and 
\mbox{$\chi_{\rm\,cr,\,e}^{*}$(HCN\,$J$\,=\,1--0)\,$<$\,$\chi_e$\,$\simeq$\,$\chi$(C$^+$)\,$\simeq$\,10$^{-4}$}, electron excitation boosts the HCN emission {(see  Fig.~\ref{fig:hcn_WTexXeMCmod})}, and thus the  
\mbox{HCN/CO\,$J$\,=\,1--0} line intensity ratio.

{Figure}~\ref{fig:CII-HCN-CO}b shows a map of the \CI~492~GHz  line emission  {around the Horsehead nebula} observed with Caltech Submillimeter Observatory \citep{Philipp2006} and  smoothed to 30$''$ resolution. The \mbox{HCN\,$J$\,=\,1--0} emission nicely follows
that of \CI~492~GHz. This agrees with the widespread nature of neutral atomic carbon, and  with the similar spatial distribution of \CI~492~GHz  and  \mbox{HCN\,$J$\,=\,1--0} emission seen at much larger scales 
(cf., the complete Orion\,B maps in Fig.~\ref{fig:fullmaps120}h). This observational result indicates that C$^0$ coexists with HCN in large areas of the cloud, which implies moderate ionization fractions, $\chi_e$\,$\gtrsim$\,10$^{-6}$ to several 10$^{-5}$, in  gas where  \CI\,492\,GHz and \mbox{HCN\,$J$\,=\,1--0} emissions  coexist.

{Since the spatial and velocity distribution of the large-scale \CI\,492\,GHz and $^{13}$CO~$J$=\,1--0 emission are very similar \citep{Ikeda2002}, the presence of C$^0$ cannot be 
restricted to cloud edges ($A_\mathrm{V}$\,$<$\,3\,mag). Otherwise \mbox{Fig.~\ref{fig:fullmaps120}h} {would} only show 
bright \CI\,492\,GHz emission parallel to the IC\,434 front associated
with the rims of all nearly edge-on PDRs such as the Horsehead (zoomed in Fig.~\ref{fig:CII-HCN-CO}). Instead, the \CI\,492\,GHz emission is widespread and extended through the cloud.
We find that \CI\,492\,GHz also {linearly} correlates 
with the HCO$^+$, HCN, and $^{13}$CO $J$\,=\,1--0 {emission} 
(Pearson coefficients of 0.80, 0.79, and 0.73 respectively).
These correlations include many positions at $A_\mathrm{V}$\,$>$\,3\,mag
({60}\% of \CI\,492\,GHz detections, {and 70\% of total \CI\,492\,GHz luminosity}, see Fig.~\ref{fig:lines-Av_hist}).
Therefore, C$^0$  must  be  abundant also toward the cloud interior. The most plausible scenario suggested by these 
maps is that GMCs are porous to FUV radiation. This {is} consistent 
with the detection of very extended 70\,$\upmu$m dust emission from
FUV-illuminated grains  (see Fig.~\ref{fig:rgb_orionb}). {This implies that} GMCs  {are inhomogeneous, }
and we are detecting \CI\,492\,GHz emission from the cloud edges as well as from \mbox{FUV-illuminated surfaces of structures} located at moderate cloud depths \citep[typically modeled as clumps,
filamentary, or fractal structures, e.g.,][]{Boisse1990, Falgarone1991,Spaans1996,Stutzki98,Barnes2013}. 
Unfortunately,  while ALMA, Keck, and JWST observations show  surprisingly rich small-scale  substructures in the  prototypical PDR the  Orion Bar \citep[][]{Goicoechea2016,Habart23a,Habart23b}, similar  \mbox{sub-arcsecond} resolution observations of the \CI\,492\,GHz are still missing. Such observations will help to constrain the small-scale origin of C$^0$.}
\subsubsection{High HCN/CO\,\textit{J}\,=\,1--0 and \texorpdfstring{\CI}{[CI]}\,492\,GHz/CO\,\textit{J}\,=\,1--0 line intensity ratios from gas at \texorpdfstring{$A_\mathrm{V}$}{Av}\,<\,3~mag}
We close our discussion of the bimodal behavior of the \mbox{$W$(HCN)/$W$(CO)\,$J$\,=\,1--0} intensity ratio  by providing more \mbox{evidence} that FUV radiation (leading to abundant C$^0$ and moderate $\chi_e$) is ultimately responsible of the increased ratios observed at $A_\mathrm{V}$\,<\,3~mag
(Fig.~\ref{fig:ratios-AV_Ifir}a). These regions correspond
to translucent gas and FUV-illuminated cloud edges. They
 coincide with enhanced \mbox{\CI\,492\,GHz/CO~$J$=1--0} line intensity ratios
(Fig.~\ref{fig:hcnratios_120}), which  traces low-density PDRs
\citep[][]{Hollenbach1991,Kaufman1999}. Hence, we expect that both ratios are related.

Figure~\ref{fig:ratios-AV_Ifir} shows the distribution of 
\mbox{\CI\,492\,GHz/CO\,$J$\,=\,1--0} and  
\mbox{HCN/CO\,$J$\,=\,1--0} intensity ratios in Orion\,B.
The upper-right panel shows all detections (at all $A_\mathrm{V}$) in the map. The running median
clearly shows that, above a threshold of \mbox{$W$(HCN)/$W$(CO)\,$J$\,=\,1--0\,$\gtrsim$\,0.02}, the ratio quickly increases with the \mbox{\CI\,492\,GHz/CO\,$J$\,=\,1--0} intensity ratio.
The lower panels in Fig.~\ref{fig:ratios-AV_Ifir} separate the HCN/CO and \CI/CO detections
in the lines of sight with $A_\mathrm{V}$\,$<$\,3\,mag (\mbox{\mbox{Fig.~\ref{fig:ratios-AV_Ifir}c}}) and $A_\mathrm{V}$\,$>$\,3\,mag (\mbox{\mbox{Fig.~\ref{fig:ratios-AV_Ifir}d}}). These
plots  show that the \mbox{HCN/CO\,$J$\,=\,1--0} line intensity ratio linearly correlates
with \mbox{\CI\,492\,GHz/CO\,$J$\,=\,1--0} at $A_\mathrm{V}$\,$<$\,3\,mag. Hence, the
ratios follow the increasing electron abundance in PDR gas.
On the other hand,  the ratios are not  correlated at higher $A_\mathrm{V}$\,>\,3~mag.
As a corollary, our observations imply that the detection of high \mbox{HCN/CO\,$J$\,=\,1--0}
line intensity ratios do not always imply the presence of dense gas. 
The existence of a low-$A_\mathrm{V}$ branch, from extended \mbox{FUV-illuminated} low-density gas, leads to 
increasing ratios with  decreasing $A_\mathrm{V}$. This cloud component cannot be overlooked, specially in the context of the very large scale emission from GMCs, otherwise the mass of the dense molecular gas can easily be overestimated.
In the next section we specifically quantify the amount of dense molecular gas traced by the \mbox{HCN~$J$\,=\,1--0} emission.

\subsection{The dense gas mass conversion factor \texorpdfstring{$\alpha$}{alpha}\,(HCN $J$=1--0)}\label{sec:alphadense}

 Using the dust SEDs {across the observed field}\footnote{{{Appendix}~\ref{app:gasmass} details how we determine the gas mass.}}, we derive the mass of the dense gas 
 in the mapped area  {\citep[represented by gas at $A_\mathrm{V}$\,$>$\,8\,mag, e.g.,][]{Lada2010}}. The likely density of this cloud component  is
$n$(H$_2$)\,$>$\,10$^4$~cm$^{-3}$ \citep[e.g.,][]{Bisbas2019}.
We obtain \mbox{$M_\mathrm{dg}$=3.1$\times$10$^3$~\Ms}, which accounts for about 20\% of the total mass  {($M_\mathrm{H_{2},tot}$\,$\sim$\,1.7$\times$10$^4$~\Ms)} in  the mapped area.  {These numbers imply a dense  gas surface density 
(\mbox{$\Sigma_{\mathrm{dg}}$\,=\,$M_\mathrm{dg}$\,/\,$A$}) and a total gas surface density (\mbox{$\Sigma_{\mathrm{H_{2},tot}}$\,=\,$M_\mathrm{H_{2},tot}$\,/\,$A$})
of  \mbox{$\Sigma_{\mathrm{dg}}$\,=\,13\,\Ms\,pc$^{-2}$} and  \mbox{{$\Sigma_{\mathrm{H_{2},tot}}$}\,=\,70\,\Ms\,pc$^{-2}$}, respectively,
where in both cases we divide by the total mapped area, \mbox{$A$\,$\simeq$\,250~pc$^2$}.}

High spatial resolution observations of the dust SED  are rarely available
in extragalactic studies. Hence, it is appropriate to calibrate the mass of the dense molecular gas with the emitted luminosity of a convenient molecular line tracer{, with \mbox{HCN $J$\,=\,1--0} being the traditional choice.} Hence, it is common to define:
\begin{equation}
M_\mathrm{dg}\,=\,\alpha(\mathrm{HCN})\,\cdot\,L'(\mathrm{HCN}),
\end{equation}
where $L'$(HCN) is the total
\mbox{HCN\,$J$\,=\,1--0} line luminosity in the mapped area
({in \mbox{\Kkms\,pc$^2$}) as  defined in Eq.~(\ref{eq:L'})}.
\cite{Gao2004a} originally estimated 
\mbox{$\alpha$(HCN)\,=\,10\,\Ms\,$/$\,\Kkms\,pc$^2$}.
Recent dust continuum and line emission surveys   
 determine $\alpha$(HCN) in a few local GMCs
\citep[e.g.,][]{Shimajiri2017,Kauffmann2017,Barnes2020}.
These studies find quite a diversity of $\alpha$(HCN) values, from 
 10 to 500~\Ms\,/\,\Kkms\,pc$^2$. However, these surveys typically map star-forming clumps
and their immediate environment (areas $<$1\,deg$^2$) but do not account for, or do not spatially resolve, 
the  extended \mbox{HCN~\,$J$\,=1--0} emission from low density and more translucent gas. This emission is weakly excited ($T_{\rm CMB}$\,$\lesssim$\,$T_{\rm ex}$\,$\ll$\,$T_{\rm k}$),
but because it covers large spatial scales, its total line luminosity 
typically exceeds the line luminosity from dense gas in star-forming cores at $A_\mathrm{V}$\,$>$\,8\,mag
($\sim$30$\%$ in Orion\,B).

Here we determine $\alpha$ for HCN, HCO$^+$, and HNC 
$J$\,=\,1--0 lines in Orion\,B.
We obtain $L'$(HCN)=110~\Kkms~pc$^2$ for \mbox{HCN~\,$J$\,=1--0}, which implies 
a dense mass conversion factor $\alpha$(HCN)\,=\,29\,~\Ms\,/\,\Kkms\,pc$^2$. Table~\ref{tab:lines_stats} summarizes the  luminosities and $\alpha$ values derived for other molecules. A recent survey of the Perseus low-mass star-forming region
(at 11\,arcmin resolution and covering \mbox{8.1~deg$^2$} or $\sim$215\,pc$^2$) finds 
$\alpha$(HCN)\,=\,92\,\Ms\,/\,\Kkms\,pc$^2$,  $L'$(HCN)\,=\,55.3~\Kkms, and $M_\mathrm{dg}$=5.1$\times$10$^3$~\Ms~\citep{Dame2023}.
{As} FUV radiation favors the formation of HCN and the excitation
of the $J$\,$=$\,1--0 line at large spatial scales (enhanced by electron collisions; Sect.~\ref{sec:HCNelectrons}),
our study suggests that the lower $\alpha$(HCN) value in Orion\,B is linked to the presence of FUV radiation from massive stars. 
Indeed, \citet{Shimajiri2017} mapped
 small areas of Orion\,B, Aquila, and Ophiuchus star-forming cores 
 ($<$10\,pc$^2$). They find
that $\alpha$(HCN) anticorrelates with $G_0$. 
We do find this tendency at the much larger spatial  scales of our maps, but not a strong anticorrelation. This can be  explained by the nonlinear dependence of the HCN abundance, \mbox{HCN\,$J$\,=\,1--0} line {emission},
and electron abundance with  $G_0$.

All in all, we conclude that there is no universal \mbox{$\alpha$(HCN $J$\,=1--0)} value, as environmental conditions and contribution of the low density extended cloud component
{at different angular scales}
likely vary from cloud to cloud. 
In  Orion\,B, the cloud mass at $A_\mathrm{V}$\,$>$\,8\,mag is similar to that at $A_\mathrm{V}$\,$<$\,3\,mag. This results in a similar value of the mass to total $L'$(HCN) ratio in both cloud components. Thus, it will not be straightforward 
to distinguish, based on the observation of {a} single  line, which component dominates the emission from spatially unresolved GMCs.

\subsection{{Schmidt-like} laws: Spatially resolved relations between molecular and atomic lines with FIR intensities} \label{sect:line-FIR}

On {more} 
global spatial scales {(hundreds of parsec to kiloparsec scales)} than those discussed in our study, 
observations of  nearby normal galaxies find a tight, close to linear, correlation between the \mbox{HCN\,$J$\,=\,1--0} line luminosity and the FIR luminosity \citep[a proxy of the SFR {when averaged on  such global scales};][]{Solomon1992, Gao2004a,Kennicutt2012}. However, 
{when considering starburst galaxies and (U)LIRGs, the relation}
often deviate{s} from linearity \cite[e.g.,][]{Gao2007,GarciaBurillo2012,Usero2015,SanchezGarcia2022HCN}.
These luminous galaxies lie above the \mbox{FIR--HCN} correlation observed in nearby normal galaxies.
They also display high \mbox{HCN/CO\,$J$\,=\,1--0} line luminosity ratios ($\sim$\,0.2) interpreted as galaxies having high fractions of dense molecular gas. 
{Our survey of Orion~B provides access to the local properties that
contribute to the large averages seen in galaxies.}
In this section we discuss the \mbox{spatially resolved}  relationships between $I_{\rm FIR}$  and  CO, HCN, HCO$^+$, \mbox{HNC~$J$=1--0}, and \CI~492~GHz line intensities  ($W$) mapped in Orion\,B.

\begin{figure}[!t]
    \centering
    \includegraphics[width=0.44\textwidth]{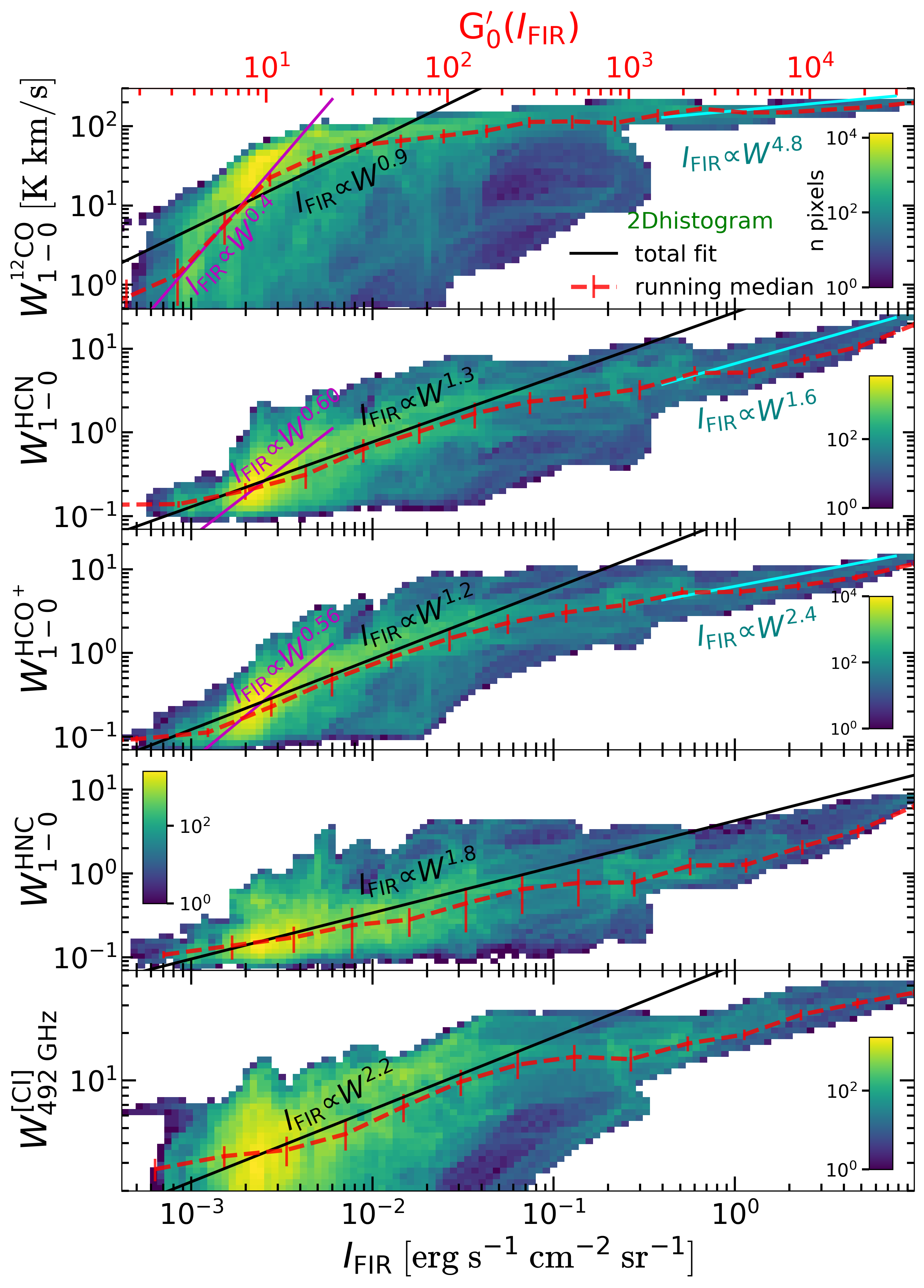}
     \caption{2D histograms of the $^{12}$CO, HCN, HCO$^+$, HNC,~$J$=1--0, and \CI~$^3P_1-^3P_0$ line intensities as a function of  FIR intensity in Orion\,B (from maps
     at 120$''$ resolution).
     The dashed red lines show the running median (median values of the integrated intensity within equally spaced log $I_\mathrm{FIR}$ bins). The error bars show their dispersion. Black lines (and associated text) show a linear fit ({orthogonal regression} in \mbox{log($y$)-log($x$)}) to all observed positions in each  map. {Magenta lines and blue lines show a linear fit to a range of $I_\mathrm{FIR}$\,<\,6$\times$10$^{-3}$~erg~s$^{-1}$~cm$^{-2}$~sr$^{-1}$ and 
     \mbox{(0.4--7.7)~erg~s$^{-1}$~cm$^{-2}$~sr$^{-1}$}, respectively.}
    We note that in each plot the number of line detections is different.} 
    \label{fig:lines-FIR_hist}
    \vspace{-0.3cm}
\end{figure}

Figure~\ref{fig:lines-FIR_hist} shows  2D histograms of the observed $W$(CO), $W$(HCN), $W$(HCO$^+$), $W$(HNC), and $W$(\CI~492~GHz) line intensities  (in 
K\,\kms) as a function of $I_\mathrm{FIR}$ (in \mbox{erg\,s$^{-1}$\,cm$^{-2}$\,sr$^{-1}$}). 
{We find that the observed line intensities $W$ scale with 
$I_\mathrm{FIR}$ as a power law. As we fit these points
using an orthogonal regression method\footnote{Using the Scipy-odr package \citep{Branham1995,Scipy2020}.
\url{https://docs.scipy.org/doc/scipy/reference/odr.html}} in \mbox{log($y$)-log($x$)} space
and we use the appropriate error bars {(the standard deviation)} in both axes, 
{we} can present the scalings as $I_\mathrm{FIR}$\,$\propto$\,$W^{N}$ (as in Fig.~\ref{fig:lines-FIR_hist}) or as $W$\,$\propto$\,$I_\mathrm{FIR}^{1/N}$ (as  in Fig.~\ref{fig:lines-FIR_hist_app}).}
{Perhaps provocatively, and in order to promote the comparison   with the extragalactic scalings
\mbox{SFR--$L_{\rm mol}$} \citep[e.g.,][]{Gao2004a,Shirley2008,Shetty13,Shetty14b,Shetty14}, here we start discussing the  power-law indexes $N$}. 

 {As discussed in Sect.~\ref{subsec-Herschel},  $I_\mathrm{FIR}$} is a surrogate of the local
 FUV radiation field, $G'_0$ (upper panel $x$-axis of Fig.~\ref{fig:lines-FIR_hist}).
{FUV photons are related to the presence of massive O and B stars that have short lifetimes. Thus, the FIR emission \mbox{from FUV-heated} grains is ultimately related to the SFR.}
{However,  the statistical connection between SFR and FIR luminosities in galaxies holds when averaging over large cloud samples \mbox{\citep[e.g.,][]{Kennicutt2012}}.}
{Therefore, the extrapolation of the local scalings in Orion~B  
 to galaxies (global averages) has to be taken with caution, bearing in mind that $I_\mathrm{FIR}$ traces the strength of the FUV radiation field, but $L_\mathrm{FIR}$ over a small region does not trace  the true SFR\footnote{The total FIR luminosity in the mapped region of Orion~B ($\sim$\,250\,pc$^2$) is \mbox{$L_{\rm FIR}$\,=\,1.5$\times$10$^5$\,\Ls} 
(or \mbox{$\Sigma_{\rm FIR}$\,=\,6$\times$10$^2$\,\Ls\,pc$^{-2}$}).
{Using the extragalactic  scalings  \mbox{\citep[e.g.,][]{Kennicutt1998a}}, these FIR luminosities  translate into a SFR of 2.6$\times$10$^{-5}$\,\Ms\,yr$^{-1}$, which is nearly an order of magnitude} {lower than the SFR estimated by counting young stellar objects (YSOs),  1.6$\times$10$^{-4}$~\,\Ms\,yr$^{-1}$ 
\citep[][]{Lada2010}. \cite{Pabst21} {find}  a similar result in Orion~A, namely that \mbox{SFR(FIR)\,$<$\,SFR(YSOs)}. }}}.

By fitting all points in {Fig.~\ref{fig:lines-FIR_hist}}, we find that $W$(CO), $W$(HCO$^+$) and $W$(HCN~$J$=1--0)   scale with $I_\mathrm{FIR}$  as a power law  with  
{\mbox{$N$\,$\sim$\,0.9--1.3}. However, $W$(HNC~$J$=1--0) and $W$(\CI\,492\,GHz) show a different  behavior, with a power-law index \mbox{$N$\,$\sim$\,1.8--2.2}}.   A closer inspection of the  \mbox{$I_\mathrm{FIR}$--$W$(CO~$J$\,=\,1--0)} running median  shows two clear tendencies. The  median toward the brightest FIR positions
\mbox{($I_\mathrm{FIR}$\,$>$\,0.4~erg~s$^{-1}$~cm$^{-2}$~sr$^{-1}$} { or $G'_0$\,$>$\,1500}, mostly arising from
NGC\,2024 star-forming clump)  shows a power-law index of   \mbox{$N$\,$\sim$\,4.8}.
On the other hand, the faintest $I_\mathrm{FIR}$ and
\mbox{$W$(CO~$J$\,=\,1--0)} positions  show   $N$\,$\sim$\,0.4 (sublinear relationship).
This faint CO emission is  associated with widespread and very extended diffuse  gas (low $A_\mathrm{V}$ and {$G'_0$\,$<$\,20}).

 The \mbox{$I_\mathrm{FIR}$--$W$(HCN~$J$\,=\,1--0)}  and \mbox{$I_\mathrm{FIR}$--$W$(HCO$^+$~$J$\,=\,1--0)} histograms are quite similar. They also reveal two different  tendencies.  Fitting the brightest $I_\mathrm{FIR}$ positions alone, most of them associated {with} dense gas in  NGC\,2024 (as demonstrated by the detection of bright \mbox{HCN\,$J$\,=\,4--3} line emission,
 Fig.~\ref{fig:hcn_multiJ_resume}),  provides \mbox{$N$$\sim$1.6} for \mbox{HCN} and 
\mbox{$N$$\sim$2.4} for \mbox{HCO$^+$} (superlinear relationships).
In contrast, the most common low surface brightness 
\mbox{HCN\,$J$\,=\,1--0}, \mbox{HCO$^+$\,$J$\,=\,1--0}, and
$I_\mathrm{FIR}$ positions {(those with \mbox{$G'_0<20$}) show  \mbox{$N$$\sim$0.6}
\mbox{(i.e., not far from the extended and diffuse CO emission index)}}.
{Interestingly}, the \mbox{$I_\mathrm{FIR}$--$W$(HNC~$J$=1--0)} 
and \mbox{$I_\mathrm{FIR}$--$W$(\CI492\,GHz)} 
histograms {show} a single superlinear tendency across the map, 
with  $N$$\sim$1.8 and 2.2, respectively. Hence, HNC  and \CI\,492\,GHz have a very different behavior compared to the other species.  The derived $N$  index for HNC resembles the index we find for HCN at the highest values of $I_\mathrm{FIR}$ ($G_0'>1500$ and dense gas), {whereas the single  index for \CI\,492\,GHz resembles that of HCO$^+$  at high $G'_0$. This similitude must reflect their related chemistry and excitation conditions}.

{From the point of view of the local gas properties, Figs.~\ref{fig:lines-FIR_hist} 
and  \ref{fig:lines-FIR_hist_app} show {that} \mbox{$W$(CO\,$J$\,=\,1--0)},
 \mbox{$W$(HCO$^+$\,$J$\,=\,1--0)}, and \mbox{$W$(HCN\,$J$\,=\,1--0)}
\mbox{intensities} 
increase with increasing  $I_\mathrm{FIR}$ up to $G'_0$\,$\simeq$\,20.
Most of these positions refer to the extended cloud component, which hosts low densities
and thus, the \mbox{HCN~$J$\,=\,1--0} line is weakly collisionally   excited
(see Sect.~\ref{sect:comp-models}), with
\mbox{$n$(H$_2$)\,$<$\,$n_{\rm cr,\,eff}$(HCN 1--0)} and thus,
 \mbox{$T_{\rm ex}$\,$<$\,$T_{\rm k}$}. Under these conditions (effectively thin emission), \mbox{$W$(HCN\,$J$\,=\,1--0)} scales with $N$(HCN) even for {large} 
 line opacities {\citep[see also ][]{Liszt2016}}. 
 {Furthermore,} our chemical analysis shows that models with {a} 
 higher  $G_0$/$n_{\rm H}$ ratio  
produce more HCN (Sect.~\ref{sec:Meudon} and Fig.~\ref{fig:hcn_Meudon}).
In addition, electron excitation contributes to enhance \mbox{$W$(HCN\,$J$\,=\,1--0)} at low densities. These conditions favor the emission of CO, HCO$^+$, {and} HCN
as $G'_0$ increases.}

{On the other hand, Figs.~\ref{fig:lines-FIR_hist} and  \ref{fig:lines-FIR_hist_app} show that \mbox{$W$(CO\,$J$\,=\,1--0)},
 \mbox{$W$(HCO$^+$\,$J$\,=\,1--0)}, and \mbox{$W$(HCN\,$J$\,=\,1--0)}  respond weakly to $I_\mathrm{FIR}$ once the FUV field becomes too intense ($G'_0$\,$>$\,1500). These other regions
at large $A_\mathrm{V}$ host denser gas, 
so that 
\mbox{$n$(H$_2$)\,$>$\,$n_{\rm cr,\,eff}$(HCN 1--0)},  
and \mbox{$J$\,=\,1--0} lines turn into very optically thick, thus becoming less sensitive to  column densities.}
{Interestingly, \mbox{$W$(HNC\,$J$\,=\,1--0)} and $W$(\CI~492\,GHz) weakly respond to $I_\mathrm{FIR}$ at all $G'_0$.
We already showed that  HNC traces slightly denser gas than
HCN   \mbox{(Fig.~\ref{fig:lines-Av}c)} and that HNC responds {less} 
to electron excitation {\mbox{(Table~\ref{tab:spec_n})}}. 
In addition, the observed \mbox{$W$(HCN 1--0)/$W$(HNC 1--0)} intensity ratio increases with 
the FUV  field for $G'_0$\,$<$\,200 (Fig.~\ref{fig:HCN-HNC_Ifir}). {Indeed,} our chemical models show
that the HNC abundance is lower in the FUV-illuminated gas (Sect.~\ref{sec:Meudon}). This gas is usually at lower density than the FUV-shielded cold gas.  
Thus, we expect
that most of the \mbox{HNC\,$J$\,=\,1--0} emission
{arises} 
from gas in which \mbox{$n$(H$_2$)\,$>$\,$n_{\rm cr,\,eff}$(HNC 1-0)}.
 These facts explain the weaker response 
of \mbox{$W$(HNC\,$J$\,=\,1--0)} to FUV radiation.
Finally, PDR models predict that the C$^0$ column density
is a weak function of gas density and especially of $G_0$ \citep[e.g.,][]{Hollenbach1991}. This is consistent with the weak
scaling we find in Orion~B.}

As a corollary, we conclude that our large-scale and spatially resolved lines maps of a local GMC show a variety of
power-law indexes, {\mbox{$I_\mathrm{FIR}$\,$\propto$\,$W^N$
(or $W$\,$\propto$\,$I_\mathrm{FIR}^{1/N}$)}. 
These $N$ indexes resemble the kind of Kennicutt-Schmidt power-law indexes, \mbox{SFR\,$\propto$\,$L_{\rm mol}^{N}$}, found  in galaxy surveys that average multiple GMCs \citep[e.g.,][]{Wu2005,Wu2010,Kennicutt2012,GarciaBurillo2012,SanchezGarcia2022HCN}.
We attribute the different scalings in Orion~B
to the different gas densities, excitation regimes, and {chemistry} of the star-forming (dense and compact) versus non-star-forming (low density, extended, {and FUV-illuminated}) environments. 
However, while it is tempting to extrapolate our results to the extragalactic  scalings \citep[as in][]{Krumholz2007,Narayanan2008}, we still need to better understand the 
 spatial scales at which $L_{\rm FIR}$  becomes a reliable  tracer
of the global SFR, as well as the connection between the extragalactic averages  versus our spatially resolved scalings.} 
 
\section{Summary and conclusions}\label{sec:conclusions}

In the context of the IRAM\,30m ORION-B  large program,
we presented a detailed analysis of 5\,deg$^2$ ($\sim$250\,pc$^2$) HCN, HNC, HCO$^+$, CO~$J$=1--0,  and \CI\,492\,GHz line emission maps of {the} Orion~B GMC.
We  complemented this dataset with new pointed  observations of rotationally excited HCN, HNC, H$^{13}$CN, and HN$^{13}$C lines.
We constructed  {integrated line intensity ($W$)}, visual extinction, and  $I_{\rm FIR}$ {(a proxy of  $G_0$)}  maps from existing dust SED observations. 
We summarize our results as follows:

-- About 70\% of the total HCN $J$=1--0 luminosity, 
\mbox{$L'$(HCN $J$\,$=$\,1--0)\,$=$\,110\,K\,km\,s$^{-1}$\,pc$^{-2}$}, arises from gas at  $A_\mathrm{V}$\,$<$\,8\,mag 
{\mbox{(Sect.~\ref{sec:HCNrelothers})}}, that is, from  gas below the common extinction threshold of star-forming cores. About  80\%\, of the total cloud mass and {50\%\, of the total FIR luminosity}
(mostly arising from \mbox{FUV-heated} dust grains)  also steams from  \mbox{$A_\mathrm{V}$\,$<$\,8\,mag}.

--  We detect anomalous \mbox{HCN $J$\,=\,1–0}  HFS  line intensity ratios
(also in the HCN \mbox{$J$\,=\,2--1} and \mbox{3--2} transitions)  almost everywhere in the cloud {\mbox{(Sect.~\ref{sect:taupopdia}}).} {That is,
HCN \mbox{$J$\,=\,1--0} \mbox{$R_{02}$\,=\,$W$($F$=0--1)/$W$($F$=2--1)} and \mbox{$R_{12}$\,=\,$W$($F$=1--1)/$W$($F$=2--1)} hyperfine line intensity ratios outside the LTE range \mbox{$R_{02}$\,=\,[0.2, 1]} and 
\mbox{$R_{12}$\,=\,[0.6, 1]}}. We also detect {anomalous HFS line width ratios. That is, 
\mbox{$\Delta v$($F$=0--1)\,$\neq$\,$\Delta v$($F$=2--1)\,$\neq$\,$\Delta v$($F$=1--1)}} {\mbox{(Sect.~\ref{sect:HCNHFS}})}.
Radiative effects induced by {moderate} {line opacities} and HFS line overlaps produce these anomalous ratios, which are inconsistent with the common assumption of the same $T_{\rm ex}$ {and line width} for all HFS lines
{of a given rotational transition}. 

-- Most of the  widespread and extended \mbox{HCN $J$\,$=$\,1--0} emission  arises from weakly collisionally excited gas with 
\mbox{$n$(H$_2$)\,$\lesssim$\,10$^4$~cm$^{-3}$}. That is, it is not line radiation emitted by dense cores that is  resonantly scattered by low density halos {\mbox{(Sect.~\ref{sec:HCNelectrons}})}.
This is demonstrated by the typical
\mbox{HCN $J$\,=\,1–0 HFS} intensity ratios \mbox{$R_{02}$\,$\gtrsim$\,0.2} and  \mbox{$R_{12}$\,<\,0.6} {observed} at large scales. Even lower  densities are possible {in FUV-illuminated gas} if  \mbox{$\chi_e$\,$\geq$\,10$^{-5}$}
and electron collisional excitation dominates
{\mbox{(Sect.~\ref{sect:comp-models}}).}

-- The \mbox{HCN/HNC\,$J$\,$=$\,1--0} line intensity ratio is sensitive to the strength of the FUV radiation field. Our chemical models and observations suggest that
the HCN/HNC abundance ratio is more sensitive to $G_0$ than to $T_{\rm k}$
{\mbox{(Sect.~\ref{sec:Meudon}}).}
In particular, HNC is {a} {slightly better tracer of} dense gas, defined as $n$(H$_2$)\,$>$\,10$^4$\,cm$^{-3}$, than HCN, because its abundance is lower in the \mbox{FUV-illuminated gas} (translucent gas and cloud edges).  This gas is usually at lower density than the \mbox{FUV-shielded} cold gas. {In addition, HNC is less sensitive to electron excitation than HCN {(Table~\ref{tab:spec_n})}}.

-- The \mbox{HCN/CO $J$=1--0} line intensity ratio 
{\mbox{(Sect.~\ref{subsec:ratios}})}, widely used as a tracer of the dense gas fraction, shows a \mbox{bimodal} behavior with respect to $A_\mathrm{V}$, with an inflection point at $A_\mathrm{V}$\,$\lesssim$\,3\,mag {\mbox{(Sect.~\ref{sec:HCN-CO_Av}})} typical of translucent gas and \mbox{FUV-illuminated} cloud edges. The  extended cloud \mbox{HCN $J$\,$=$\,1--0} emission {\mbox{(Sect.~\ref{sect:comp-models}})} explains the  low $A_\mathrm{V}$ branch of the  observed distribution of the  \mbox{HCN/CO\,$J$\,$=$1--0} line intensity ratio. The highest \mbox{HCN/CO\,$J$\,$=$1--0} line intensity ratios ($\sim$\,0.1) at  $A_\mathrm{V}$\,$<$\,3\,mag correspond to regions displaying high \mbox{\CI\,492\,GHz/CO\,$J$\,$=$1--0} intensity ratios too ($>$\,1). These values are characteristic of low-density PDRs  and $\chi_e$\,$\gtrsim$\,10$^{-5}$. Therefore, we conclude that the detection of high \mbox{HCN/CO\,$J$\,=\,1--0} intensity ratios does not always imply the presence of dense gas.

-- {Given the widespread and extended nature of the \CI492\,GHz emission
(a {typical} tracer of PDR gas), and its spatial correlation with  \mbox{$W$(HCO$^+$\,$J$\,$=$\,1--0)}, \mbox{$W$(HCN\,$J$\,$=$\,1--0)}, and \mbox{$W$($^{13}$CO\,$J$\,$=$\,1--0)}}  (see \mbox{Sect.~\ref{subsec:porosity}}),  the extended component  of Orion\,B (and likely in most GMCs), must be porous to FUV radiation from nearby massive stars.  {Indeed, 70\% of the total \CI\,492\,GHz luminosity}
arises from lines of sight with  $A_\mathrm{V}$\,$>$\,3\,mag
{(i.e., not exactly from the cloud surface). In addition, the 70\,$\upmu$m continuum emission
from FUV-illuminated dust grains is very extended}.
The {enhanced} FUV  field
favors the formation of HCN and the excitation of the \mbox{$J$\,=\,1--0} line at large scales, not only in dense star-forming cores. This is exemplified by the relatively low value of the dense gas mass to \mbox{the HCN~$J$\,$=$\,1--0} line luminosity ratio, \mbox{$\alpha$\,(HCN)\,$=$\,29\,M$_\odot$\,/\,\Kkms\,pc$^{2}$}, in Orion\,B
{\mbox{(Sect.~\ref{sec:alphadense})}}.
The existence of a widespread \mbox{HCN $J$\,=\,1--0} emission component associated with low density gas  affects the interpretation of the extragalactic relationship 
$L_{\rm HCN}$ versus SFR.

-- {The low-surface brightness and extended  
\mbox{HCN\,$J$\,$=$\,1--0} and
\mbox{HCO$^+$\,$J$\,$=$\,1--0}
emissions ($\lesssim$\,1~\Kkms)
scale with $I_{\rm FIR}$ with a similar power-law
index  (Sect.~\ref{sect:line-FIR}). Together with \mbox{CO\,$J$\,$=$\,1--0}, these lines 
respond to the increasing $I_{\rm FIR}$  up to $G'_0$\,$\simeq$\,20. 
On the other hand, the  bright HCN emission ($>$\,6~\Kkms) from dense gas in star-forming clumps  weakly responds to $I_{\rm FIR}$  once the FUV radiation
field becomes too intense ($G'_0$\,$>$\,1500).
 \mbox{HNC\,$J$\,$=$\,1--0} and \CI\,492\,GHz lines  weakly respond  to $I_{\rm FIR}$ at all $G'_0$.}

-- {Our large-scale and spatially resolved lines maps of a local GMC show a variety of power-law indexes, \mbox{$I_\mathrm{FIR}$\,$\propto$\,$W^N$} (from sublinear to superlinear), that resemble the kind of \mbox{Kennicutt-Schmidt} power-law indexes, \mbox{SFR\,$\propto$\,$L_{\rm mol}^{N}$}, found  in surveys of different galaxy types that {spatially} average multiple GMCs \citep[{e.g.,}][]{Kennicutt2012}. 
We attribute the different scalings in Orion~B to the different gas densities, excitation regimes, and {chemistry} of the star-forming (compact) versus non-star-forming (extended) environments (Sect.~\ref{sect:line-FIR}).}

Our study stresses the major contribution of the extended and low density component of GMCs to the total  CO, HCO$^+$, and \mbox{HCN\,$J$\,=\,1--0} line luminosity. {It also enables us to remark that there is a need to \mbox{carry out}} sensitive wide field surveys of galactic GMCs in multiple molecular lines. 
This will allow us to determine the properties of the star formation environment and to  better understand  the origin {of the extragalactic Kennicutt-Schmidt scalings on global galaxy averages}. In Orion\,B, the \mbox{HCN\,$J$\,=\,1--0} line intensity at any position of the extended cloud component 
is obviously much fainter than that arising from dense star-forming clumps such as NGC~2024. However, the much larger area of the  extended cloud component at low $A_\mathrm{V}$ implies that the  emission arising from dense cores does not dominate the HCN~$J$=1--0 line luminosity from GMCs \citep[see also][]{Santa-Maria2021}.  
Finally, better knowledge of the rate coefficient of some
critical gas-phase reactions, namely reaction 
\mbox{NCO\,+\,N\,$\rightarrow$\,HCN\,+\,O} and reactions of HNC  with H, C, and O atoms, will help us to refine our abundance estimations from chemical models.

\begin{acknowledgements}
{We are very grateful to our referee for a very detailed and constructive report that allowed us to improve the presentation of our results.}
MGSM and JRG thank the Spanish MICINN for funding support under grant \mbox{PID2019-106110GB-I00}. This work was supported by the French Agence Nationale de la Recherche through the DAOISM grant 
\mbox{ANR-21-CE31--0010}, and by the Programme National “Physique et Chimie du Milieu Interstellaire” (PCMI) of CNRS/INSU with INC/INP, co-funded by CEA and CNES. Part of this research was carried out at the Jet Propulsion Laboratory, California Institute of Technology, under a contract with the National Aeronautics and Space Administration (80NM0018D0004). We thank \mbox{A. Zanchet} for useful discussions on the energy barrier of the HNC~+~H and  HNC~+~O reactions and \mbox{A. Faure} for providing the HFS-resolved rate coefficients for \mbox{HCN-$e^-$} inelastic collisions.  
This work is based on observations carried out under project number 019-13, 022-14, 145-14, 122-15, 018-16, the large program number 124-16, 130-21, and 127-22 with the IRAM~30m telescope. IRAM is supported by INSU/CNRS (France), MPG (Germany), and Spain (ING). This research  made use of data from the Herschel Gould Belt survey (HGBS) project (http://gouldbelt-herschel.cea.fr). The HGBS is a Herschel Key Programme jointly carried out by SPIRE Specialist Astronomy Group 3 (SAG 3), scientists of several institutes in the PACS Consortium (CEA Saclay, INAF-IFSI Rome and INAF-Arcetri, KU Leuven, MPIA Heidelberg), and scientists of the Herschel Science Center (HSC).
\end{acknowledgements}

%
%

\bibliographystyle{aa}
\bibliography{references3}

\begin{appendix}

\section{Main regions in Orion~B}
\label{App:regions}

{In this Appendix we provide more details about the properties
of the main regions in Orion~B discussed in this work 
(see Fig.~\ref{fig:fullmaps120})}.

    --- NGC 2024: also known as the Flame nebula, is located east of the belt star Alnitak ($\zeta$~Ori). This
    is an active massive star-forming region, with the highest H$_2$ column density (\mbox{$\sim$5$\times$10$^{23}$~cm$^{-2}$}) and star formation efficiency  \citep[SFE$\sim$30\%, ][]{Lada1997} in Orion~B. It is composed of an embedded stellar cluster and associated \HII$\;$region \citep{Barnes1989}.     Inside the molecular ridge behind the \HII\;region, FIR and radio observations reveal the presence of embedded dense cores, protostars, and YSOs \citep{Mezger1988,Gaume1992,Chandler1996,Choi2015,Ren2016, Konyves2020}. 
    The bulk of the FIR line emission arises from an extended PDR, with  $T_\mathrm{k}\!\simeq$75-100~K and \mbox{$n$(H$_2$)$\simeq$10$^6$~cm$^{-3}$}. This region is illuminated by
    ionizing and dissociating UV photons from a massive star 
    \citep[e.g.,][]{Giannini2000,Emprechtinger2009}.
    The dominant ionizing source is likely the late-O or early-B star IRS2b \citep{Bik2003,Meyer2008}. 
    
     --- NGC 2023: is a reflection nebula located 20$'$ south of NGC~2024 \citep{Meyer2008}. It is illuminated by a B1.5V star \citep{Abt1977}. This is a filamentary massive star-forming region \citep{Konyves2020,Gaudel2022}, with a FUV radiation flux
    equivalent to  a few 10$^4$ times the mean interstellar radiation field ($G_0$)   toward the \CII~158~$\upmu$m emission peak  \mbox{\citep[e.g.,][]{Sandell2015}}.

    --- The Horsehead nebula\footnote{Discovered  by Williamina Fleming \citep{Pickering1908}.}: also known as Barnard 33,  is a dense pillar seen projected against the bright \HII\;region and ionization front IC~434. The multiple stellar system, \mbox{$\sigma$~Orionis}, formed by an O9.5V and a B0.5V binary, photoionizes the region \citep{Walter2008}.  The Horsehead and all the western rim of the Orion~B cloud is a large scale PDR  eroded by UV radiation from $\sigma$~Ori. The molecular cloud is located at a projected distance of $\sim$4~pc from the ionizing stars, resulting in a moderate incident FUV flux, $G_0\!\simeq$100 \citep{Abergel2003}.
    The expansion of the \HII\;region likely triggers gas compression along the cloud rim and perhaps star-formation \citep{Bally2018}.
     The Horsehead pillar points radially to the ionizing source, and its western edge is a PDR observed nearly edge-on.
    This region hosts a few dense cores, protostars and YSOs \citep{Bowler2009,Konyves2020}. The gas temperature ranges from $\sim$100\,K (at the \mbox{FUV-illuminated} edge) to 10-20~K deeper inside the cloud
    \citep[e.g.,][]{Habart2005,Pety2007,Goicoechea2009}.

    --- Orion~B9: 
    is an active low-mass star-forming region composed of supercritical filaments hosting pre- and proto-stellar cores \citep{Miettinen2009,Miettinen2010}. 
    Line observations resolve two different velocity  components at about \mbox{$v_\mathrm{LSR}$=0-4~\kms}\;and 8-10~\kms, respectively \citep[e.g.,][]{Gaudel2022}.

    --- The Flame filament: is physically connected with NGC\,2024, at the southeast of NGC\,2024. This is a very structured filament  \citep[][]{Orkisz2019,Gaudel2022}.
    
    --- The Hummingbird filament: is one of the longest isolated filaments in the observed field. Only a few embedded  YSOs exist in this region \citep{Orkisz2019}.

     --- The Cloak: is a filamentary structure that crosses Orion~B from east to west \citep{Gaudel2022}. Embedded starless and prestellar cores exist in the region \citep{Konyves2020}.

\clearpage
\section{{Line intensities,} line intensity ratios, and their relation to \texorpdfstring{$A_\mathrm{V}$}{Av} and \texorpdfstring{$I_\mathrm{FIR}$}{Ifir}}

In this Appendix we show 2D histograms that display the relationship between line intensity ratios, $A_\mathrm{V}$, and  $I_\mathrm{FIR}$. In Sect.~\ref{sec:HCN-CO_Av} we discussed the behavior of the HCN/CO~$J$=1--0 line intensity ratio with $A_\mathrm{V}$ and its relation to the \CI/CO~$J$=1--0 line intensity ratio. In Sect.\ref{sec:isomerization} we discussed  the HCN/HNC abundance ratio and its possible relation with $G_0$. {In addition, Fig.~\ref{fig:lines-FIR_hist_app} shows 2D histograms of the relation between line intensities and $I_\mathrm{FIR}$, showing the exponents $W$\,$\propto$\,$I_\mathrm{FIR}^{1/N}$ (see Sect.~\ref{sect:line-FIR}).}

\begin{figure}[!h]
    \centering
    \includegraphics[width=0.49\textwidth]{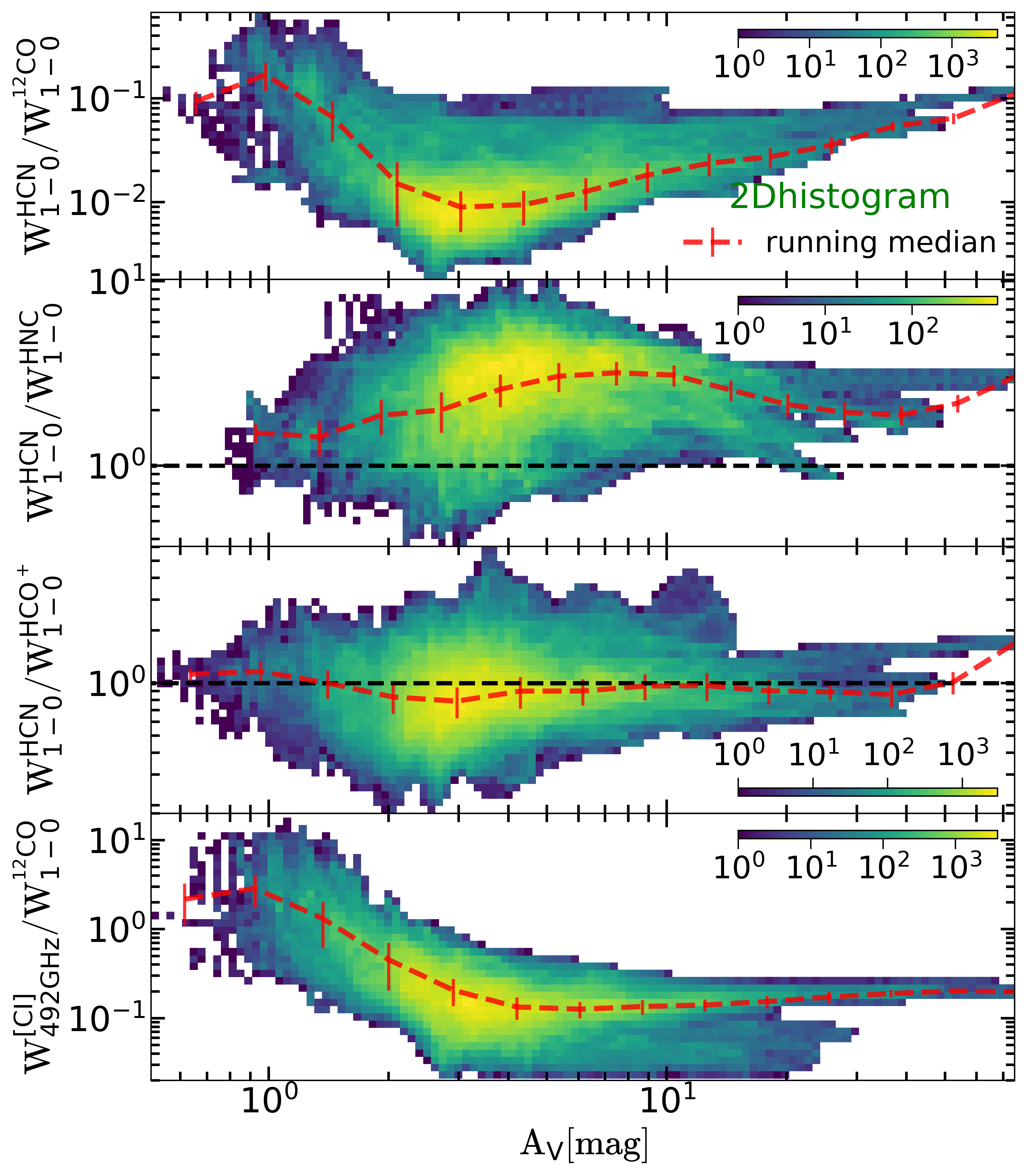}
    \caption{2D histograms showing  the line intensity ratios HCN/CO, HCN/HNC, HCN/HCO$^+$~$J$=1--0, and \CI~492~GHz/CO~$J$=1--0 with respect to the visual extinction $A_\mathrm{V}$.  The dashed red line marks the running median. The error bars mark the standard deviation. The dashed black line marks the intensity ratio equal to one.} 
    \label{fig:ratiosAv}
\end{figure}
\begin{figure}[!h]
    \centering
    \includegraphics[width=0.49\textwidth]{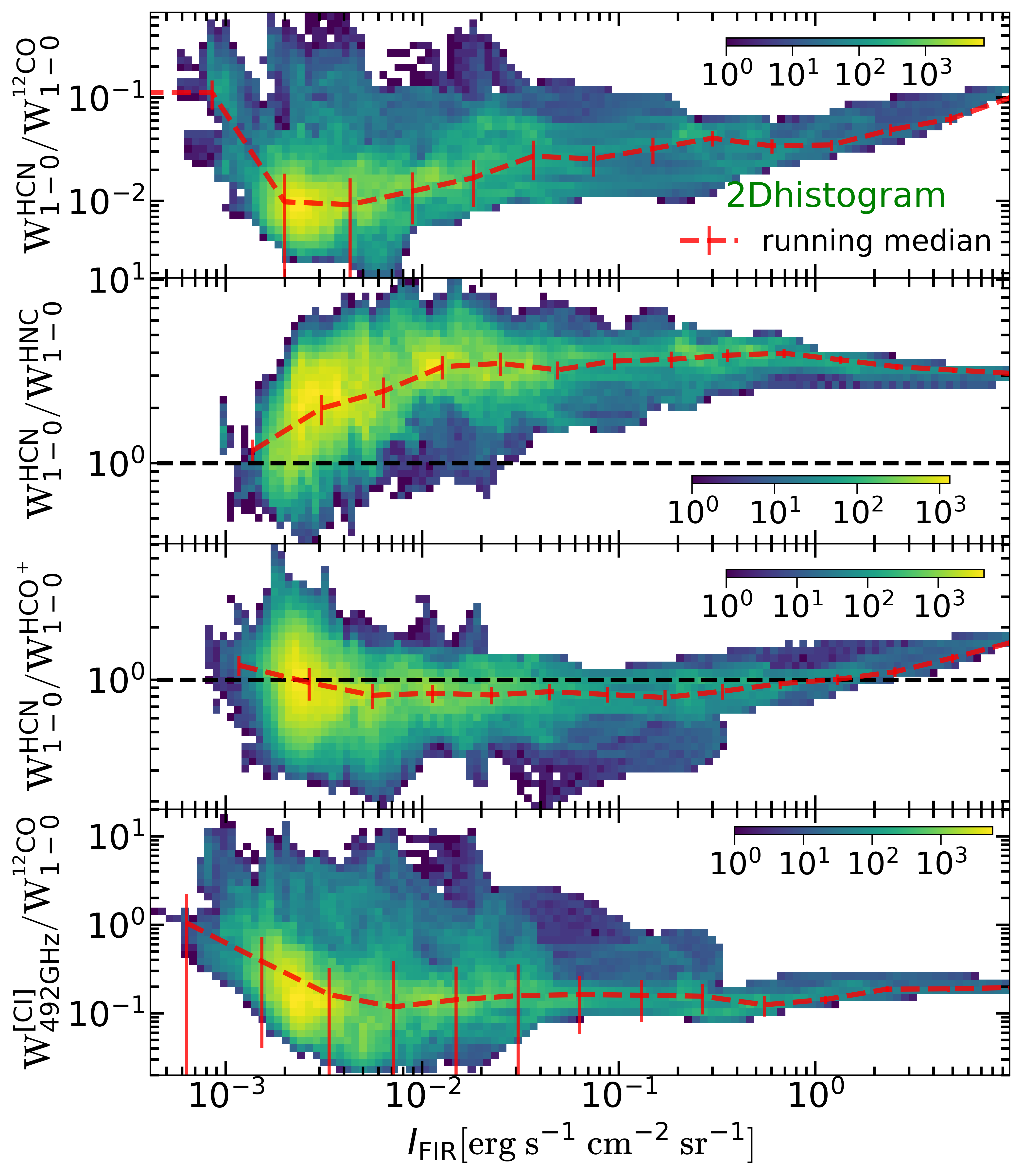} 
        \caption{Same as Fig.~\ref{fig:ratiosAv} but as a function of $I_{\rm FIR}$.} 
    \label{fig:ratiosFIR}
\end{figure}

\begin{figure}[!h]
    \centering
    \includegraphics[width=0.49\textwidth]{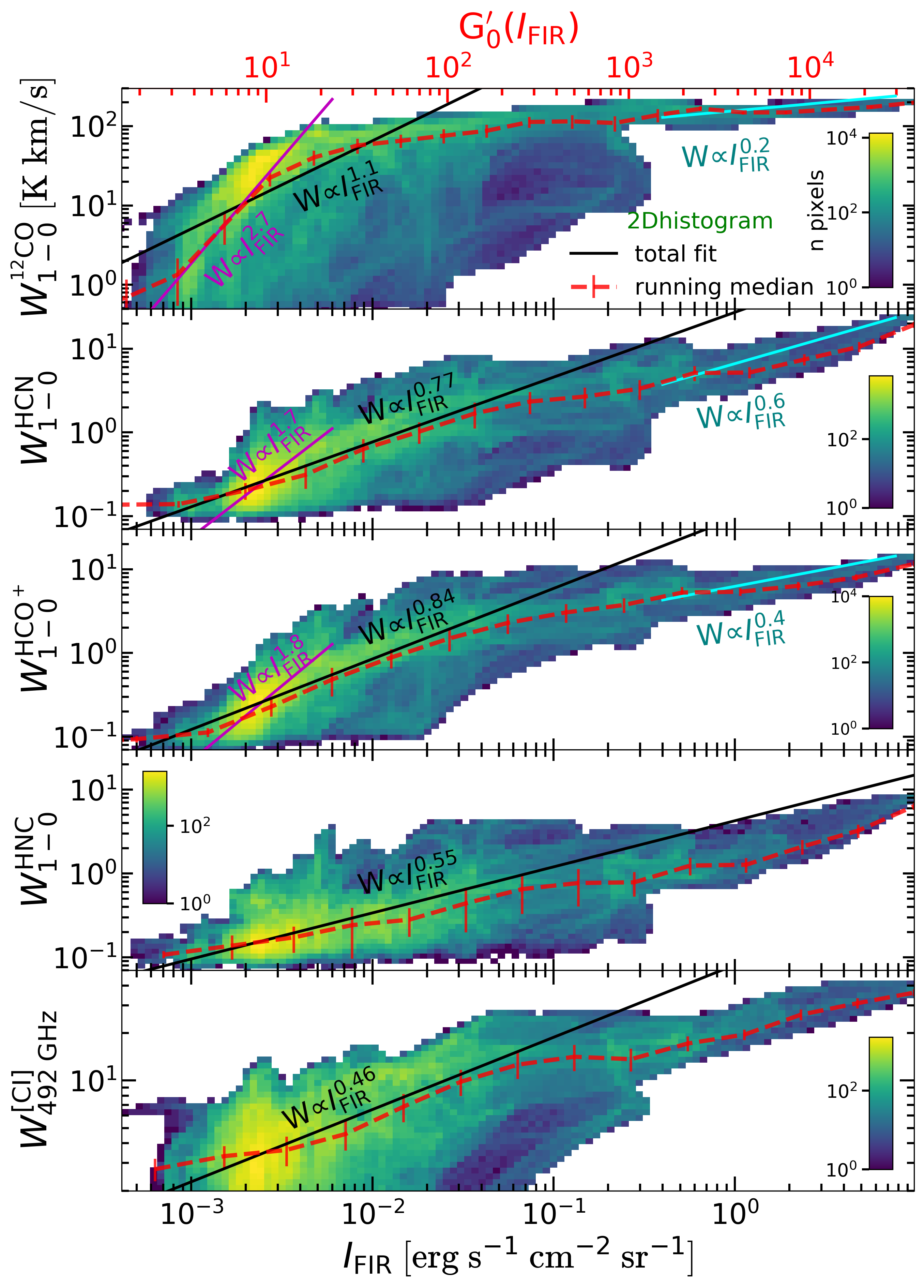}
    \caption{{Same as Fig.~\ref{fig:lines-FIR_hist} but showing
    the exponents  $W$\,$\propto$\,$I_{\rm FIR}^{1/N}$ (see text).}} 
    \label{fig:lines-FIR_hist_app}
\end{figure}

\clearpage

\section{SED-derived parameters \label{Sect:App-SED}}

\subsection{Dust temperature and column density maps}

In previous papers \citep[e.g.,][]{Pety17a}, we have used dust temperature and column density maps derived from the maps published by \citet{Lombardi2014}, based on \textit{Herschel} and \textit{Planck} data. More recently, similar maps have been published by \citet{Konyves2020} as part of the HGBS \citep[][]{Andre2010,Schneider2013}\footnote{The HGBS data can be found at \url{http://www.herschel.fr/cea/star-formation/en/Phocea/Vie_des_labos/Ast/ast_visu.php?id_ast=66}}.

\begin{figure}[!t]
    \centering
    \includegraphics[width=0.47\textwidth]{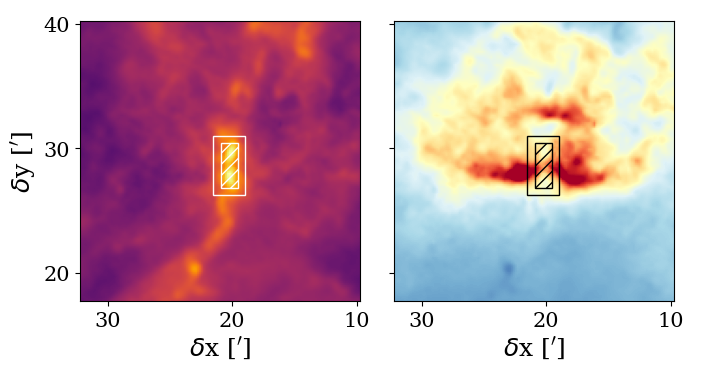}
    \caption{Compositing geometry in the NGC\,2024 region between data from \citet{Lombardi2014} and \citet{Konyves2020}, for column density (\textit{left}) and dust temperature (\textit{right}). The outer contour is the one in which the compositing is executed. The ``selection area'' over which the K20 datasets are rescaled to match the L14 ones corresponds to the outer contour minus the inner, hashed area (affected by the artifact in L14 and masked out). {The full maps corresponding to these zoom-ins can be found in Fig. \ref{fig:SED_TAI}}} 
    \label{fig:inpaint}
\end{figure}

The \citet{Konyves2020} maps benefit from additional \textit{Herschel} observations, which were used to correct a small patch of pixels where the SPIRE detectors are saturated in the NGC\,2024 region in the original data (Fig. \ref{fig:inpaint}). This saturation creates an artifact in the maps derived by \citet{Lombardi2014}. However, the two datasets, which in principle use essentially the same observational data and the same derivation method (modified black-body SED fitting) display some discrepancies (typically a factor of \mbox{1.2 -- 1.5} in temperature and a factor \mbox{1.2 -- 2} in column density), which can partly be attributed to the fact that \citet{Konyves2020} use a fixed value of zero-point calibrations for the \textit{Herschel} bands and a fixed $\beta$ index for the modified black body, whereas \citet{Lombardi2014} use spatially varying values for these parameters based on previous SED fitting of \textit{Planck} and IRAS data. 

For the sake of consistency with our previous work and due to the level of detail of its SED fitting, we chose to keep the \citet{Lombardi2014} {data} as our reference, but we incorporated the newly available reobserved patch in NGC\,2024 by rescaling the \citet{Konyves2020} maps and compositing {them} into the \citet{Lombardi2014} maps (hereafter K20 and L14 maps respectively), as illustrated in Fig. \ref{fig:inpaint}.

{The compositing procedure for both column density and temperature was the following:}

\begin{enumerate}
    \item {The saturated patch in NGC\,2024 is masked out as tightly as possible, with a $81''\times216''$ ($9\times24$ pixels) mask.}
    \item {A $36''$ wide (4 pixels, about one beam) border is selected around the masked patch, yielding a $153''\times288''$ selection region.
    \item From this we obtain a correction constant $A_X$ corresponding to the local ratio between the datasets:}
    \begin{equation}
        A_X = \left\langle X_\mathrm{K20}/X_\mathrm{L14} \right\rangle_\textrm{selection}
    \end{equation}
     {where $X$ corresponds to $N$ or $T$.}
    \item {The $153''\times288''$ binary mask is then smoothed with a Hann window of radius 45" (4/3 beam) to avoid compositing artifacts. This yields a gray-scale mask $M$, with value 1 in most of the map, 0.5 at the edge of the compositing region, and 0 at the center of the NGC\,2024 region.}
    \item {The two datasets are finally merged into a final column density map $N$ or a final temperature map $T$:}
    \begin{equation}
        X = M \cdot X_\mathrm{L14} + \left( 1 - M \right) \cdot \frac{X_\mathrm{K20}}{A}
    \end{equation}
   {where again $X$ corresponds to $N$ or $T$.}
\end{enumerate}

\begin{figure}[!h]
    \centering
    \includegraphics[height=0.3\textwidth]{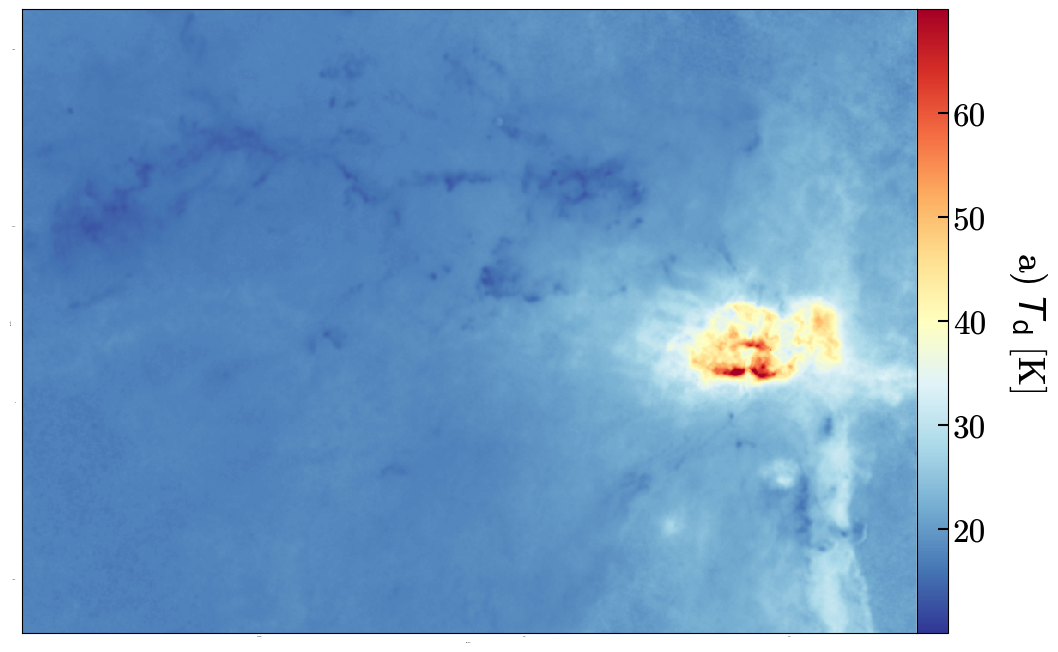}
    \includegraphics[height=0.3\textwidth]{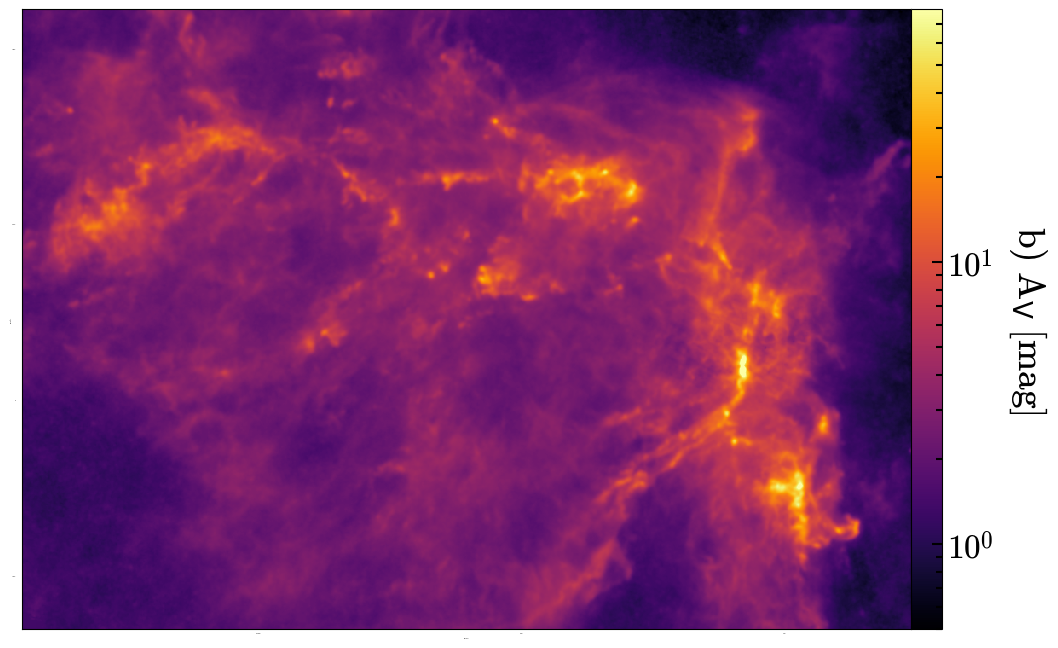}
    \includegraphics[height=0.3\textwidth]{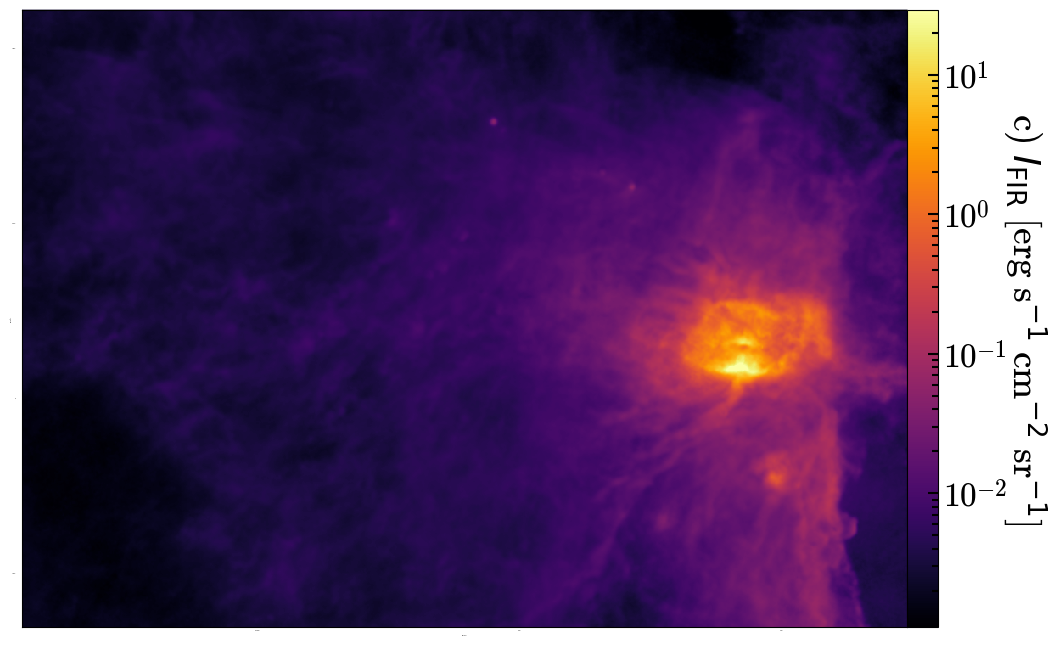}
    \caption{Maps of dust  SED derived parameters at  30$''$ angular resolution. (a)~Dust temperature, $T_\mathrm{d}$. (b)~Visual extinction, $A_\mathrm{V}$. (c)~FIR surface brightness (integrated from 40 to 500~$\upmu$m).} 
    \label{fig:SED_TAI}
\end{figure}

No significant discontinuity is visible in the combined datasets. Figure~\ref{fig:SED_TAI} shows the spatial distribution of the dust temperature, the visual extinction, and integrated FIR intensity which are derived from these maps.

\subsection{Determination of gas masses}\label{app:gasmass}

We determined the  mass of molecular gas as:
\begin{equation}
    M_\mathrm{H_2} = \mu\;m_\mathrm{H}\;A_\mathrm{pixel}\;\sum\;N(\mathrm{H_2})
\end{equation}
where \mbox{$N$(H$_2$)=1.9$\times$10$^{21}$~$A_\mathrm{V}$},  $\upmu$=2.8 is the molecular weight per H$_2$, $m_\mathrm{H}$ is the hydrogen atom mass, and \mbox{$A_\mathrm{pixel}$} is the area of each pixel in cm$^2$. 
In Orion~B, the total H$_2$ mass is \mbox{$M_\mathrm{H_{2},tot}$\,$\sim$1.7$\times$10$^4$~\Ms}. The dense gas mass, $M_\mathrm{dg}$,
was computed for  visual extinctions \mbox{$A_\mathrm{V}>$8~mag} \citep[e.g.,][]{Lada2010,Shimajiri2017}.

\section{{HCN hyperfine structure  analysis}}
\subsection{HCN $J$=\,1--0 LTE-HFS fitting method}
\label{App:HFS fit}
\begin{figure}[!h]
    \centering
    \includegraphics[width=0.49\textwidth]{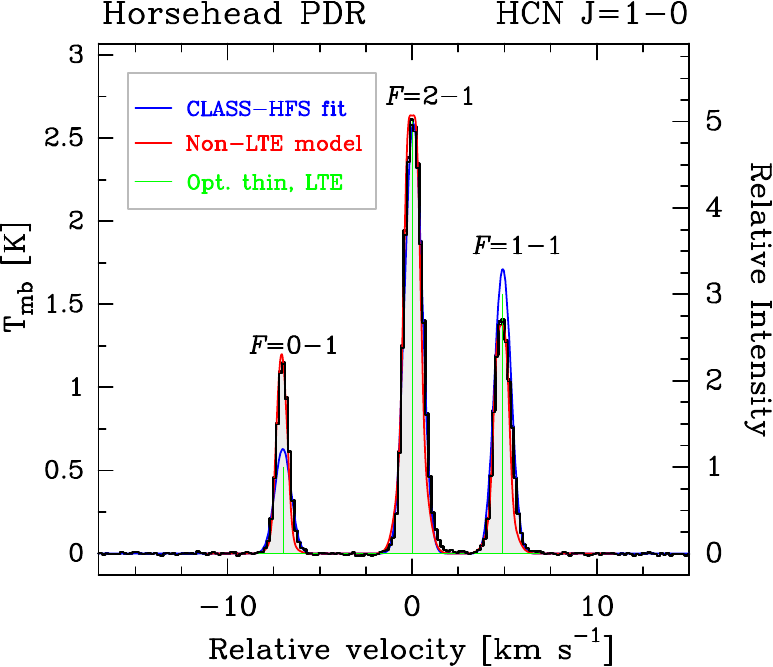}
    \caption{HCN\,$J$\,=\,1--0 HFS lines toward the Horsehead PDR. The right axis shows the normalized line intensity to make clear that the observed HFS emission differs from the optically thin LTE line ratios 1:5:3 (green lines).
    Red curves show the results of a non-LTE radiative transfer model including line overlaps
    \citep[for details, see  Sect.~5.1 in][]{Goicoechea2022}.
    Blue curves show the best LTE-HFS fit using \texttt{CLASS}\cref{footn:class}.}
    \label{fig:HFS-gauss}
\end{figure}
In Sect.~\ref{sect:res-HFSfitting} we determined $T_\mathrm{ex}$($J$=1--0) and the opacity corrected column density $N^{\tau,\mathrm{corr}}$(HCN) by applying the LTE-HFS fitting method implemented in \texttt{CLASS}{\cref{footn:class}}. This method uses as input the line separation and the intrinsic line strengths of the $J$=1--0 HFS components, 1:5:3 {(so that $S$\,=\,9)}. The method assumes that the three HFS lines have the same $T_\mathrm{ex}$ and linewidth $\Delta v$. The output parameters  are the linewidth $\Delta v${\,=\,$p_3$}, the velocity of the reference component $v_{\mathrm{LSR}}${\,=\,$p_2$}, the sum of all the line center opacities 
{such as $p_4$\,S\,=\,$\sum \tau_{\rm HFS,\,i}$}, 
and the product of the antenna temperature and the line opacity 
{$p_1$\,=\,$T_{a}^*$\,$\tau$}. This  procedure allows one to derive the {(LTE)} excitation temperature as:
\begin{equation}
{T_\mathrm{ex}=T_\mathrm{bg}+\frac{F_\mathrm{eff}}{B_\mathrm{eff}}\frac{p_1}{p_4}} 
\end{equation}
where $T_{\mathrm{bg}}$ is the background temperature ($\sim$2.73~K), and $F_\mathrm{eff}/B_\mathrm{eff}$ is the ratio of the telescope forward and beam efficiencies  (see \texttt{CLASS}\cref{footn:class} documentation).
From the opacities and $T_\mathrm{ex}$, we derive the opacity corrected column density, assuming Boltzmann populations at a single $T_{\rm ex}$ value:
\begin{equation}
    N^{\tau,\mathrm{corr}} = \frac{8\pi\nu^3}{A_{\mathrm{ul}}~c^3} \frac{Q(T_{\mathrm{ex}})}{g_{\mathrm{u}}}\frac{e^{E_{\mathrm{u}}/kT_{\mathrm{ex}}}}{e^{h\nu/kT_{\mathrm{ex}}}-1}    \frac{W_{F=2-1}}{[J_\nu(T_{\mathrm{ex}})-J_\nu(T_{\mathrm{bg}})]}\frac{{\tau_{F=2-1}}}{1-e^{-{\tau_{F=2-1}}}},
    \label{eq:HCN-N_hfs}
\end{equation}
where $W_{F=2-1}$ is the integrated intensity of the {main HFS} component \mbox{$F$=2--1}, $\tau_{F=2-1}$ is the line center opacity, $Q(T_\mathrm{ex})$ is the partition function at a temperature of $T_\mathrm{ex}$, $g_\mathrm{u}$ is the statistical weight of the transition upper level, and $E_\mathrm{u}/k$ is the upper level energy. {The rotational partition function can be approximated with precision as:}
\begin{equation}
    {Q(T_{\mathrm{ex}})\simeq \frac{k\;T_{\mathrm{ex}}}{h\;B_0}\;e^{h\;B_0/3kT_{\mathrm{ex}}}.}
    \label{eq:partfunc}
\end{equation}
{where $B_0$ is the rotational constant \citep{McDowell1988}.}
{We took the HCN HFS spectroscopic parameters compiled in CDMS
\citep[][and references therein]{Endres16}.}

This {fitting} method works better on high S/N spectra. Thus, we only applied {it}  to the main cloud velocity  component ($v_{\rm LSR}\simeq$10~\kms) where S/N>5$\sigma$. {\mbox{Figure~\ref{fig:HFS-gauss}}} 
shows the anomalous HCN~$J$=1--0 spectrum  observed (at  by the IRAM\,30m telescope toward the Horsehead PDR position $\delta v$\,$\simeq$0.16~\kms~resolution). This figure compares the expected 
HFS line strengths in the LTE and optically thin limit (green lines), 
the result of the LTE-HFS fit in \texttt{CLASS} (blue curve), and a non-LTE radiative transfer model {\citep[red curve,][]{Goicoechea2022}.}

\subsection{{HCN \textit{J}\,=\,2--1 and 3--2 HFS line ratios}}

{The HCN~$J$=2--1 transition has six HFS lines that blend
into three lines with relative intensity ratios  \mbox{$\sim$1:9:2} in the LTE and optically thin limit.
The HCN~$J$=3--2 transition also has six HFS lines. Only the central ones are blended and cannot be spectrally resolved. This gives the impression of three lines with relative intensity ratios \mbox{1:25:1} in the LTE and optically thin limit \mbox{\citep[e.g.,][]{Ahrens02,Loughnane2012}}. 
Here we term these  three apparent components (blueshifted, central, and redshifted) of 
 the \mbox{$J$=2--1}  and \mbox{$J$=3--2} rotational lines 
 as ``satellite($L$),'' ``main,'' and ``satellite($R$),'' respectively.
\mbox{Table~\ref{tab:HCN-HFS_multiJ}} provides the entries used
to construct \mbox{Fig.\ref{fig:hcn_hfs_21_32}}.}

\begin{table*}[!h]
\caption{{Observed HCN~$J$\,=\,2--1 opacities (see \mbox{Sect.~\ref{app:opacities}}  for their definition), and HCN~$J$\,=\,2--1 and $J$\,=\,3--2 HFS line intensity ratios \mbox{satellite($R$)/main} and
    \mbox{satellite($B$)/main} (see \mbox{Sect.~\ref{sect:taupopdia}}  for their definition) in the sample of representative positions.}}
    \label{tab:HCN-HFS_multiJ}
    \centering
    \begin{tabular}{lccccc}
        \toprule
        Pos & HCN $J$=2$-$1 & \multicolumn{2}{c}{$J$=2$-$1} & \multicolumn{2}{c}{$J$=3$-$2} \\
         & $\tau$ &  Ratio satellite($B$)/main & Ratio satellite($R$)/main & Ratio satellite($B$)/main & Ratio satellite($R$)/main \\ 
         \midrule
         \#1 & 13.9 &  1.1$\pm$0.2 & 1.5$\pm$0.2 & -- & -- \\
         \#2 &  1.7 &  0.15$\pm$0.02 & 0.38$\pm$0.38 & 0.12$\pm$0.006 & 0.14$\pm$0.004 \\
         \#3 & 5 &  0.31$\pm$0.01 & 0.51$\pm$0.02 & 0.28$\pm$0.02 & 0.50$\pm$0.03 \\
         \#4 & 2.6 &  0.34$\pm$0.03 & 0.67$\pm$0.05 & 0.29$\pm$0.04 & 0.23$\pm$0.04 \\
         HH PDR & 1.8 &  0.30$\pm$0.03 & 0.60$\pm$0.05 & 0.10$\pm$0.02 & 0.24$\pm$0.03 \\
         \#6 & -- &  -- & -- & -- & -- \\
         \#7 & 2 &  0.4$\pm$0.1 & 0.8$\pm$0.1 & 0.11$\pm$0.02 & 0.23$\pm$0.02 \\
         \#8 & 3 &  0.55$\pm$0.08 & 1.1$\pm$0.1 & 0.13$\pm$0.04 & 0.29$\pm$0.05 \\
         HH CORE &  1.7 &  0.25$\pm$0.02 & 0.63$\pm$0.04 & -- & -- \\
         \#10 & -- &  0.21$\pm$0.04 & 0.51$\pm$0.04 & -- & -- \\
         \#11 & -- &  -- & -- & -- & -- \\
         \#12 & -- &  0.24$\pm$0.04 & 0.51$\pm$0.05 & -- & -- \\
         \#13 & -- &  0.17$\pm$0.04 & 0.41$\pm$0.04 & -- & -- \\
         \#14 & -- &  -- & -- & -- & -- \\
         \bottomrule
    \end{tabular}

\end{table*}

\section{{HCN and HNC rotational diagrams}}
\label{app:rot-diag}
In Sect.~\ref{sect:taupopdia}, we analyzed multiple-$J$ HCN and {HNC} line observations. Here we detail how we obtained the opacity corrected population diagrams.
{We took the HCN and HNC rotational spectroscopic parameters from CDMS
\citep[][and references therein]{Endres16}.}
\subsection{Estimation of line opacities:}\label{app:opacities}
To estimate the opacity of a given rotational transition \mbox{HCN $J_\mathrm{u}\rightarrow J_\mathrm{l}$}, we used the observed HCN/H$^{13}$CN line ratio. We assumed that the H$^{13}$CN $J_\mathrm{u}\rightarrow J_\mathrm{l}$ line is optically thin, that $T_\mathrm{rot}$ is the same for the two isotopologues, and that HCN and H$^{13}$CN $J_\mathrm{u}\rightarrow J_\mathrm{l}$ emit from the same gas volume \citep{Goldsmith1984}. Hence,
\begin{equation}
    \frac{W_\mathrm{HCN}}{W_\mathrm{H^{13}CN}} \approx \frac{1-e^{\tau_\mathrm{HCN}}}{\tau_\mathrm{HCN}} \frac{[^{12}\mathrm{C}]}{[^{13}\mathrm{C}]},
    \label{eq:tauC}
\end{equation}
where $\tau$ is the line opacity, $W$ is the integrated line intensity, and [$^{12}$C]/[$^{13}$C] is the isotopic ratio, around 60 in Orion \citep{Langer1990}.

The H$^{13}$CN $J$=1--0 line is detected toward NGC 2024 cores, as well as toward the Horsehead PDR and Core positions in the higher sensitivity WHISPER survey \citep{Gerin2009,Pety2012}. We detect HCN and H$^{13}$CN $J$=2--1 toward several  positions (\#1, \#2, \#3, \#4, \#7, \#8, HH-PDR and HH-Core). {Table~\ref{tab:HCN-HFS_multiJ} shows the estimated HCN~$J$\,=\,2$-$1 opacities following Eq.~(\ref{eq:tauC}).}  We computed the opacities of the rotational lines $J$=1--0 
{(for positions with no H$^{13}$CN $J$=1--0 detections)},
3--2, and 4--3 as a function of $J$=2--1 {line opacity, as}:
\begin{equation}
    \tau_\nu = \frac{A_{\mathrm{ul}}\; g_{\mathrm{u}}}{8\pi\;\Delta{v}} \left (\frac{c}{\nu}\right )^3 \frac{N_{\mathrm{tot}}}{Q(T_{\mathrm{rot}})}\frac{e^{h\nu_{\mathrm{ul}}/kT_{\mathrm{rot}}}-1}{e^{E_{\mathrm{u}}/kT_{\mathrm{rot}}}},\mathrm{and}
    \label{eq:taunu}
\end{equation}
\begin{equation}
    \frac{\tau_{\mathrm{ul}}}{\tau_{2-1}} = \frac{A_{\mathrm{ul}}\;g_{\mathrm{u}}\;\Delta {v}_{2-1}}{A_{\mathrm{2-1}}\;g_{\mathrm{2}}\;\Delta {v}_{\mathrm{ul}}} \left (\frac{\nu_{2-1}}{\nu_{\mathrm{ul}}}\right )^3 \frac{e^{E_2/kT_{\mathrm{rot}}}}{e^{E_{\mathrm{u}}/kT_{\mathrm{rot}}}}\frac{e^{h\nu_{\mathrm{ul}}/kT_{\mathrm{rot}}}-1}{e^{h\nu_{2-1}/kT_{\mathrm{rot}}}-1},
    \label{eq:taunu21}
\end{equation}
where $T_\mathrm{rot}$ is the rotational temperature, ul refers to the transition from the upper to lower level, and $\Delta v_\mathrm{ul}$ is the linewidth. For simplicity, we assume a linewidth ratio $\Delta v_{2-1}/\Delta v_{\mathrm{ul}}=1$. For the fainter emitting positions, where we do not detect H$^{13}$CN, we assume that the HCN emission is optically thin.

\subsection{Opacity corrected population diagram}\label{app:taupopdia}
In order to determine  $T_\mathrm{rot}$ and the column density toward each observed position, we computed  rotational diagrams assuming a single $T_\mathrm{rot}$(HCN) \citep{Goldsmith_1999},
\begin{equation}
    \mathrm{{ln}}\left(\frac{N_\mathrm{u}}{g_\mathrm{u}}\right) = \mathrm{{ln}}\left (\frac{N_\mathrm{tot}}{Q(T_\mathrm{rot})}\right)-\frac{E_\mathrm{u}}{kT_\mathrm{rot}},
    \label{eq:rotdiagthin}
\end{equation}
where $N_\mathrm{u}$ is the level $u$ population, $N_\mathrm{tot}$ is the total column density. We  iteratively applied the opacity correction $\left (C_\tau=\frac{\tau}{1-e^{-\tau}}\right )$ to the population diagram until a solution for $T_\mathrm{rot}$ and $N$ converged. For the first iteration we use eq.~(\ref{eq:rotdiagthin}) and compute the line opacities from eq.~(\ref{eq:taunu21}). From the second iteration to convergence, we implement the opacity correction as
\begin{equation}
    \mathrm{{ln}}\left (\frac{N_\mathrm{u}}{g_\mathrm{u}}\right ) = \mathrm{ln}\left(\frac{N_\mathrm{tot}}{Q(T_\mathrm{rot})}\right )-\frac{E_\mathrm{u}}{kT_\mathrm{rot}}-\mathrm{{ln}}~C_\tau.
    \label{eq:rotdiagthick}
\end{equation}

\begin{figure}[!h]
    \centering
    \includegraphics[width=0.4999\textwidth]{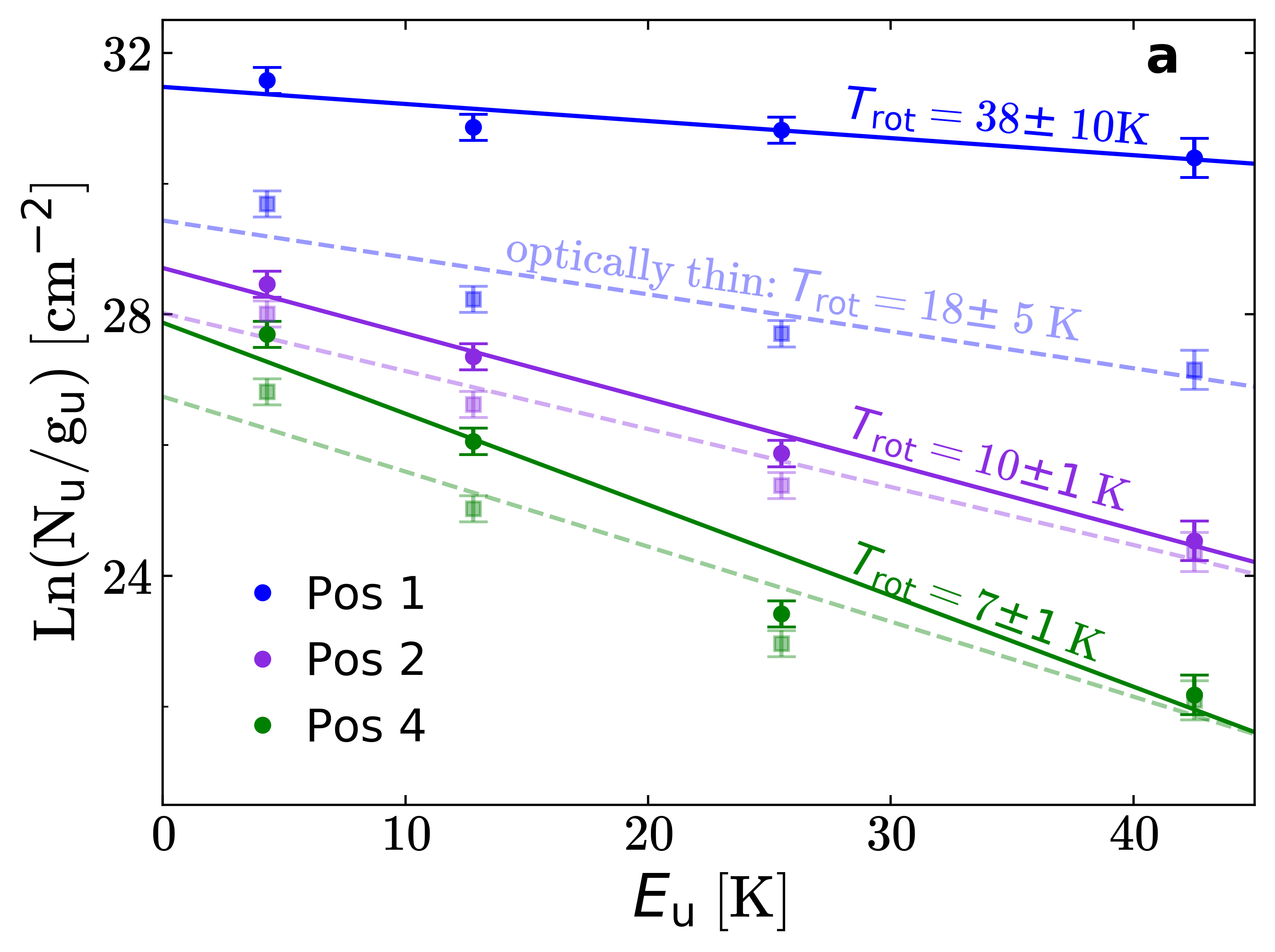}
    \includegraphics[width=0.4999\textwidth]{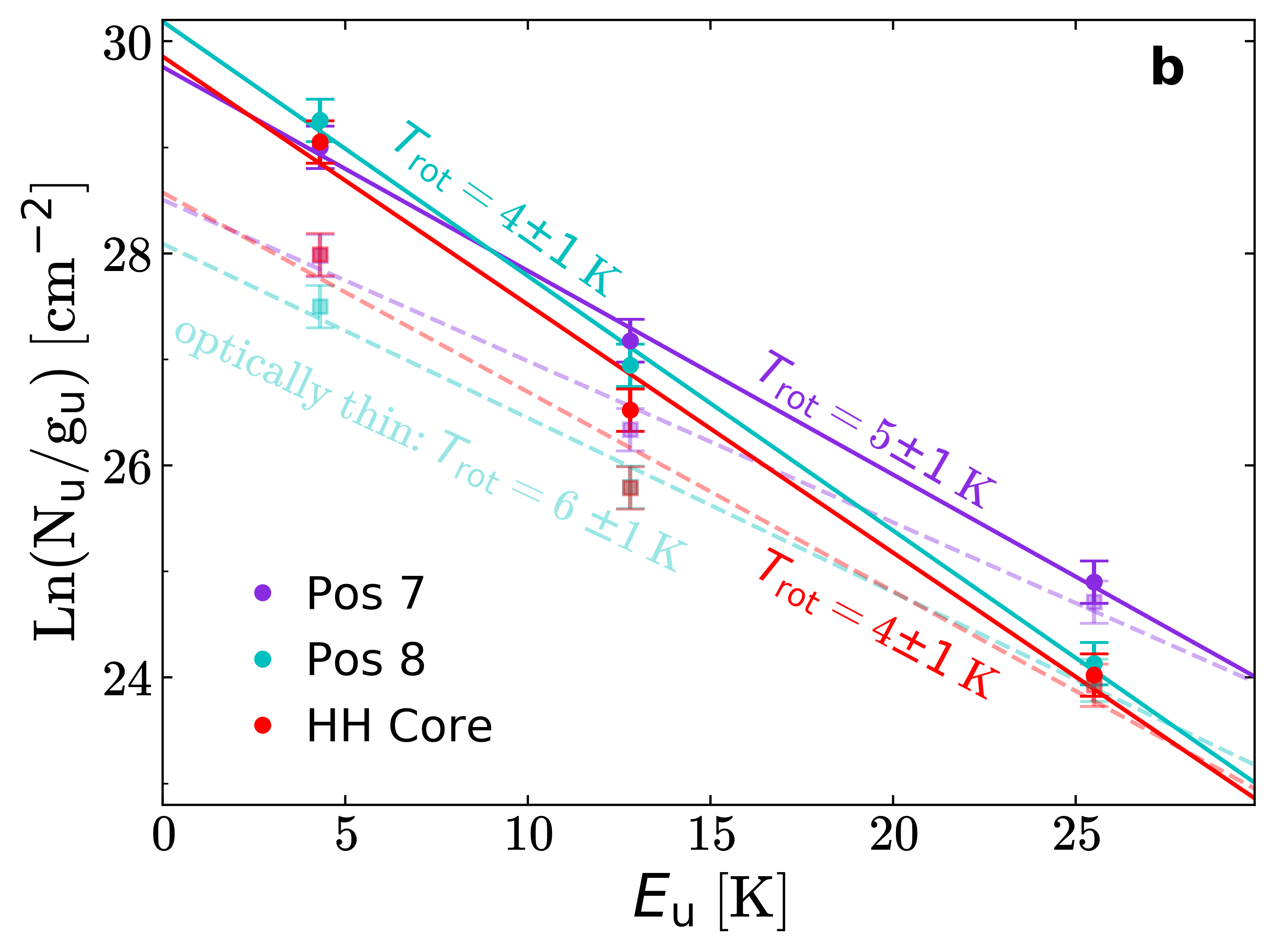}

    \caption{HCN rotational diagrams. (a) Opacity corrected (straight lines) population diagrams for positions \#1, \#2, \#4. Lighter color dashed lines and squares show the optically thin population diagrams.
    (b) The same but for positions \#7, \#8, and the HH-Core.} 
    \label{fig:hcn_taus_pos}
\end{figure}

The uncertainties are $\Delta \left ( \mathrm{{ln}}\left(\frac{N_u}{g_u}\right)\right) = \frac{\Delta W}{W}$, where $\Delta W$ is the uncertainty of the integrated intensity, $\sim$20\% of $W$. Figure~\ref{fig:hcn_taus_pos} shows a comparison between the optically thin (squares and dashed lines), and opacity-corrected HCN population diagrams (circles and straight lines), for positions \#1, \#2, and  \#4 in Fig.~\ref{fig:hcn_taus_pos}a, and for positions \#7, \#8, and Core, in the Horsehead, in Fig.~\ref{fig:hcn_taus_pos}b.

\subsection{{HNC rotational population diagrams}}\label{sect:HNC_dr}
{Table~\ref{tab:HNC_TexN} shows the rotational temperatures and column densities obtained from rotational diagrams constructed with the observed
HNC \mbox{$J$\,=\,1--0} and \mbox{$J$\,=\,3--2} lines. This table also shows the estimated \mbox{$J$\,=\,1--0} line opacities toward positions with HNC and HN$^{13}$C~$J$=1$-$0 detections (assuming that both emission lines steam from the same gas)}.

\begin{table}[!h]
\caption{HNC excitation temperature and column densities obtained from rotational diagrams{, as well as HNC~$J$=1$-$0 line opacities for  positions in which we  detected HN$^{13}$C~$J$=1$-$0}.}
\label{tab:HNC_TexN}  
\centering
\begin{threeparttable}
\resizebox{0.32\textwidth}{!}{%
\begin{tabular}{lccc@{\vrule height 7.5pt depth 5pt width 0pt}}
\toprule
     & $T^\mathrm{thin}_\mathrm{rot}$     & $N^\mathrm{thin}$     & $\tau$(HNC 1--0) \\
     & {[}K{]} &  10$^{13}$ {[}cm$^{-2}${]}     &      \\
     \midrule
\#1 & 11 &  1.6   & 1.2  \\
\#2  & 9    & 0.2   &      \\
\#3  & 7    & 0.3    &      \\
\#4  & 6 & 0.3   &    \\
PDR HH  & 5 &  0.3    & 1.5     \\
\#6 & 6    & 0.4   &    \\
\#7  & 8    & 0.9   &     \\
\#8  & 8    & 0.6    &      \\
Core HH & 6 & 0.4    & 2.8       \\
\#10  & 6    & 0.3     &       \\
\#11  &   &     &      \\
\#12 & 6    & 0.4   &     \\
\#13 & 5    & 0.3     &     \\
\#14 & 5    & 0.1   &     \\

\bottomrule \vspace{-0.6cm}      
\end{tabular}} 
\end{threeparttable}
\end{table}

\newpage

\section{Raster crossmap observation strategy}

\begin{figure*}[!t]
    \centering
    \includegraphics[height=0.25\textwidth]{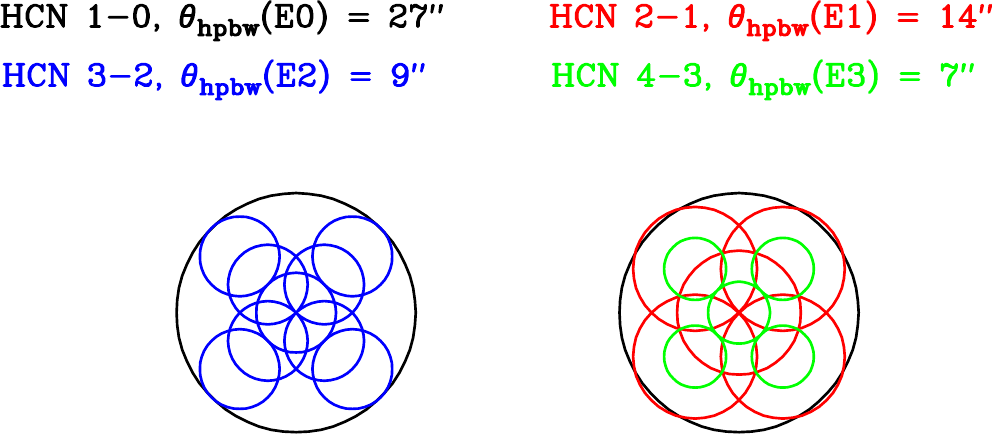}
    \caption{Pointing strategy of raster crossmaps designed to obtain multi frequency line observations at the $\sim$30$''$ angular resolution of the Orion-B $J$\,=\,1--0 maps.} 
    \label{fig:raster-map-post}
\end{figure*}

Because the telescope beam size changes with frequency, we split the observation of each position  into a small raster crossmap of $\sim$30'' extent (the beam size in the  3\,mm band). Figure~\ref{fig:raster-map-post} shows the target positions. This way, the raster averaged spectra from the 2 and 1\,mm bands ($J$=2--1 and 3--2 lines), and to a lesser extend the 0.8\,mm band ($J$=4--3), can directly be compared with the $J$=1--0 observations.

\begin{table}[!h]
\centering
\caption{Observed frequency ranges and telescope parameters}
\label{tab:obs_bands}
\vspace{-0.2cm}
\begin{threeparttable}
\resizebox{0.47\textwidth}{!}{
\begin{tabular}{lcccc}
\toprule

Rec. \& Back. & Freq. range & $\delta v$  & HPBW  & Pointings  \\ 
     &   [GHz]  &  [\kms]    & [arcsec] &   [beams]         \\\midrule
E1-FTS200   &  171.7 -- 179.8  & 0.33  & 14 &  5 \\
\multirow{2}{*}{E2-FTS200}   &   249.2 -- 253.3 & 0.22 & \multirow{2}{*}{9} &  \multirow{2}{*}{9}  \\
     &  253.3 -- 257.3  & 0.23  &  &  \\
E3-FTS200   &  349.7 -- 357.8  & 0.17  & 7 & 5    \\
\bottomrule
\end{tabular}}
\tablefoot{Col 1. EMIR receiver and FTS backend. Col 2. Observed frequency range. Col 3. $\delta v$ is the observed spectral resolution in velocity units in the observed frequency range. HPBW is the angular resolution of the telescope. Col 4. Number of pointings in the raster map, around each target position, for each spectral band (see Fig.~\ref{fig:raster-map-post}).}
\end{threeparttable}
\vspace{-0.2cm}
\end{table}
\clearpage

\newpage

\section{Complementary figures and tables}

\begin{figure*}[!h]
    \centering
    \includegraphics[height=1.\textheight]{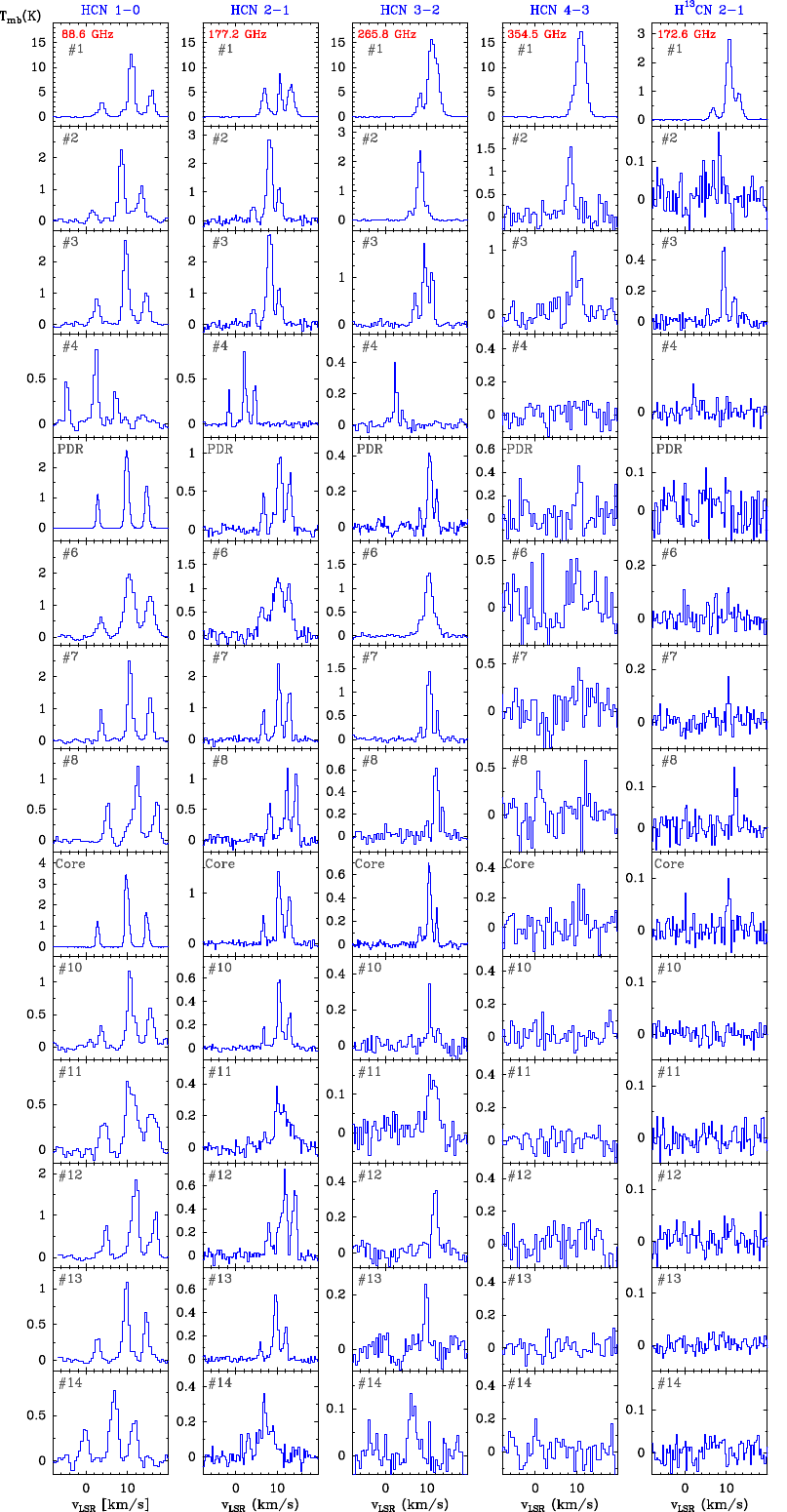} 
    \caption{HCN $J$=1--0 to $J$=4--3, and H$^{13}$CN $J$=2--1 pointed observations. The velocity resolution is $\sim$0.5~\kms. } 
    \label{fig:HCN-multiJ}
\end{figure*}

\begin{figure*}[!h]
    \vspace{-0.3cm}
    \centering
    \includegraphics[height=1.\textheight]{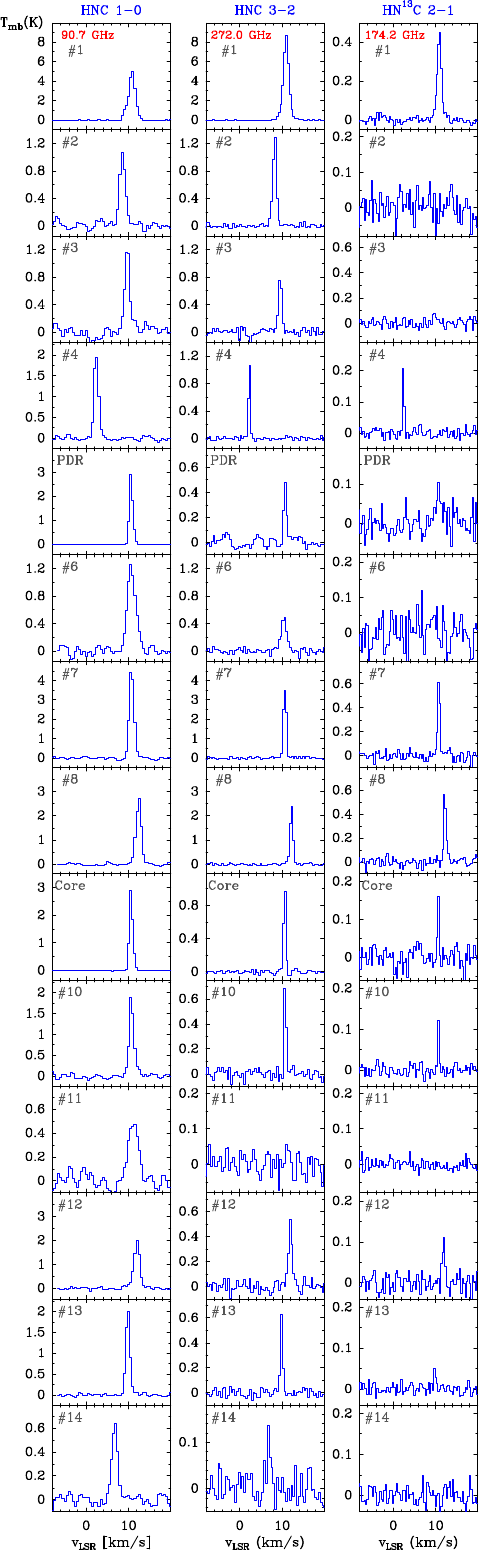}
    \caption{HNC \textit{J}=1--0 and \textit{J}=3--2, and HN$^{13}$C $J$=2--1 pointed observations. The velocity resolution is $\sim$0.5~\kms. } 
    \label{fig:HNC-multiJ}
\end{figure*}

{In this section we provide figures (Figs.~\ref{fig:HCN-multiJ} and \ref{fig:HNC-multiJ})  with all detected
HCN, HNC, H$^{13}$CN, and HN$^{13}$C line spectra toward the sampled of selected positions in Orion\,B (pointed observations). The following tables summarize their observed spectroscopic parameters: integrated line intensities, peak LSR velocity, line width
and, peak temperature.}

\newpage
\begin{sidewaystable*}[!h]

\caption{Line spectroscopic parameters obtained from Gaussian fits to the observed HCN lines.\vspace{-0.3cm}} 
\label{tab:HCNJfit}

    \begin{threeparttable}
    \resizebox{\textheight}{!}{%
    \begin{tabular}{lcccccccccccccccc}
    \toprule
    
        &   \multicolumn{4}{c}{HCN \textit{J}=1--0} & \multicolumn{4}{c}{HCN \textit{J}=2--1} &    \multicolumn{4}{c}{HCN \textit{J}=3--2} & \multicolumn{4}{c}{HCN \textit{J}=4--3}    \\
    Freq. [MHz] & \multicolumn{4}{c}{88631.6} & \multicolumn{4}{c}{177261.1} &    \multicolumn{4}{c}{265886.4} & \multicolumn{4}{c}{354505.5} \\
    Pos.    & $W$ & $v_\mathrm{LSR}$ & $\Delta v$($F$=0-1) & $T_\mathrm{mb}$ & $W$ & $v_\mathrm{LSR}$ & $\Delta v^\dagger$ & $T_\mathrm{mb}$ & $W$ & $v_\mathrm{LSR}$ & $\Delta v^\dagger$ & $T_\mathrm{mb}$ & $W$ & $v_\mathrm{LSR}$ & $\Delta v^\ddagger$ & $T_\mathrm{mb}$ \\
       & [\Kkms] & [\kms] & [km s$^{-1}$] & [K] & [\Kkms] & [\kms] & [km s$^{-1}$]  & [K] & [\Kkms] & [\kms] & [km s$^{-1}$] & [K] & [\Kkms] & [\kms] & [km s$^{-1}$] & [K] \\
    
    \midrule
    \vspace{0.05cm} \#1 & 37.2  & 11.0 & 1.6 & 12.8 & 34.5 & 10.6 & 6.6 & 9.7  & 45.9 & 11.1 & 3.8 & 14.8  & 46.9  & 11.0 & 2.8 & 16.4       \\ 

     \#2 & 6.9 & 8.5 & 1.8 & 2.3 & 6.9  & 8.2 & 4.0 & 3.1  & 4.5 & 8.4 & 2.8 & 2.3  & 2.9  & 8.3 & 3.1 &  1.7     \\

     \#3 & 8.1 & 9.4 & 1.5 & 2.7  & 8.4 & 9.2 & 4.9 & 3.2 & 4.0  & 9.5 & 3.3 & 1.5 & 2.1 & 9.7 & 2.4 & 1.0 \\

     \#4 & 2.1  & 2.6 & 1.0 & 0.8  & 1.4 & 2.0 & 4.8 & 0.9  & 0.4 & 2.3 & 1.9 & 0.4 & 0.3 & 0.8 & 2.0 & 0.2 \\

     PDR HH & 5.6 & 9.6 & 1.2 & 2.3 & 2.7 & 10.6 & 6.0 & 1.0 & 0.7 & 10.5 & 2.3 & 0.4 & 0.23 & 10.5  & 0.9 & 0.4 \\

     \#6 & 10.3 & 10.4 & 2.0 & 2.0  & 6.2 & 11.3 & 5.4 & 1.2 & 3.6 & 10.5 & 2.7 & 1.3  &  & & & \\
    
    \#7 & 6.7 & 10.5 & 0.9 & 2.5 & 5.2 & 10.8 & 2.5 & 2.4 & 2.3 &  10.4 & 2.6 & 1.6 &  & & &   \\
    
    \#8 & 4.2 & 12.6 & 1.7 & 1.2 & 3.0 & 12.2 & 5.5 & 1.2 & 1.1 & 12.1 & 2.6 & 0.6 &  & & & \\
    
    Core HH & 6.8 & 9.6 & 0.8 & 3.2 & 3.4 & 10.8 & 4.7 & 1.4 & 1.1 & 10.5 & 2.5 & 0.7 &  & & & \\
    
    \#10 & 3.4 & 10.5 & 1.1 & 1.2 & 1.3 & 10.7 & 4.4 & 0.6 & 0.4 & 10.6 & 2.5 & 0.3 &  & & &  \\

    \#11  & 4.0 & 9.9 & 2.1 & 0.8 & 1.4  & 10.0 & 5.0 & 0.4 & 0.4 & 10.5 & 2.5 & 0.1 & --  & -- & -- & --       \\

    \#12 & 6.9 & 12.1 & 1.5 & 1.9 & 2.2 & 11.8 & 4.8 & 0.7 & 0.7 & 12.3 & 3.7 & 0.3  &  & & & \\
    
    \#13 & 3.3 & 10.0 & 1.7 & 1.1  & 1.1 & 9.8 & 4.2 & 0.6 & 0.3 & 9.7 & 2.8 & 0.2 &  & & & \\
    
    \#14 & 3.2  & 7.0 & 1.8 & 0.8  & 1.0 & 6.9 & 4.8 & 0.4 & 0.2 & 6.2 & 2.0 & 0.1 &  & & & \\    
    
\bottomrule \vspace{-0.6cm}
   
    \end{tabular}}
    \tablefoot{$W$ refers to velocity-integrated line intensity, $\int T_\mathrm{mb}\;dv$.. $v_\mathrm{LSR}$ is the local standard of rest velocity. $T_\mathrm{mb}$ is the peak temperature: for $J$=1--0 is the peak of the F=2--1 component; for $J$=2--1, the central line are the blended HFS components F=3--2 and F=2--1; for $J$=3--2, the central line are the blended HFS components F=4--3, 3--2, 2--1; for $J$=4--3 the HFS is not spectrally resolved. $\Delta$v($F$=0-1) is the linewidth of the hyperfine component $J$=1--0 $F$=0--1. $\Delta v^\dagger$ refers to Moment 2 of the $J$=2--1 line emission (the individual HFS component overlap). $\Delta v^\ddagger$ is the $J$=4-3 total linewidth from a gaussian fit (HFS complents overlap).  }
    \end{threeparttable}

\end{sidewaystable*}

\begin{table*}[!h]
\caption{Line spectroscopic parameters obtained from Gaussian fits to the observed HNC lines. \vspace{-0.3cm}}  
\label{tab:HNCJfit}

    \centering
   
    \begin{threeparttable}

    \begin{tabular}{lcccccccc}
    
    \toprule
    
        &   \multicolumn{4}{c}{HNC \textit{J}=1--0}  &    \multicolumn{4}{c}{HNC \textit{J}=3--2}    \\
    Freq. [MHz] & \multicolumn{4}{c}{90663.6} &    \multicolumn{4}{c}{271981.1}  \\
    Pos.    & $W$ & $v_\mathrm{LSR}$ & $\Delta v$ & $T_\mathrm{mb}$  & $W$ & $v_\mathrm{LSR}$ & $\Delta v$  & $T_\mathrm{mb}$  \\
       & [\Kkms] & [\kms] &  [\kms]  & [K]  & [\Kkms] & [\kms] & [\kms]  & [K] \\
    \midrule 
    \#1    & 9.7$\pm$0.1  & 10.8$\pm$0.1 & 1.9$\pm$0.1 & 4.8 & 12.4$\pm$0.1  & 11.0$\pm$0.1 & 1.5$\pm$0.1 & 7.8 \\ 

    \#2     & 1.7$\pm$0.1 & \phantom{0}8.6$\pm$0.1 & 1.5$\pm$0.1 & 1.1 & 1.3$\pm$0.1 & \phantom{0}8.4$\pm$0.1 & 1.0$\pm$0.1 & 1.3 \\ 

    \#3  & 1.9$\pm$0.1 & \phantom{0}9.7$\pm$0.1 & 1.4$\pm$0.1 & 1.2  & 0.8$\pm$0.1 & \phantom{0}9.3$\pm$0.1 & 0.9$\pm$0.1 & 0.8  \\ 

    \#4     & 2.6$\pm$0.1 & \phantom{0}2.4$\pm$0.1 & 1.2$\pm$0.1 & 2.1 & 0.7$\pm$0.1  & \phantom{0}2.3$\pm$0.1 & 0.5$\pm$0.1 & 1.0  \\

     PDR HH   & 2.4$\pm$0.1 & 10.7$\pm$0.1 & 1.1$\pm$0.1 & 2.0 & 0.4$\pm$0.1  & 10.6$\pm$0.1 & 0.8$\pm$0.1 & 0.5  \\

     \#6    & 3.0$\pm$0.1 & 10.7$\pm$0.1 & 2.3$\pm$0.1 & 1.2 & 0.8$\pm$0.1 & 10.6$\pm$0.1 & 1.5$\pm$0.1 & 0.5   \\ 

     \#7    & 6.5$\pm$0.1 & 10.7$\pm$0.1 & 1.3$\pm$0.1 & 4.7 & 3.7$\pm$0.1 & 10.6$\pm$0.1 & 1.0$\pm$0.1 & 3.5    \\ 

     \#8    & 4.2$\pm$0.1 & 12.4$\pm$0.1 & 1.5$\pm$0.1 & 2.7 & 2.4$\pm$0.1 & 12.2$\pm$0.1 & 1.0$\pm$0.1 & 2.4  \\
    
    Core HH    & 3.1$\pm$0.1 & 10.6$\pm$0.1 & 1.0$\pm$0.1 & 2.9  & 0.7$\pm$0.1 & 10.6$\pm$0.1 & 0.6$\pm$0.1 & 1.0   \\ 
    
    \#10    & 2.6$\pm$0.1 & 10.7$\pm$0.1 & 1.3$\pm$0.1 & 1.9  & 0.5$\pm$0.1 & 10.5$\pm$0.1 & 0.5$\pm$0.1 & 0.7  \\
    
    \#11     & 1.4$\pm$0.1  & 11.1$\pm$0.1 & 2.7$\pm$0.1 & 0.5  & --  & -- & -- & -- \\

    \#12    & 3.2$\pm$0.1 & 12.0$\pm$0.1 & 1.6$\pm$0.1 & 1.9  & 0.6$\pm$0.1 & 11.8$\pm$0.1 & 1.1$\pm$0.1 & 0.5  \\
    
    \#13   & 2.7$\pm$0.1  & \phantom{0}9.9$\pm$0.1 & 1.2$\pm$0.1 & 2.1 & 0.4$\pm$0.1 & \phantom{0}9.8$\pm$0.1 & 0.6$\pm$0.1 & 0.6  \\
    
    \#14   & 1.1$\pm$0.1 & \phantom{0}6.8$\pm$0.1 & 1.7$\pm$0.1 & 0.6 & 0.1$\pm$0.1 & \phantom{0}6.6$\pm$0.1 & 0.7$\pm$0.1 & 0.1  \\

\bottomrule \vspace{-0.6cm}
    \end{tabular}
    \tablefoot{$W$ refers to velocity-integrated line intensity, $\int T_\mathrm{mb}\;dv$.  $T_\mathrm{mb}$ is the line peak temperature.}
    \end{threeparttable}
\end{table*}


\begin{table*}[!h]

\caption{Line spectroscopic parameters obtained from Gaussian fits to the observed H$^{13}$CN and HN$^{13}$C lines.\vspace{-0.3cm}} 
\label{tab:isotoJfit}

    \centering
   
    \begin{threeparttable}
   
    \begin{tabular}{lcccccccc}
    \toprule
        &   \multicolumn{4}{c}{H$^{13}$CN \textit{J}=2--1} & \multicolumn{4}{c}{HN$^{13}$C \textit{J}=2--1} \\  
    Freq. [MHz] & \multicolumn{4}{c}{172677.9} & \multicolumn{4}{c}{174179.4} \\  
    Pos.    & $W$ & $v_\mathrm{LSR}$ & $\Delta v^\dagger$ & $T_\mathrm{mb}$ & $W$ & $v_\mathrm{LSR}$ & $\Delta v$ & $T_\mathrm{mb}$  \\  
       & [\Kkms] & [\kms] & [km s$^{-1}$] & [K] & [\Kkms] & [\kms] & [km s$^{-1}$]  & [K]  \\  
    \midrule
    \vspace{0.05cm} \#1 & 8.0 & 10.7 & 4.3 & 3.3 & 0.6$\pm$0.1 & 10.9$\pm$0.1 & 1.2$\pm$0.1 & 0.5   \\

     \#2 & 0.2 & 7.4 & 3.7 &  &  & & &  \\  
     
     \#3 & 0.67 & 9.72 & 3.6 & &  & & &   \\  
     
     \#4 & 0.06 & 2.4 & 3.0 & 0.05 & 0.1$\pm$0.1 & 2.4$\pm$0.1 & 0.4$\pm$0.1 & 0.2  \\  
     
     PDR HH  & 0.08$\pm$0.02 & 10.4$\pm$0.1 & 0.9$\pm$0.2 & 0.08  & 0.1 & 10.7$\pm$0.2 & 1.2$\pm$0.8 & 0.1   \\  
     
     \#6 & 0.08 & 11.3 & 2.1 & 1.0 &  & & &  \\  
     
     \#7 & 0.18 & 9.7 & 5.8 &  & 0.44$\pm$0.02 & 10.80$\pm$0.02 & 0.64$\pm$0.04 & 0.6 \\  
     
     \#8 & 0.15 & 12.1 & 3.9 & & 0.45$\pm$0.02 & 12.2$\pm$0.02 & 0.71$\pm$0.04 & 0.59 \\ \
    
     Core HH & 0.1 & 11.0 & 4.0 &  & 0.07$\pm$0.01 & 10.62$\pm$0.09 & 0.4$\pm$0.4 & 0.2 \\  
    
     \#10 &  & & & & 0.049$\pm$0.008 &  & &  \\  
    
     \#11 & --  & -- & -- & -- & --  & -- & -- & --  \\

     \#12  &  & & & & 0.09$\pm$0.01 & 11.94$\pm$0.06 & 0.8$\pm$0.1 & 0.11 \\  
     \#13 &  & & & & 0.03$\pm$0.006 & 9.8$\pm$0.03 & 0.3$\pm$0.1 & 0.08 \\  
     \#14 &  & & & &  & & & \\
\bottomrule \vspace{-0.6cm}  
    \end{tabular}
    
    \tablefoot{$W$ refers to velocity-integrated line intensity, $\int T_\mathrm{mb}\;dv$.  $T_\mathrm{mb}$ is the line peak temperature. 
    $\Delta v^\dagger$ refers to Moment 2 of the $J$=2--1 line emission (the individual HFS component overlap).}
    \end{threeparttable}
\end{table*}


\end{appendix}

\end{document}